\documentclass{JHEP3}
\pdfoutput=1
\usepackage[english]{babel}
\usepackage{amssymb}
\usepackage{amsfonts}
\usepackage{amsmath}
\usepackage{cite}
  \newcommand{\cL}{{\cal L}}
  
\newcommand{\cO}{{\cal O}}

\newcommand{\be}{\begin{equation}} \newcommand{\ee}{\end{equation}}
\newcommand{\bea}{\begin{eqnarray}} \newcommand{\eea}{\end{eqnarray}}
\newcommand{\beann}{\begin{eqnarray*}}  \newcommand{\eeann}{\end{eqnarray*}}
\newcommand{\bfig}{\begin{figure}} \newcommand{\efig}{\end{figure}}
\newcommand{\ba}{\begin{array}} \newcommand{\ea}{\end{array}}
\newcommand{\bcen}{\begin{center}} \newcommand{\ecen}{\end{center}}
\newcommand{\btab}{\begin{tabular}} \newcommand{\etab}{\end{tabular}}

\def\tr{\operatorname{tr\:}}

\newtheorem{Proposition}{Proposition}[section]

\newtheorem{Theorem}{Theorem}[section]
\newtheorem{Lemma}{Lemma}[section]
\newtheorem{Corrolary}{Corrolary}[section]

\newcommand{\bp}{\begin{Proposition}}	\newcommand{\ep}{\end{Proposition}}
\newcommand{\bt}{\begin{Theorem}}	\newcommand{\et}{\end{Theorem}}
\newcommand{\bl}{\begin{Lemma}}		\newcommand{\el}{\end{Lemma}}
\newcommand{\bc}{\begin{Corrolary}}	\newcommand{\ec}{\end{Corrolary}}
\title{Holographic Gravitational Anomaly in First and Second Order Hydrodynamics}

\author{Eugenio Meg\'{\i}as$^{a}$,  
Francisco Pena-Benitez$^{b,c}$\\
$^{a}$Grup de F\'{\i}sica Te\`orica and IFAE, Departament de F\'{\i}sica, \\
Universitat Aut\`onoma de Barcelona, Bellaterra E-08193 Barcelona, Spain\\
E-mail: \email{emegias@ifae.es}\\
$^{b}$Instituto de F\'{\i}sica Te\'orica UAM/CSIC, C/ Nicol\'as Cabrera 13-15,\\
Universidad Aut\'onoma de Madrid, Cantoblanco E-28049 Madrid, Spain\\
$^{c}$Departamento de F\'isica Te\'orica, Universidad Aut\'onoma de Madrid, Cantoblanco E-28049 Madrid, Spain\\
E-mail: \email{fran.penna@uam.es}
}
\abstract{We compute, in the framework of the fluid/gravity
  correspondence, the transport coefficients of a relativistic fluid
  affected by chiral and gauge-gravitational anomalies, including
  external electromagnetic fields. The computation is performed at
  first and second order in the hydrodynamical expansion. We use a
  5-dim holographic model with pure gauge and mixed
  gauge-gravitational Chern-Simons terms in the action. We reproduce
  at first order previous results on the anomaly induced current of a
  magnetic field and a vortex in a relativistic fluid, and compute at
  second order the anomalous and non anomalous transport coefficients
  by using a Weyl covariant formalism. We find a dissipative and
  anomalous correction to the chiral magnetic conductivity due to the
  time dependence of the magnetic field. We also find a new
  contribution from the mixed gauge-gravitational anomaly to the shear
  waves dispersion relation. The role played by the chiral and
  gravitational anomalies in other transport coefficients is
  discussed.}

\preprint{FTUAM-13-9\\IFT-UAM/CSIC-13-039\\UAB-FT-734}

\begin{document}

\section{Introduction}
\label{sec:intro}

Standard thermodynamics assumes thermodynamical equilibrium, implying that the intensive parameters (pressure, temperature and chemical potential)  are constant along the volume of the system. Furthermore it is always possible to find a frame in which the total momentum of the system vanishes. In order to study systems in more interesting regimes one can allow  the thermodynamical parameters to vary in space and time taking the system out of equilibrium. However, we assume local thermodynamical equilibrium which means that the variables vary slowly in space and time. This approximation, also called hydrodynamical approach, makes sense when the mean free path of the particles is much shorter than the characteristic size or length of the system $l_{mfp}\ll L$ \cite{Kovtun:2012rj}.

The modern understanding of hydrodynamics is based on the effective field theory formalism. The hydrodynamical systems should obey the (anomalous) conservation laws of the spin one currents and the energy-momentum tensor, which are supplemented by expressions of the current and the energy-momentum tensor in terms of the quantities in the fluid, the so-called constitutive relations. These relations can be written as
\begin{eqnarray}
\langle T^{\mu\nu} \rangle &=& \left(\epsilon + p\right) u^\mu u^\nu + p g^{\mu\nu} + \langle T^{\mu\nu} \rangle_{\textrm{\scriptsize diss \& anom}} \,, \label{eq:cfT} \\
\langle J^\mu \rangle &=& n u^\mu + \langle J^\mu \rangle_{\textrm{\scriptsize diss \& anom }} \,. \label{eq:crJ}
\end{eqnarray}
Here $\epsilon$ is the energy density, $p$ the pressure, $n$ the charge density and $u^\mu$ the local fluid velocity. In addition to the equilibrium  contributions, there are extra terms in the constitutive relations which lead to dissipative and anomalous effects. These terms are usually computed in the long wavelength approximation, so that they are organized in a derivative expansion, also called hydrodynamical expansion.  Some examples of dissipative coefficients are the shear viscosity~$\eta$, bulk viscosity~$\zeta$ and electric conductivity~$\sigma$ (see e.g.\cite{Kovtun:2012rj,Policastro:2001yc,Arnold:2006fz} and references therein).

During the past few years a new set of transport coefficients has been discovered as a consequence of chiral anomalies. The axial anomaly of QED is responsible for two particularly interesting effects of strong magnetic fields in dense strongly interacting matter. At large quark chemical potential $\mu$, chirally restored quark matter gives rise to an axial current parallel to the magnetic field \cite{Son:2004tq,Metlitski:2005pr,Newman:2005as}
\be\label{J5}
\mathbf J_5=\frac{eN_c}{2\pi^2}\mu \,\mathbf B \,,
\ee
which may indeed lead to observable effects in strongly magnetized neutron stars and heavy ion collisions \cite{Charbonneau:2009ax,KerenZur:2010zw}. This phenomena is known as {\it chiral separation effect} (CSE).

In the context of heavy ion collisions it was argued in
\cite{Kharzeev:2007jp,Fukushima:2008xe} that the excitation of
topologically non-trivial gluon field configurations in the early
non-equilibrium stages of a heavy ion collision might lead to an
imbalance in the number of left- and right-handed quarks. This
situation can be modelled by an axial chemical potential.~\footnote{As
  soon as thermal equilibrium is reached, this imbalance is frozen and
  it is modelled by a chiral chemical potential, at least as long as the
  electric field is zero.} During the collision one expects the
generation of magnetic fields that momentarily exceed even those found
in magnetars. It has been proposed by Kharzeev et
al.\ \cite{Kharzeev:2004ey,Kharzeev:2007tn,Kharzeev:2007jp,Fukushima:2008xe,Kharzeev:2009pj}
that the analogous effect, so-called {\it chiral magnetic effect}
(CME)~\cite{Alekseev:1998ds} \be\label{CME} \mathbf
J=\frac{e^2N_c}{2\pi^2}\mu_5 \,\mathbf B \,, \ee where $\mathbf J$ is
the electromagnetic current and $\mu_5$ the axial chemical potential,
could render observable event-by-event P and CP violations. Indeed,
there is recent experimental evidence for the CME in the form of
charge separation in heavy ion collisions with respect to the reaction
plane \cite{Abelev:2009uh,Voloshin:2008jx}, and more recently from LHC
data \cite{Abelev:2012pa} (see however
\cite{Wang:2009kd,Asakawa:2010bu}). For lattice studies of this effect,
see for example \cite{Buividovich:2009wi,Buividovich:2010tn}. In the context of holography the CME was under an intense discussion to confirm its presence at strong coupling \cite{Yee:2009vw,Lifschytz:2009si,Rebhan:2009vc,Gynther:2010ed,Brits:2010pw}.

The fluid/gravity correspondence~\cite{Bhattacharyya:2008jc} is a very powerful tool to understand the hydrodynamic regime of quantum field theories with holographic dual. This technique has contributed to the understanding of the positivity of the entropy production using techniques of black hole thermodynamics \cite{Bhattacharyya:2008ks,Loganayagam:2008is,Chapman:2012my}. It is also very useful for the computation of transport coefficients. The application of the fluid/gravity correspondence to theories including chiral anomalies~\cite{Erdmenger:2008rm,Banerjee:2008th} lead to another surprise: it was found that not only a magnetic field induces a current but that also a vortex in the fluid leads to an induced current, the latter is called {\it chiral vortical effect} (CVE).~\footnote{A generalization of the model of \cite{Erdmenger:2008rm,Banerjee:2008th} to a Maxwell-Gauss-Bonnet gravity has been done in~\cite{Hu:2011qa}, where some corrections of the transport coefficients induced by the Gauss-Bonnet coupling have been computed.} Again it is a consequence of the presence of  chiral anomalies.  It was later realized that the chiral magnetic and vortical conductivities are almost completely fixed in the hydrodynamic framework by demanding the existence of an entropy current with positive definite divergence~\cite{Son:2009tf}. That this criterion did not fix completely the anomalous transport coefficients was noted in~\cite{Neiman:2010zi}, and various terms depending on the temperature instead of the chemical potentials were shown to be allowed as undetermined integration constants. The contributions from pure gauge anomalies is fixed uniquely by this method and provides therefore a non-renormalization theorem~(see however \cite{Gorbar:2013upa} for a discussion on radiative corrections to the CSE). 

Using a Kubo formula for the chiral vortical conductivity in a system of fermions at the weakly coupled regime, a purely temperature dependent contribution was found. This contribution was consistent with the integration constants found in \cite{Neiman:2010zi} and it was shown to arise if and only if the system of chiral fermions features a mixed gauge-gravitational anomaly \cite{Landsteiner:2011cp}. The gravitational anomaly contribution to the chiral vortical effect was also established in a strongly coupled AdS/CFT approach and precisely the same result as at weak coupling was found \cite{Landsteiner:2011iq}. Some evidence of this effect has been found recently also from lattice studies~\cite{Braguta:2013loa}.
  
Some very recent attempts to establish a non-renormalization theorem for anomalous conductivities lead to the fact that the chiral vortical conductivity indeed renormalizes due to gluon fluctuations \cite{Golkar:2012kb,Hou:2012xg}. On the other hand it has been studied in~\cite{Bai:2012ci,Landsteiner:2012dm} the ultraviolet cutoff dependence of the anomalous transport coefficients and their holographic flow. 

In~\cite{Jensen:2012kj} the authors claim that the gravitational anomaly produces a Casimir momentum in the cone formed by the space-time with imaginary time, which breaks the derivative counting and it is the responsible that first order transport coefficients being fixed by the mixed gravitational anomaly.  The gravitational anomaly contribution was confirmed also in a fluid/gravity context~\cite{Chapman:2012my}, in a weakly coupled gas of Weyl fermions in arbitrary dimensions~\cite{Loganayagam:2012pz}, and it was found in~\cite{Loganayagam:2011mu} that the anomalous conductivities can be obtained directly from the anomaly polynomial substituting the field strength with the chemical potential and the first Pontryagin density by the negative of the temperature squared. Recently the anomalous conductivities have also been obtained in effective action approaches~\cite{Sadofyev:2010is,Lin:2011aa,Dubovsky:2011sk,Jensen:2012jy,Banerjee:2012iz,Jain:2012rh,Valle:2012em,Kirilin:2012mw,Banerjee:2013qha} and using group theory techniques \cite{Nair:2011mk}.

Stability and causality issues of the hydrodynamic equations demand the knowledge of second order hydrodynamics~\cite{Israel:1976tn,Israel:1979wp,Hiscock:1983zz}. A classification of the terms contributing to this order was presented in~\cite{Kharzeev:2011ds}. In this work we compute within the fluid/gravity correspondence the transport coefficients at first and second order in the hydrodynamical expansion, using an holographic model which includes both gauge and mixed gauge-gravitational anomalies and external electromagnetic fields.

The manuscript is organized as follow. In section~\ref{sec:holo_model} we define our holographic model and present the renormalized action and the equations of motion. In section~\ref{sec:holo_higher_der} we perform a formal derivation of the one point functions for a general Lagrangian, either in consistent and in covariant form. We review in section~\ref{sec:constitutives} the first and second order hydrodynamical Weyl covariant formalism, and present our main results.

In section~\ref{sec:fluid_gravity} we explain the method to compute the transport coefficients  within the fluid/gravity formalism. We present our result of the transport coefficients to first and second order in sections~\ref{sec:first_order} and~\ref{sec:second_order} respectively. Finally we conclude with a discussion of our results and an outlook towards possible future directions in section~\ref{sec:discussion}. The full expressions for the sources and transport coefficients at second order are collected in the appendices.

\section{Holographic Model}
\label{sec:holo_model}

The model we will use here was presented
in~\cite{Landsteiner:2011iq}. We will fix first our conventions. We
choose the five dimensional metric to be of signature
$(-,+,+,+,+)$. The epsilon tensor has to be distinguished from the
epsilon symbol, the latter being defined by $\epsilon(rtxyz) = +1$
whereas the former is defined by
$\epsilon_{ABCDE}=\sqrt{-g}\,\epsilon(ABCDE)$. Five dimensional
indices are denoted with upper case latin letters. We define an
outward pointing normal vector $n_A \propto g^{AB} \frac{\partial
  r}{\partial x^B}$ to the holographic boundary of an asymptotically
$AdS$ space with unit norm $n_A n^A =1$. The action is given by
\bea\nonumber S &=& \frac{1}{16\pi G} \int d^5x \sqrt{-g} \left[ R +
  12 - \frac 1 4 F_{MN} F^{MN} \right.\\ &&\left.+ \epsilon^{MNPQR}
  A_M \left( \frac\kappa 3 F_{NP} F_{QR} + \lambda R^A\,_{BNP}
  R^B\,_{AQR} \right) \right] + S_{GH} + S_{CSK}
\,, \label{eq:S}\\ S_{GH} &=& \frac{1}{8\pi G}
\int_{\partial_\epsilon} d^4x \sqrt{-h} \, K
\,, \label{eq:S_GH}\\ S_{CSK} &=& - \frac{1}{2\pi G}
\int_{\partial_\epsilon} d^4x \sqrt{-h} \, \lambda n_M
\epsilon^{MNPQR} A_N K_{PL} D_Q K_R^L \,, \label{eq:S_CSK} \eea where
$S_{GH}$ is the usual Gibbons-Hawking boundary term and $D_A$ is the
induced covariant derivative on the four dimensional cut-off surface. The
second boundary term $S_{CSK}$ was motivated in
\cite{Landsteiner:2011iq} (see also \cite{Landsteiner:2012dm}).
Notice that the action is diffeomorphism invariant, the Chern Simons
terms are well formed volume forms and as such they are diffeomorphism
invariant. They do depend however explicitly on the gauge connection
$A_M$.  Under gauge transformations $\delta A_M = \nabla_M \xi$ they
are therefore invariant only up to a boundary term. This model needs a
counterterm in order to make the on-shell boundary action well
defined~\footnote{It has been proved in \cite{Landsteiner:2011iq} that
  the gravitational Chern-Simons term does not introduce new
  divergences in the system if the space is asymptoticaly~$AdS$.}
\begin{equation}
S_{ct} = - \frac{1}{16\pi G} \int_{\partial_\epsilon} d^4x \sqrt{-h} \bigg[
6 + 3P  - \left( P^\mu_\nu P^\nu_\mu - P^2 -  \frac{1}{4} \hat{F}_{\mu\nu} \hat{F}\,^{\mu\nu} \right)\log \epsilon \bigg] \,, \label{eq:Sct}
\end{equation}
%la relacion entre el r del paper anterior y en este paper es asi $r_N=e^{r_{old}}$, por eso desaparece el factor 1/2
where
\begin{equation}
P = \frac{\hat{R}}{6} \,, \qquad  P^\mu_\nu = \frac{1}{2} \left[ \hat{R}^\mu_\nu - P \delta^\mu_\nu \right] \,. \label{eq:PPij} 
\end{equation}
Quantities with hat ($\hat{F}, \hat{R}, \dots$) refer to their induced four dimensional objects at the cut-off surface, which is located at the radius $r \sim 1/\epsilon$. So taking the limit $\epsilon\to 0$, one takes the surface to the $AdS$ boundary.

The bulk equations of motion $E_{MN}=0$ and $M^D=0$ are
\begin{eqnarray}
G_{MN}  + \left(\frac 1 8 F^2- 6\right) g_{MN} - \frac 1 2 F_{ML} F_N\,^L  - 2 \lambda \epsilon_{LPQR(M} \nabla_B\left( F^{PL} R^B\,_{N)}\,^{QR} \right) &=& 0 \,, \label{eq:Gbulk}\\
\nabla_NF^{ND} + \epsilon^{DNPQR} \left( \kappa F_{NP} F_{QR} + \lambda  R^A\,_{BNP} R^B\,_{AQR}\right) &=&0\,,  \label{eq:Abulk}
\end{eqnarray}
and they are gauge and diffeomorphism covariant.

\section{One point functions and Ward identities}
\label{sec:holo_higher_der}

After an ADM decomposition it is possible realize that the action~(\ref{eq:S}) is third order in $r$ derivatives (see Ref. \cite{Landsteiner:2011iq}), so in order to get the correct one point functions we have to take into account this fact and the assumption that the bulk space is asymptotically anti-de Sitter. Asymptotically $AdS$ is enough to get a well defined boundary value problem just in terms of the field boundary theory sources. Let us analyze now what this implies for a general Lagrange density.

\subsection{The holographic dictionary with higher derivatives}

Let us assume a general renormalized Lagrangian for an arbitrary set  of fields that we will call $\phi$.  After the four dimensional ADM decomposition one has, 
$$
S = \int  \, d^4x\, dr\,\cL ( \phi,  \dot\phi, D_\mu \phi,D_\mu\dot\phi,\ddot{\phi} ) \,,
 $$
where dot indicates derivative with respect to the radial coordinate. A general variation of the action leads now to
\be
\delta S = \int \,d^4x\, dr\,\left[ \frac{\partial \cL}{\partial \phi} \delta \phi + \frac{\partial \cL}{\partial \dot\phi} \delta \dot\phi+ 
\frac{\partial \cL}{\partial (D_\mu\phi)} \delta(D_\mu
\phi) 
+ \frac{\partial \cL}{\partial (D_\mu\dot\phi)} \delta(D_\mu\dot
\phi)
+ \frac{\partial \cL}{\partial \ddot\phi} \delta \ddot\phi \right] \,.
\ee
Through a series of partial integrations we can bring this into the
following form,
\begin{equation}
\delta S = \int\, d^4x\,dr\, \textrm{E.O.M.} \, \delta\phi + \int_{\partial_\epsilon} \,d^4x\, \left[ \left( 
\frac{\partial \cL}{\partial \dot\phi} -  D_\mu \left( \frac{\partial \cL}{\partial
      (D_\mu\dot\phi)} \right) - \left( \frac{\partial
      \cL}{\partial \ddot\phi} \right)^.
\right) \delta\phi 
+   \frac{\partial
      \cL}{\partial \ddot\phi}\delta\dot{\phi}  \right]\,.
\end{equation}
The bulk terms are the equations of motion. For a generic boundary, the
form of the variation shows that Dirichlet boundary conditions can not
be imposed.  Vanishing of the action rather imposes a relation between
$\delta\phi$ and $\delta\dot\phi$.

If we have applications of holography in mind, there is however another
way of dealing with the boundary term. We suppose now that we are
working in an asymptotically anti-de Sitter space. The field $\phi$
has therefore a boundary expansion 
$$
\phi = e^{(\Delta-4)r} \phi^{(0)} + \mathrm{subleading}\,,
$$
here $\Delta$ is the dimension
(conformal weight) of the operator that is sourced by
$\phi^{(0)}$. Since this is a generic property of holography in asymptotically  $AdS$
spaces, we can relate the derivative of the variation to the variation
itself,
$$
\delta\dot\phi = (\Delta-4) e^{(\Delta-4)r}\delta \phi^{(0)} +
\mathrm{subleading}\,.
$$ 
Using this and the fact that the one point function of the
consistent operator $\cO_ \phi$ is defined as the variation of the
on-shell action with respect to the source $\phi^{(0)}$, we find \be
\label{curr}
\sqrt{-h^{(0)}}\cO_\phi = \lim_{\epsilon\rightarrow 0} \epsilon^{(\Delta-4)r} \left[ 
\frac{\partial \cL}{\partial \dot\phi} -  D_\mu \left( \frac{\partial \cL}{\partial
      (D_\mu\dot\phi)} \right) - \frac{d}{dr}\left( \frac{\partial
      \cL}{\partial \ddot\phi} \right)
+(\Delta-4) \left( \frac{\partial
      \cL}{\partial \ddot\phi}\right) 
\right] \,.
\ee
Without loss of generality we can evaluate this in Gaussian normal coordinates where the metric takes the form $ds^2 = dr^2 + h_{\mu\nu}
dx^\mu dx^\nu$, and in the gauge $A_r=0$.  The gauge variation of the
action depends only on the intrinsic four dimensional curvature of the
boundary. From this we can compute the ``bare'' consistent U(1) current and energy-momentum tensor, and the result is
\begin{eqnarray}
 16\pi G J_{(c)}^\mu &=&  -\frac{\sqrt{-h}}{\sqrt{-h^{(0)}}}\left[ F^{r\mu} + \frac{4}{3}\kappa\epsilon^{\mu\nu\rho\lambda} A_\nu \hat F_{\rho\lambda} \right]_\epsilon\,, \label{eq:Jc}\\
 8\pi G T^{\mu\nu} _{(c)} &=& \frac{\sqrt{-h}}{\sqrt{-h^{(0)}}}\left[ K^{\mu\nu} - K \gamma^{\mu\nu} + 4\lambda \epsilon^{(\mu \alpha\beta\rho} \left( \frac{1}{2}\hat{F}_{\alpha\beta}\hat{R}^{\nu)}_\rho + D_\delta(A_{\alpha} \hat{R}^{\delta\nu)}\,_{\beta\rho}) \right) \right] _\epsilon \,. \label{eq:Tc}
\end{eqnarray}
Now taking the divergence of these expressions and using the equations of motion, we get the anomalous charge conservation and the energy-momentum conservation relations respectively, 
\begin{eqnarray}
\label{eq:dJ} D_\mu J_{(c)}^\mu &=& -\frac{1}{16\pi G} \epsilon^{\mu\nu\rho\lambda} \left( \frac\kappa 3\hat F_{\mu\nu}\hat F_{\rho\lambda}+ \lambda \hat R^\alpha\,_{\beta\mu\nu}\hat R^\beta\,_{\alpha\rho\lambda} \right) \,,\\
\label{eq:dT} D_\mu T_{(c)}^{\mu\nu} &=& -J_{(c)\mu} \hat F^{\mu\nu} + A^\nu D_\mu J_{(c)}^\mu \, .
\end{eqnarray}
These are precisely the consistent Ward identities for a theory
invariant under diffeomorfisms with a mixed gauge gravitational
anomaly. A good general reference for anomalies is Bertlmann's book
\cite{Bertlmann:1996xk} where the consistent form of the anomaly for
chiral fermions transforming under a $U(1)_{L}$ symmetry group is
quoted as
\begin{equation}\label{eq:consistent}
D_\mu J_{(c)}^\mu  =\frac{1}{96\pi^2} \epsilon^{\mu\nu\rho\lambda} F_{\mu\nu} F_{\rho\lambda} + \frac{1}{768\pi^2}\epsilon^{\mu\nu\rho\lambda} \hat R^\alpha\,_{\beta\mu\nu}\hat R^\beta\,_{\alpha\rho\lambda}  \,.
\end{equation}
We use this to fix $\kappa$ and $\lambda$ to the anomaly coefficients for a single chiral fermion transforming under a $U(1)_L$ symmetry, therefore
\begin{equation}
-\frac{\kappa}{48 \pi G} = \frac{1}{96 \pi^2} \qquad \,, \qquad -\frac{\lambda}{16\pi G} = \frac{1}{768 \pi^2} \,.
\end{equation}

\subsection{Covariant form of the current and energy-momentum tensor}

We have computed the currents as the derivative of the field theory
quantum action, and the anomaly is therefore in the form of the
consistent anomaly. Since we are dealing only with a single $U(1)$
symmetry, the (gauge) anomaly is automatically expressed in terms of
the field strength. However it is always possible to add a
Chern-Simons current and to redefine the charge current $J^\mu
\rightarrow J^\mu + c \epsilon^{\mu\nu\rho\lambda}A_\nu
F_{\rho\lambda}$, and the energy-momentum tensor $T^{\mu\nu}
\rightarrow T^{\mu\nu} + c' \epsilon^{\alpha(\mu \rho\lambda}D_\beta
\left( A_\alpha R^{\beta \nu)}\,_{\rho\lambda}\right)$. These redefined
quantities can not be expressed as the variation of a local functional
of the fields with respect to the gauge and metric fields
respectively. In particular the so-called covariant form of the
anomaly differs precisely in such a redefinition of the
current.~\footnote{Note that the  approaches used in~\cite{Son:2009tf,Neiman:2010zi} and in subsequent works,  typically make use of the covariant form of the anomaly.} 

Adding such a terms to the consistent current and energy-momentum tensor~(\ref{eq:Jc})-(\ref{eq:Tc}), we can write the covariant expressions for these quantities which are the ones we will use to construct the hydrodynamical constitutive relations in the fluid/gravity approach,
\begin{eqnarray}
16\pi G J^\mu &=&  -\frac{\sqrt{-h}}{\sqrt{-h^{(0)}}} F^{r\mu} |_\epsilon \,, \label{eq:Jcov}\\
 8\pi G T^{\mu\nu} &=& \frac{\sqrt{-h}}{\sqrt{-h^{(0)}}}\left[ K^{\mu\nu} - K h^{\mu\nu} + 2\lambda \epsilon^{(\mu \alpha\beta\rho} \hat{F}_{\alpha\beta}\hat{R}^{\nu)}_\rho  \right]_\epsilon  \,. \label{eq:Tcov}
\end{eqnarray}

\section{Constitutive relations, derivative expansion and Weyl covariance}
\label{sec:constitutives}

Some notions on conformal/Weyl covariant formalism are needed to construct the constitutive relations up to second order (for a detailed explanation see~\cite{Loganayagam:2008is}). A conformal fluid has to be invariant under the change
\begin{equation}
g_{\mu\nu}\to e^{-2\phi(x)}g_{\mu\nu} \,,
\end{equation}
where $\phi(x)$ is an arbitrary function. We will say that a tensor is Weyl convariant with weight $w$ if it transforms as \begin{equation}
Q^{\alpha\beta\ldots}_{\mu\nu\ldots}\to e^{w\phi(x)} Q^{\alpha\beta\ldots}_{\mu\nu\ldots} \,.
\end{equation}

The consequences of conformal symmetry on hydrodynamics is that the energy momentum tensor and (non)-conserved currents have to be covariant under Weyl transformations and the energy momentum has to be traceless modulo contributions from Weyl anomaly. To construct Weyl covariant quantities it is necessary to introduce the Weyl connection
\begin{equation}
\mathcal{A}_\mu=u^\nu D_\mu u_\nu -\frac{1}{3}D_\nu u^\nu \,,
\end{equation}
and the Weyl covariant derivative
\begin{eqnarray}
\nonumber\mathcal{D}_\lambda Q^{\mu\ldots}_{\nu\ldots} &=& D_\lambda Q^{\mu\ldots}_{\nu\ldots} - w \mathcal{A}_\lambda Q^{\mu\ldots}_{\nu\ldots} +\\
\nonumber &&+\left[g_{\lambda\alpha}\mathcal{A}^\mu - \delta^\mu_\lambda\mathcal{A}_\alpha - \delta^\mu_\alpha\mathcal{A}_\lambda\right]Q^{\alpha\ldots}_{\nu\ldots}+\ldots \\
&&-  \left[g_{\lambda\nu}\mathcal{A}^\alpha - \delta^\alpha_\lambda\mathcal{A}_\nu - \delta^\alpha_\nu\mathcal{A}_\lambda\right]Q^{\mu\ldots}_{\alpha\ldots}+\ldots \,.
\end{eqnarray}
We show in table~\ref{tab:pesos} the Weyl weights of some of the hydrodynamical variables. It is possible to reduce in a systematic way the number of independent sources contributing to the constitutive relations by imposing Weyl covariance and the hydrodynamical equations of motion (Ward idetities). A classification in the so called Landau frame  of all the possible terms that can appear in the energy-momentum tensor and U(1) current has been done up to second order in~\cite{Kharzeev:2011ds,Erdmenger:2008rm,Banerjee:2008th}. The Ward identities  in four dimensions in presence of quantum anomalies  are shown in (\ref{eq:dJ}) and (\ref{eq:dT}). The curvature part has been usually neglected in the literature as it is fourth order in derivatives and the expansion is usually done up to second order. But it was shown in~\cite{Landsteiner:2011cp,Landsteiner:2011iq} that the gravitational anomaly indeed fixes part of the transport coefficients at first order. Actually in \cite{Jensen:2012kj} it was understood why the derivative expansion breaks down in presence of the gravitational anomaly.
\begin{table}[h]
\begin{center}
\begin{tabular}{|c|c|}
\hline
Field & weight \\
\hline
$\mu$, $T$, $u^\mu$ & 1 \\
$g_{\mu\nu}$ & -2 \\
$p$ & 4 \\
$n$, $E^\mu$, $B^\mu$  & 3 \\
\hline
\end{tabular}
\caption{Weyl weights for the chemical potential, temperature, fluid velocity, metric, pressure, charge density, electric field and magnetic field.}
\label{tab:pesos}
\end{center}
\end{table}

With these ingredients we can write down the constitutive relations in the Landau frame
\begin{eqnarray}
T^{\mu\nu} &=& p(4u^\mu u^\nu +\eta^{\mu\nu}) + \tau_{(1)}^{\mu\nu}  + \tau_{(1)\textrm{ano}}^{\mu\nu} + \tau_{(2)}^{\mu\nu} + \tau_{(2)\textrm{ano}}^{\mu\nu} \,, \label{eq:fullconstiT}\\
J^\mu &=& n u^\mu + \nu_{(1)}^{\mu}  + \nu_{(1)\textrm{ano}}^{\mu} + \nu_{(2)}^{\mu} + \nu_{(2)\textrm{ano}}^{\mu} \,, \label{eq:fullconstiJ}
\end{eqnarray}
where we have split the expressions in the equilibrium, first order and second order (anomalous + non anomalous) parts. Weyl invariance implies the equation of state $\epsilon = 3p$ and the vanishing of the bulk viscosity $\zeta=0$.~\footnote{As it has been discussed in~\cite{Kharzeev:2011ds} the anomalous terms in the constitutive relations are those whose transport coefficients have $(C, P) = (\pm 1, -1)$, and they correspond to the ones containing odd powers in the anomaly coefficients $\kappa$ and $\lambda$. See this reference for a systematic classification of anomalous and non anomalous terms.} The subindex in parenthesis indicates the order in the derivative expansion. The ambiguity in the definition of temperature, chemical potential and fluid velocity which appears when the system is slightly out of equilibrium is fixed by using the Landau frame, in which it is demanded that $u^\mu\tau_{(n)\mu\nu}=0$ and $u^\mu\nu_{(n)\mu}=0$. Up to first order, the most general contributions to the conformal energy momentum tensor and $U(1)$ current are
\begin{eqnarray}
\tau_{(1)}^{\mu\nu}  &=& -2\eta\sigma^{\mu\nu}\,,  \qquad\qquad\qquad\qquad\, \tau_{(1)\textrm{ano}}^{\mu\nu}  =  0 \,, \\
 \nu_{(1)}^{\mu}  &=&  -\sigma\left(T P^{\mu\nu}\mathcal{D}_\nu\bar{\mu}-E^\mu\right)\,,  \quad \quad\, \nu_{(1)\textrm{ano}}^{\mu} =  \xi_V\omega^\mu + \xi_B B^\mu\,,
\end{eqnarray}
where we have defined the Weyl invariant quantity $\bar{\mu}=\mu/T$. In these expressions $\eta$, $\sigma$, $\xi_V$ and $\xi_B$ are the shear viscosity, electrical conductivity, chiral vortical and chiral magnetic conductivities respectively, while $\sigma^{\mu\nu}$, $\omega^\mu$, $E^\mu$ and $B^\mu$ are the shear tensor, vorticity, electric and magnetic field respectively, defined as
\begin{eqnarray}
\sigma_{\mu\nu} &=& \frac{1}{2}\left( \mathcal{D}_\mu u_\nu + \mathcal{D}_\nu u_\mu \right) \,, \\
\omega_{\mu\nu} &=&  \mathcal{D}_\mu u_\nu - \mathcal{D}_\nu u_\mu  \,, \\
\omega^{\mu} &=& \frac{1}{2}\epsilon^{\mu\nu\alpha\beta}u_\nu \omega_{\alpha\beta} \,,\\
E^\mu &=& F^{\mu\nu}u_\nu \,, \\
B^\mu &=& \frac{1}{2}\epsilon^{\mu\nu\alpha\beta}u_\nu F_{\alpha\beta} \,.
\end{eqnarray}
Finally $P^{\mu\nu}=h^{\mu\nu}+u^\mu u^\nu$ is the projector in the space orthogonal to the velocity field. From previous definitions one can easily prove that the strength tensor decomposes in the following way:
\be
F_{\alpha\beta} = u_\alpha E_\beta - u_\beta E_\alpha -\epsilon_{\alpha\beta\rho\mu}u^\rho B^\mu   \,.
\ee
The second order contributions to the constitutive relations are
\begin{eqnarray}
 \tau_{(2)}^{\mu\nu}  &=& \sum_{a=1}^{a=15} \Lambda_a \mathcal{T}^{(a)\mu\nu} \quad , 	\quad \tau_{(2)ano}^{\mu\nu}  = \sum_{a=1}^{a=8} \tilde\Lambda_a  \mathcal{\tilde T}^{(a)\mu\nu} \,, \label{eq:tau2}\\
 j_{(2)}^{\mu} &=&  \sum_{a=1}^{a=10}\xi_a\mathcal{J}^{(a)\mu}\quad , 	\quad   j_{(2)ano}^{\mu} = \sum_{a=1}^{a=5}\tilde\xi_a\mathcal{\tilde J}^{(a)\mu}\label{eq:nu2} \,,
 \end{eqnarray}
with the second order tensors defined as
\be
\begin{array}{ccc}
 \mathcal{T}^{(1)\mu\nu} = u^\alpha\mathcal{D}_\alpha\sigma^{\mu\nu} \,, & 
  \mathcal{T}^{(2)\mu\nu} = \sigma^{\langle\mu}\,_\gamma \sigma^{\nu\rangle\gamma}\,, &
   \mathcal{T}^{(3)} = \sigma^{\langle\mu}\,_\gamma \omega^{\mu\rangle\gamma}\,, \\
  \mathcal{T}^{(4)\mu\nu} = \omega^{\langle\mu}\,_\gamma \omega^{\nu\rangle\gamma} \,, &  \mathcal{T}^{(5)\mu\nu} = \mathcal{D}^{\langle\mu}\mathcal{D}^{\nu\rangle} \bar{\mu}\,, & \mathcal{T}^{(6)\mu\nu} = \mathcal{D}^{\langle\mu}\bar{\mu}\mathcal{D}^{\nu\rangle} \bar{\mu} \,,\\
 \mathcal{T}^{(7)\mu\nu} = \mathcal{D}^{\langle\mu} E^{\nu\rangle}\,,  &
 \mathcal{T}^{(8)\mu\nu} = E^{\langle\mu} \mathcal{D}^{\nu\rangle}\bar{\mu} \,,&
 \mathcal{T}^{(9)\mu\nu} = E^{\langle\mu} E^{\nu\rangle}  \,,\\
 \mathcal{T}^{(10)\mu\nu} = B^{\langle\mu} B^{\nu\rangle}  \,,&
  \mathcal{T}^{(11)\mu\nu} = \epsilon^{\gamma\delta\eta\langle\mu}u_\gamma B_\delta\sigma^{\nu\rangle}\,_\eta  \,,& 
  \mathcal{T}^{(12)\mu\nu}=\omega^{\langle\mu} B^{\nu\rangle}\,, \\
  \mathcal{T}^{(13)\mu\nu} = C^{\mu\alpha\nu\beta} P_{\alpha\beta} \,, &
  \mathcal{T}^{(14)\mu\nu} = \epsilon^{\mu\alpha\beta\gamma}\epsilon^{\nu\delta\eta\lambda}C_{\alpha\beta\delta\eta} u_\gamma u_\lambda \,, & 
  \mathcal{T}^{(15)\mu\nu}=\epsilon^{\langle\mu\gamma\delta\eta}C_{\gamma\delta}\,^{\nu\rangle\lambda}u_\eta u_\lambda\,,
\end{array}
\ee

\be\begin{array}{ccc}
\mathcal{\tilde T}^{(1)\mu\nu} = \mathcal{D}^{\langle\mu}\omega^{\nu\rangle} \,, &
\mathcal{\tilde T}^{(2)\mu\nu} = \omega^{\langle\mu} \mathcal{D}^{\nu\rangle}\bar{\mu}\,,  &
\mathcal{\tilde T}^{(3)\mu\nu} = \epsilon^{\gamma\delta\eta\langle\mu}\sigma^{\nu\rangle}\,_\eta u_\gamma \mathcal{D}_\delta \bar{\mu}  \,,\\ 
\mathcal{\tilde T}^{(4)\mu\nu} = \mathcal{D}^{\langle\mu} B^{\nu\rangle}  \,,&
\mathcal{\tilde T}^{(5)\mu\nu} = B^{\langle\mu} \mathcal{D}^{\nu\rangle}\bar{\mu}   \,,&
\mathcal{\tilde T}^{(6)\mu\nu} = E^{\langle\mu} B^{\nu\rangle} \,,\\
\mathcal{\tilde T}^{(7)\mu\nu} = \epsilon^{\gamma\delta\eta\langle\mu}\sigma^{\nu\rangle}\,_\eta u_\gamma E_\delta  \,,&
\mathcal{\tilde T}^{(8)\mu\nu}= \omega^{\langle\mu} E^{\nu\rangle} \,,& 
\end{array}
\ee
where $C^{\mu\alpha\nu\beta}$ is the conformal Weyl curvature tensor, $\Pi^{\mu\nu}\,_{\alpha\beta}= \frac{1}{2}\left(P^\mu_\alpha P^\nu_\beta + P^\nu_\alpha P^\mu_\beta -\frac{2}{3}P^{\mu\nu}P_{\alpha\beta}\right)$ is a transverse traceless projector and we use the notation $X^{\langle \mu \nu\rangle} = \Pi^{\mu\nu}\,_{\alpha\beta}X^{\alpha\beta}$. The second order vectors are
\be
\begin{array}{ccc}
 \mathcal{J}^{(1)\mu}=\sigma^{\mu\nu}\mathcal{D}_\nu\bar{\mu} \,, & \mathcal{J}^{(2)\mu}=\omega^{\mu\nu}\mathcal{D}_\nu\bar{\mu}  \,,& \mathcal{J}^{(3)\mu}=P^{\mu\nu}\mathcal{D}^\alpha\sigma_{\nu\alpha}\,, \\ \mathcal{J}^{(4)\mu}=P^{\mu\nu}\mathcal{D}^\alpha\omega_{\nu\alpha}  \,,&
\mathcal{J}^{(5)\mu}=\sigma^{\mu\nu}E_\nu \,,&
\mathcal{J}^{(6)\mu}=\omega^{\mu\nu}E_\nu \,,\\
 \mathcal{J}^{(7)\mu}=u^\nu\mathcal{D}_\nu E^\mu\,, &
 \mathcal{J}^{(8)\mu}=\epsilon^{\mu\nu\alpha\beta}u_\nu B_\alpha\mathcal{D}_\beta \bar{\mu} \,,  &
 \mathcal{J}^{(9)\mu}=\epsilon^{\mu\nu\alpha\beta}u_\nu E_\alpha B_\beta \,, \\
  \mathcal{J}^{(10)\mu}= \epsilon^{\mu\nu\alpha\beta}u_\nu \mathcal D_\alpha B_\beta \,,& &
\end{array}
\ee
\be
\begin{array}{ccc}
\mathcal{\tilde J}^{(1)\mu} = \sigma^{\mu\nu}\omega_\nu \,,&
 \mathcal{\tilde J}^{(2)\mu}=\sigma^{\mu\nu}B_\nu  \,,&
  \mathcal{\tilde J}^{(3)\mu}=\omega^{\mu\nu}B_\nu  \,,\\
 \mathcal{\tilde J}^{(4)\mu} =\epsilon^{\mu\nu\alpha\beta}u_\nu E_\alpha\mathcal{D}_\beta \bar{\mu} \,,& \mathcal{\tilde J}^{(5)\mu}=\epsilon^{\mu\nu\alpha\beta}u_\nu \mathcal{D}_\alpha E_\beta \,.& 
\end{array}
\ee
This classification of independent possible terms has been previously presented in~\cite{Kharzeev:2011ds}. We consider in this work a flat background metric, so we are neglecting those terms proportional to the conformal Weyl tensor. 

Here we write for completeness the first order transport coefficients for the anomalous holographic plasma computed in the literature,
\begin{eqnarray}
\eta &=& \frac{r_+^3}{16\pi G} \,, \qquad \sigma =\frac{\pi  r_+^7 T^2}{16 G m^2} \,,\\
\xi_B &=& -\frac{\sqrt{3} q \left(m+3 r_+^4\right) \kappa }{8\pi  G m  r_+^2}+\frac{\sqrt{3} \pi  q T^2 \lambda }{G m} \,, \quad  \xi_V = -\frac{3 q^2 \kappa }{4\pi G m  }+\frac{2 \pi  \left(2 q^2-r_+^6\right) T^2 \lambda }{G m r_+^2} \,,
\end{eqnarray}
where $m$, $q$ and $r_+$ are the mass, charge and radius of the outer horizon of the black hole, which we will define in more detail in section~\ref{sec:fluid_gravity}. We write also here our main results corresponding to the second order transport coefficients that are completely new. We express some of them as an expansion in the parameter $\bar{\mu}$ assuming that $\bar{\mu} \ll 1$,
\begin{eqnarray}
\tau_E &=& -\frac{\Lambda_7}{2\eta} =  \frac{\eta^2}{4p^2}\left(\frac{ \mu}{r_+}  + \frac{5  }{6}\left(\frac{\mu}{r_+}\right) ^3\right) \,,\\ 
 \Lambda_8 &=& -\frac{\eta}{r_+^2  } \left(\frac{1}{2\pi}-\frac{1}{12\pi^3}    (14 \log 2 -1)\bar{\mu}^2 + \mathcal{O}(\bar{\mu}^4)\right)  \, ,\\
 \Lambda_9 &=& \frac{\eta}{r_+^3}\left(\frac{11}{6 }+\frac{1}{2}\log 2 + \frac{1}{12\pi^2 }(7-8 \log 2)\bar{\mu}^2 + \mathcal{O}(\bar{\mu}^4)\right) \, ,\\
\nonumber \Lambda_{12} &=& -\frac{\eta \bar\mu }{r_+^2}\left(\frac{T}{r_+}  - \frac{16\kappa T^3\bar{\mu}^2}{Mr_+^3}+ \frac{128}{3\pi} \kappa  \lambda  (5-12\log 2)+\frac{256}{15\pi} \lambda ^2 (29+60\log 2) + \mathcal{O}(\bar{\mu}^2) \right) \, ,\\
\end{eqnarray}
\begin{eqnarray}
\tilde l_B &=&  -\frac{\Lambda_4}{2\eta} =\frac{1}{2\pi p G }\left(   \kappa\frac{  \mu ^2}{8 } + \lambda(\pi T)^2\right)\,,\\ 
 \tilde \Lambda_5 &=& -\frac{4\eta \bar{\mu}}{ \pi^2 r_+^2 }\left(\kappa(1-  2\log 2) + 2\lambda  (1+2\log 2) + \mathcal{O}(\bar{\mu}^2)\right)   \, ,\\
\tilde \Lambda_6 &=&  \frac{8\eta\bar\mu }{\pi r_+^3}\left(-\kappa  \log 2+\lambda  (5+2\log 2)  + \mathcal{O}(\bar{\mu}^2)\right)\, ,\\
\tilde \Lambda_{7} &=& -\frac{32\eta  }{r_+^2 }\lambda \,,\\
\tilde \Lambda_8 &=& \frac{\eta}{p^2}\left[\frac{r_+^4\bar\mu^2}{64(\pi G)^2 }\left(-   T^2 + \frac{ T^4 \bar\mu ^2}{3r_+^2 }\right)\kappa\right.\\
\nonumber && \left.+\frac{ r_+^6\lambda}{8\pi^2 G^2 } \left(-\log 2+\frac{5}{18\pi^2}  \bar\mu^2 (5+6 \log 2)+\frac{1}{216
\pi ^4}  \bar\mu^4 (199-726 \log 2) + \mathcal{O}(\bar{\mu}^6) \right)\right] \,,
\end{eqnarray}
\begin{eqnarray}
\xi_5 &=& - \frac{1}{16\pi G}\left(\frac{1}{2 }\log 2 + \frac{1}{12\pi^2}   \bar{\mu}^2 (11- 24 \log 2 ) + \mathcal{O}(\bar{\mu}^4) \right)  \, ,\\
\nonumber \xi_6 &=& -\frac{1}{16  \pi G}-\frac{17 \bar\mu^2}{192  \pi ^3G}+\frac{\kappa ^2 \bar\mu^2}{ \pi ^3G}-\frac{\log 2}{64  \pi G}+\frac{\bar\mu^2 \log 2}{48  \pi ^3G} +\lambda ^2 \left(\frac{1456 \bar\mu^2}{15  \pi ^3G}-\frac{16 \bar\mu^2 \log 2}{ \pi ^3G}\right)\\
 &&+\kappa  \lambda  \left(\frac{4}{G \pi }-\frac{130 \bar\mu^2}{9 G \pi ^3}+\frac{8 \log 2}{G \pi }-\frac{20 \bar\mu^2 \log 2}{3 G \pi ^3}\right)+ \mathcal{O}(\bar{\mu}^4) \, ,\\
 \xi_7 &=& -\frac{1}{16\pi G}\left( \frac{4 +\log 2}{2 } + \frac{1}{12\pi^2}   \bar{\mu}^2 (1-8 \log 2) \right) + \mathcal{O}(\bar{\mu}^4)\, ,\\
 \nonumber \xi_8 &=& -\frac{\bar\mu}{\pi^3 G}\Big(\frac{7 }{192 }-2 \kappa ^2   (2-3\log 2)+\frac{4 \kappa  \lambda   (14-27 \log 2)}{3 }-\frac{4 \lambda ^2  (31-60 \log 2)}{5 }\Big)+ \mathcal{O}(\bar{\mu}^3)\,,\\
 &&\\
 \xi_9 &=& \frac{1}{\pi^2 G r_+}\left( \frac{11 }{192 }+ 2\kappa ^2  \log 2  - 2 \kappa  \lambda  (1+2\log 2) +39 \lambda ^2\right)\bar{\mu} + \mathcal{O}(\bar{\mu}^3)\,,\\
 \xi_{10} &=& \frac{1}{\pi G} \left( \frac{1}{16 } + \frac{\bar{\mu}^2   }{64\pi^2} +\frac{2\bar{\mu}^2}{\pi^2}\kappa ^2 \left(-1 +2  \log 2\right)   + \frac{2\bar{\mu}^2}{\pi^2}\kappa  \lambda  \left(5 -12  \log 2\right)  \right.\\
 &&\left.-\frac{2}{\pi^2}\lambda ^2 \left(9\pi^2 - \bar{\mu}^2\left(\frac{116 }{15}+16  \log 2\right) \right) \right) + \mathcal{O}(\bar{\mu}^4)\,,
\end{eqnarray}
\begin{eqnarray}
\nonumber \tilde \xi_2 &=& \frac{\bar\mu}{144\pi^4 G} \Big[ 
3\kappa( 24 \pi^2 \log{2} + \bar\mu^2(33 - 68\log{2}) ) - 8\lambda ( 18 \pi^2 (1 + \log{2}) + \bar\mu^2 (5 - 51 \log{2}) ) \Big] +  \mathcal{O}(\bar{\mu}^5)\,,\\
&&\\
\tilde \xi_3 &=& -\frac{  \mu\sigma}{\pi  G p }\left( \kappa\frac{   \mu ^2}{24 }+\lambda (\pi T)^2   \right) \,, \\
\nonumber \tilde \xi_4 &=& \frac{1}{8 G \pi ^2}\left(\kappa  \left(2\log 2 + \frac{3 \bar\mu^2 (2-5\log 2)}{\pi ^2}\right) - 4 \lambda  \left(1+2 \log 2 -\frac{\bar\mu^2 (5+90 \log 2)}{6 \pi ^2}\right)\right) + \mathcal{O}(\bar{\mu}^4)\,,\\ 
&&\\
 \tilde l_E &=& \frac{\tilde \xi_5}{\sigma} =  -\frac{8 \bar\mu }{\pi  r_+}(\kappa  \log 2 - 2 \lambda  (1+2\log 2))+  \mathcal{O}(\bar{\mu}^3)\,. \label{eq:lE}
\end{eqnarray}
As we shall see most of the coefficients computed receive $\lambda-$corrections. Indeed not only the anomalous transport coefficients are sensitive to the presence of the anomaly, but also the non-anomalous ones get corrected as well. The latter had been computed in the past without including the mixed gravitational anomaly, so these corrections were neglected.

\section{Fluid/Gravity Computation}
\label{sec:fluid_gravity}

The system of bulk equations of motion~(\ref{eq:Gbulk}) and (\ref{eq:Abulk}) admits an $AdS$ Reissner-Nordstr\"om black-brane solution of the form
\begin{eqnarray}
ds^2 &=& -r^2f(r)dt^2+\frac{dr^2}{r^2f(r)}+dx^idx^i\,,\\
A &=&  \phi(r)dt  \,,
\end{eqnarray}
with $f(r)=1-m/r^4+q^2/r^6$ and $\phi(r)=-\sqrt{3}q/r^2$ . The real and positive zeros of $f(r)$ are
\begin{eqnarray}
r_+ &=& \frac{\pi T}{2}\left(1 + \sqrt{1 + \frac{2}{3\pi^2}\bar{\mu}^2}\right)\,, \label{eq:r+} \\
r_-^2 &=& \frac{1}{2}r_+^2\left( -1 + \sqrt{9 - \frac{8}{\frac{1}{2}\left( 1 + \sqrt{1 + \frac{2}{3\pi^2}\bar{\mu}^2} \right)}}\right)\,,
\end{eqnarray}
where $r_+$ is the outer horizon and $r_-$ the inner one. The mass of the black hole can be written in terms of hydrodynamical variables as
\begin{equation}
m = \frac{\pi^4T^4}{2^4}\left(1 + \sqrt{1 + \frac{2}{3\pi^2}\bar{\mu}^2}\right)^3\left(-1 + 3\sqrt{1 + \frac{2}{3\pi^2}\bar{\mu}^2}\right).
\end{equation}
The boosted version of this blackhole in Eddington-Finkelstein coordinates looks like
\begin{eqnarray}
ds^2 &=& -r^2f(r)u_\mu u_\nu dx^\mu dx^\nu + r^2P_{\mu\nu}dx^\mu dx^\nu-2u_\mu dx^\mu dr\,, \label{eq:ds2Eddington_Finkelstein}\\
A &=& -\phi(r)u_\mu  dx^\mu\,, \label{eq:A0}
\end{eqnarray}
with the normalization condition $u_\mu u^\mu=-1$.  (\ref{eq:ds2Eddington_Finkelstein}) and (\ref{eq:A0}) is a solution of the equations of motion as long as $m$, $q$ and $u_\mu$  are independent of the space-time coordinates $x^\mu$. The fluid/gravity approach tells us that we have to promote all the parameters to slow varying functions of the space time coordinates, and include corrections to the metric in order to make it a solution of the equations of motion again. 

\subsection{Weyl covariant ansatz}

In order to follow the fluid/gravity techniques \cite{Bhattacharyya:2008jc,Haack:2008cp,Banerjee:2008th,Erdmenger:2008rm}  we will use a Weyl invariant formalism \cite{Bhattacharyya:2008dl} in which the identification of the transport coefficients is direct. We start with the ansatz
\begin{eqnarray}\nonumber
ds^2 &=& -2W_1(\rho)u_\mu dx^\mu \left(dr^2+r\mathcal{A}_\nu dx^\nu\right) + r^2\left[W_2(\rho)\eta_{\mu\nu}+W_3(\rho)u_\mu u_\nu + 2\frac{W_{4\sigma}(\rho)}{r_+}  P^\sigma_{\mu}u_\nu \right. \label{eq:ds2} \\
&& \left. +\frac{W_{5\mu\nu}(\rho)}{r_+^2}\right] dx^\mu dx^\nu \,,\\
A &=& \left(a^{(b)}_\mu+ a_\nu(\rho)P^\nu_\mu  + r_+ c(\rho) u_\mu\right)dx^\mu \,, \label{eq:A}
\end{eqnarray}
where now $r_+$ is an unknown function of the space-time coordinates, $r_+(x^\mu)$, and coincides with the radius of the outer horizon of the black hole~(\ref{eq:r+}) only when the $x^\mu$ dependence is gone. $a_\mu^{(b)}=a_\mu^{(b)}(x^\mu)$ is a boundary background gauge field satisfying $a_\mu^{(b)}(x^\mu_0)=0$. Notice that $\eta_{\mu\nu}$ is the Minkowski metric, so we will look for metric solutions with flat boundary. The $r$-coordinate has Weyl weight $+1$, and in consequence $r_+$ has the same property. By construction the $W$ functions are Weyl invariant, so that they will depend on~$r$ only in a Weyl invariant way, i.e, $W(r)\equiv W(\rho)$ with $\rho=r/r_+$.  $W_{5\mu\nu}(r)$ obeys the traceless and transversality conditions $W_{5\mu}^\mu(r) =0$, $u^\mu W_{5\mu\nu}(r)=0$. All these scalars, vectors and tensors will be understood in term of a derivative expansion in the transverse coordinates, i.e. $F(\rho)=F^{(0)}(\rho)+ \varepsilon F^{(1)}(\rho)+\varepsilon^2 F^{(2)}(\rho) +O(\varepsilon^3)$ for a generic function $F$, with $\varepsilon$ a parameter counting the number of boundary space-time derivatives.~\footnote{This expansion of $F(\rho)$ is basically a Taylor expansion around the point $x_0^\mu$.} This solution leads to the current and energy momentum tensor after using the AdS/CFT dictionary (see~(\ref{eq:Jcov}) and (\ref{eq:Tcov}))
\begin{eqnarray}
\label{eq:Jreno} J_\mu  &=& \frac{1}{8\pi G}\lim_{\epsilon\to 0} \left(r^3_+c^{(\bar{2,}\epsilon)}u_\mu+ r_+^2 a_\mu^{(\bar{2},\epsilon)} + J^{ct}_\mu\right) \,, \\
\label{eq:Treno}\nonumber T_{\mu\nu}  &=& \frac{ 1}{16\pi G}\lim_{\epsilon\to 0}\left( r_+^4(W_2+W_3)^{(\bar{4},\epsilon)}( 4u_\mu u_\nu +\eta_{\mu\nu} ) + 4r_+^2W_{5\mu\nu}^{(\bar{4},\epsilon)} + 8r_+^3W^{(\bar{4},\epsilon)}_{4\sigma} P^\sigma_{(\mu} u_{\nu)} + T^{ct}_{\mu\nu}\right) ,\\
\end{eqnarray}
where $F^{(\bar{n},\epsilon)}$ denotes the coefficient of the term $(\rho^{-1}-\epsilon)^{n}$ in an expansion around the regularized boundary, and $\epsilon$ determines the position of the cut-off surface $\rho = 1/\epsilon$. The counterterms in the current $J^{ct}_\mu$  and energy momentum tensor $T^{ct}_{\mu\nu}$ are needed to make the expressions finite, and they follow from the counterterm of the action~(\ref{eq:Sct}). They write
\begin{eqnarray}
J^{ct}_\mu &=&  \frac{1}{2}\log\epsilon\left[ (\mathcal{D}_\nu E^\nu - 2\omega_\nu B^\nu )u_\mu + \mathcal{J}^{(5)}_\mu - \frac{1}{2}\mathcal{J}^{(6)}_\mu -\mathcal{J}^{(7)}_\mu  + \mathcal{J}^{(10)}_\mu \right]\,,  \label{eq:Jct}\\
T^{ct}_{\mu\nu}  &=& \log\epsilon \Big[ -\frac{1}{6}(B_\beta B^\beta +E_\beta E^\beta)P_{\mu\nu} -\frac{1}{2} ( E_\alpha E^\alpha + B_\alpha B^\alpha )u_\mu u_\nu  +  \mathcal{T}^{(9)}_{\mu\nu} +  \mathcal{T}^{(10)}_{\mu\nu} \nonumber \\
&&\qquad - (\mathcal{J}^{(9)}_\mu u_\nu + \mathcal{J}^{(9)}_\nu u_\mu )   \Big] \,. \label{eq:Tct} 
\end{eqnarray}

We are considering a flat background metric, and so the divergences appear only through terms involving electromagnetic fields, in addition to the cosmological constant contribution which was already taken into account in (\ref{eq:Treno}).

The functions at zeroth order in the derivative expansion correspond to the boosted charged blackhole, i.e.~\footnote{Following the notation in~\cite{Erdmenger:2008rm}, barred superscripts $(\bar{n})$ should not be confused with superscripts~$(n)$, where the latter refers to the order in the hydrodynamical expansion.}
\begin{eqnarray}
c^{(0)}(\rho)&=& -\frac{\phi(\rho)}{r_+} \,, \\
W^{(0)}_1(\rho)&=& 1 = W^{(0)}_2(\rho) \,, \\
W^{(0)}_3(\rho)&=&1-f(\rho) \,, \\
W^{(0)}_{4\mu}(\rho)&=& 0 = W^{(0)}_{5\mu\nu}(\rho) \,, \\
a^{(0)}_\mu(\rho)&=&0 \,.
\end{eqnarray}
Then the charge current and energy momentum tensor at this order read
\begin{equation}
 J^{(0)}_\mu  = \frac{\sqrt{3}q}{8\pi G} u_\mu \,, \qquad T^{(0)}_{\mu\nu}  = \frac{m}{16\pi G} \left( 4 u_\mu u_\nu +\eta_{\mu\nu} \right)  \,.
\end{equation}
From this we obtain the equilibrium pressure and charge density $p=\frac{m}{16\pi G}$ and $n=\frac{\sqrt{3}q}{8\pi G}$. For computational reasons it is convenient to define a Weyl invariant charge  $Q\equiv q/r_+^3$ and mass $M\equiv m/r_+^4=1+Q^2$. In terms of these redefined parameters, the black hole temperature and chemical potential read
\begin{equation}
T=\frac{r_+}{2\pi}(2-Q^2) \,, \qquad \mu = \sqrt{3}r_+ Q \,.
\end{equation}
We also define the inner horizon in the $\rho$-coordinate, $\rho_2 \equiv r_{-}/r_+\,$.

\subsection{Einstein-Maxwell equations of motion and Ward identities}

Inserting the ansatz~(\ref{eq:ds2})-(\ref{eq:A}) into the Einstein-Maxwell system of equations we find a set of $(2\times\mathbf{1}+2\times\mathbf{3}+\mathbf{5})$ differential equations and $(2\times\mathbf{1}+\mathbf{3})$ constraints relating the allowed $Q(x^\mu)$, $r_+(x^\mu)$, $u^\nu(x^\mu)$ and $a^{(b)}_\nu(x^\mu)$ \cite{Banerjee:2008th}.~\footnote{$\mathbf{1}$, $\mathbf{3}$ and $\mathbf{5}$ denote the $SO(3)$ scalars, vectors and tensors in which the fields are decomposed.} We need to solve the equations of motion around  a certain point $x_0^\mu$ that we choose to be $x_0^\mu=0$. At such point we sit in a frame in which $u^\mu(0)=(1,0,0,0)$ and $a^{(b)}_\mu(0)=0$.

The scalar sector is obtained from the $rr$, $rv$ and $vv$ components of the Einstein equations, and the $r$ and $v$ components of the  Maxwell equations. One finds two constraints 
\begin{eqnarray}
\label{eq:dtexp}E^{(n)}_{vv} + r^2f(r) E^{(n)}_{rv} &=& 0 \quad \Rightarrow\quad \left(\mathcal{D}_\mu T^\mu_v=F_{v\alpha}J^\alpha\right)^{(n-1)} \,, \\
\label{eq:djexp}M^{(n)}_v + r^2f(r)M^{(n)}_r &=& 0\quad \Rightarrow\quad \left(\mathcal{D}_\mu J^{\mu}=c_1F\wedge F\right) ^{(n-1)} \,, 
\end{eqnarray}
which, as indicated, correspond to the energy-momentum and current non-con\-ser\-va\-tion relations at order~$n-1$.~\footnote{Notice that there are no curvature terms in (\ref{eq:djexp}) and (\ref{eq:dtexp}) because we are working with a flat boundary.} The combinations $E_{rr}=0$, $E_{rv} + r^2f(r) E_{rr} = 0$ and $M_r=0$ leads respectively to the set of differential equations
\begin{eqnarray}
\label{ES1} 3\partial_\rho W_1^{(n)}(\rho) - \frac{3}{2}\rho^{-1}\partial_\rho\left(\rho^2\partial_\rho W^{(n)}_{2}(\rho)\right) &=&\mathbb S^{(n)}(\rho) \,, \quad \\
\label{ES2} \partial_\rho \left(\rho^4 W_3^{(n)}\right) + 8\rho^3W_1^{(n)} - \frac{2}{\sqrt{3}}Q\partial_\rho  c^{(n)} +(1-4\rho^4)\partial_\rho W^{(n)}_2 -4\rho^3 W_2^{(n)} &=&\mathbb K^{(n)}(\rho) \,, \\
\label{MS}  \partial_\rho\left(\rho^3\partial_\rho c^{(n)} \right) - 2\sqrt{3}Q \partial_\rho W^{(n)}_1(\rho) +3\sqrt{3}Q\partial_\rho W_2^{(n)} &=&\mathbb C^{(n)}(\rho) \,.
\end{eqnarray}
At this stage there is still some gauge freedom in the metric. There are  three (metric) scalar fields~$(W_1\,, W_2$ and $ W_3 )$ but the Einstein's equations give two differential equations, (\ref{ES1}) and (\ref{ES2}). We choose the gauge $W_2(\rho)=1$ in which the system partially decouples and can be solved as
\begin{eqnarray}
W_1^{(n,\epsilon)}(\rho) &=& -\frac{1}{3}\int_\rho^{\frac{1}{\epsilon}} \mathrm{d}x  \,\mathbb S^{(n)}(x)\,,\\
W_3^{(n,\epsilon)}(\rho) &=&  \frac{C_0}{\rho^4} -  \frac{1}{\rho^4}\int_\rho^{\frac{1}{\epsilon}} \mathrm{d}x   \left(\mathbb K^{(n)}(x) - 8 x^3 W_1^{(n,\epsilon)}(x) + \frac{2Q}{\sqrt{3}}\partial_x c^{(n,\epsilon)}(x) \right)\,,\\
c^{(n,\epsilon)}(\rho) &=& c_0\frac{(1-\epsilon^2\rho^2)}{\rho^2} - \int_\rho^{\frac{1}{\epsilon}}\mathrm{d}x \, x^{-3}\int_1^x \mathrm{d}y \left(\mathbb C^{(n)}(y)+\frac{2Q}{\sqrt{3}}\mathbb S^{(n)}(y)\right)  \,. 
\end{eqnarray} 
These solutions have been constructed by requiring Dirichlet boundary conditions at the cut-off surface and demanding regularity at the interior of the bulk. The remaining integration constants $C_0$  and $c_0$ are associated to the freedom of choosing a frame in the hydrodynamic set up.

In a similar way, the vector sector is constructed with the components of the equations of motion $E_{ri}$, $E_{vi}$ and $M_i$. They lead to a constraint equation,
\begin{equation}
E^{(n)}_{vi} + r^2f(r) E^{(n)}_{ri} = 0\quad \Rightarrow\quad \left(\mathcal{D}_\mu T^\mu_i=F_{i\alpha}J^\alpha\right)^{(n-1)} \,,
\end{equation}
implying the energy conservation equation, and the two dynamical equations
\begin{eqnarray}
\partial_\rho\left(\rho^5\partial_\rho W^{(n)}_{4i} + 2\sqrt{3}Q a^{(n)}_i(\rho)\right) &=&\mathbb J^{(n)}_{i}(\rho) \,, \\
\partial_\rho\left(\rho^3f(\rho)\partial_\rho a^{(n)}_{i}(\rho) +2\sqrt{3}Q\partial_\rho W^{(n)}_{4i}(\rho)\right)&=&\mathbb  A^{(n)}_{i}(\rho) \,, 
\end{eqnarray}
corresponding to $E_{ri} = 0$ and $M_i = 0$ respectively. The general solution of this system in the Landau frame has been found in \cite{Erdmenger:2008rm} with neither background fields at the boundary nor gravitational anomaly. It is straightforward to generalize the solution to the case in which electromagnetic sources are included. In this case new divergences arise that need to be regulated with the cut-off $1/\epsilon$, and then to be substracted with the corresponding counterterms~(\ref{eq:Jct})-(\ref{eq:Tct}). Having found in~\cite{Erdmenger:2008rm} the general solution of the system, the contribution to the current coming from the vector sector writes (cf.~(\ref{eq:Jreno})) 
\begin{equation}
\label{a2} a_\nu^{(\bar{2},\epsilon)} = \frac{1}{2}\left(\frac{1}{\epsilon}\mathbb A_\nu\left(1/\epsilon\right) - \int_1^{\frac{1}{\epsilon}}\mathrm{d}x \, \mathbb A_{\nu}(x)\right) - \frac{\sqrt{3}Q}{M}C^{(\epsilon)}_\nu - \frac{\sqrt{3}Q}{4M}D^{(\epsilon)}_\nu,
\end{equation}
where the integration constants $C_\nu^{(\epsilon)}$ and $D_\nu^{(\epsilon)}$ are determined by fixing the Landau frame ($W^{(\bar{4},\epsilon)}_{4\sigma} = 0$), and demanding regularity at the outer horizon respectively. These constants write
\begin{eqnarray}
4C^{(\epsilon)}_\nu &=& - \sum_{m=0}^2\frac{(-1)^m\partial^m_\rho \mathbb J_\nu(1/\epsilon)}{\epsilon^{m+1}(m+1)!} + \int_1^{\frac{1}{\epsilon}} \mathrm{d}x \, \mathbb J_\nu(x) - \mathcal J_\nu^{(9)}\log\epsilon  \,, \label{eq:4Cnu}\\
D^{(\epsilon)}_\nu &=&  - \sqrt{3}Q\int_1^{\frac{1}{\epsilon}} \mathrm{d}x \, \frac{\mathbb A_\nu(x)}{x^2} - M\int_1^{\frac{1}{\epsilon}} \mathrm{d}x \,  \frac{\mathbb J_\nu(x)}{x^4} + Q^2\int_1^{\frac{1}{\epsilon}} \mathrm{d}x \, \frac{\mathbb J_\nu(x)}{x^6}\,. \label{eq:Dnu}
\end{eqnarray}

Finally the tensor equations are the combination $E_{ij}-\frac{1}{3}\delta_{ij}\tr(E_{kl})=0$, which leads to the dynamical equation
\begin{equation}
\partial_\rho\left(\rho^5f(\rho)\partial_\rho W^{(n)}_{5ij}(\rho)\right)=\mathbb P^{(n)}_{ij}(\rho) \,.
\end{equation}
The solution of this equation that satisfies the Dirichlet boundary  and regularity conditions writes
\begin{equation}
W^{(n,\epsilon)}_{5\mu\nu}(\rho) = -\int_\rho^{\frac{1}{\epsilon}} \mathrm{d}x \, \frac{\int_1^x \mathrm{d}y \, \mathbb P^{(n)}_{\mu\nu}(y)}{x^5f(x)}\,.
\end{equation}
After doing an asymptotic expansion of this solution around the regularized boundary surface, one can extract the tensor contribution to the energy momentum tensor (cf.~(\ref{eq:Treno})),
\begin{equation}\label{eq:W45}
4W^{(\bar{4},\epsilon)}_{5\mu\nu} = - \sum_{m=0}^2\frac{(-1)^m\partial^m_\rho\mathbb P_{\mu\nu}(1/\epsilon)}{\epsilon^{m+1}(m+1)!} - \int_1^{\frac{1}{\epsilon}} \mathrm{d}x \,\mathbb  P_{\mu\nu}(x) \,.
\end{equation}

Note that the form of the homogeneous part in the dynamical equations in the scalar, vector and tensor sectors is the same at any order in the derivative expansion. Each order~$n$ is then characterized by the specific form of the sources. In the next two sections we will compute the sources, and integrate them according to the formulae presented above to get the transport coefficients at first and second order.

\section{First Order Transport Coefficients}
\label{sec:first_order}

The technology presented in Sec.~\ref{sec:fluid_gravity} can be used to construct the solutions of the system at any order in a derivative expansion. As it has been already explained, the solution at zeroth order trivially leads to the charged blackhole with constant parameters~(\ref{eq:ds2Eddington_Finkelstein})-(\ref{eq:A0}). In this section we will solve the system up to first order. The transport coefficients at this order have been obtained previously in the literature using different methods in field theory and holography. In particular, they have been computed within the fluid/gravity approach, but not including external electric fields in this formalism, see eg.~\cite{Erdmenger:2008rm,Banerjee:2008th,Son:2009tf,Chapman:2012my}.

\subsection{Scalar sector}
In the scalar sector, the first order sources look like
\begin{equation}
\mathbb S^{(1)}(\rho) =\mathbb K^{(1)}(\rho) =\mathbb C^{(1)}(\rho) = 0 \,.
\end{equation} 
This very simple situation leads to the solution
\begin{eqnarray}
W_1^{(1,\epsilon)}(\rho) &=& 0 \,, \\
c^{(1,\epsilon)}(\rho) &=& c_0\frac{(1-\epsilon^2\rho^2)}{\rho^2}  \,, \\
W_3^{(1,\epsilon)}(\rho) &=&  \frac{C_0}{\rho^4} + \frac{2Qc_0}{\sqrt{3}}\frac{(1-\epsilon^2\rho^2)}{\rho^6}\,.
\end{eqnarray}
The integration constants $c_0$ and $C_0$ can be fixed to zero because they just redefine the charge and mass of the black hole respectively.

\subsection{Vector and tensor sector}
The first order sources  are given by
\begin{eqnarray}
\mathbb J^{(1)}_{\mu} &=&  -\lambda\frac{96}{\rho^3}\left(\frac{5 Q^2  }{ \rho ^2} - M\right)\frac{B_\mu}{r_+} - \sqrt{3} Q\lambda\left(\frac{1008 Q^2  }{\rho ^7}-\frac{320 M }{\rho ^5}\right) \omega_\mu \,, \\
\nonumber\mathbb A^{(1)}_{\mu} &=& - \frac{\sqrt{3} \pi  T}{M r_+ \rho ^2}P_\mu^{\nu}\mathcal{D}_\nu Q  - \left(1+\frac{9 Q^2}{2 M \rho ^2}\right)\frac{E_\mu}{r_+} - \frac{16\sqrt{3}\kappa Q}{\rho^3}  \frac{B_\mu}{r_+}  - \frac{48\kappa Q^2}{\rho^5} \omega_\mu \\
&&-\frac{48 \lambda  \left(15 Q^4-16 M Q^2 \rho ^2+4 M^2 \rho ^4\right)}{\rho ^{11}} \omega_\mu  \,, \\
\mathbb P^{(1)}_{\mu\nu} &=& -6r_+\rho^2\sigma_{\mu\nu}  \,,
\end{eqnarray}
where $\mathcal{D}_\mu$ is the Weyl covariant derivative and $\mathcal{D}_\alpha Q =\frac{2\pi T^2}{\sqrt{3}r_+^2(1+M)}\mathcal{D}_\alpha\bar{\mu}$. Using equations (\ref{eq:Jreno}),  (\ref{eq:Treno}), (\ref{a2}), (\ref{eq:4Cnu}) and (\ref{eq:Dnu}) it is straightforward to find the first order transport coefficients shown at the end of section~\ref{sec:constitutives}. We write the result again for completeness,
\begin{eqnarray}
\eta &=& \frac{r_+^3}{16\pi G} \,, \qquad \sigma =\frac{\pi  r_+^7 T^2}{16 G m^2} \,,\\
\xi_B &=& -\frac{\sqrt{3} q \left(m+3 r_+^4\right) \kappa }{8 G m \pi  r_+^2}+\frac{\sqrt{3} \pi  q T^2 \lambda }{G m} \,, \qquad  \xi_V = -\frac{3 q^2 \kappa }{4 G m \pi }+\frac{2 \pi  \left(2 q^2-r_+^6\right) T^2 \lambda }{G m r_+^2} \,.
\end{eqnarray}
Chiral magnetic $\xi_B$ and vortical $\xi_V$ conductivities have been computed at first order in holography within the Kubo formulae formalism in~\cite{Landsteiner:2011iq,Landsteiner:2011tg,Landsteiner:2011tf,Landsteiner:2012kd,Landsteiner:2012hc}, including chiral and gauge gravitational anomalies. Here we reproduce the same result within the fluid/gravity approach.~\footnote{The gauge gravitational anomaly contribution to the chiral vortical conductivity was also computed recently within an holographic setup in~\cite{Chapman:2012my,Megias:2013xla}.}

Note that to compute the first order transport coefficients one needs only the terms $a_\mu^{(\bar{2},\epsilon)}$ and $W^{(\bar{4},\epsilon)}$ in the near boundary expansion. However, in order to go to the next order in the derivative expansion, we need to know the exact solutions, which can be written in terms of the sources as
\begin{eqnarray}\label{eq:W4}
W_{4\mu}^{(1)}(\rho) &=& F_1 [\rho]P_\mu^\nu\mathcal{D}_\nu Q(x) + F_2[\rho]\omega_\mu(x) + F_3[\rho]\frac{E_\mu(x)}{r_+} +  F_4[\rho]\frac{B_\mu(x)}{r_+} \,,\\
W_{5\mu\nu}^{(1)}(\rho) &=&  F_5[\rho]\, r_+ \, \sigma_{\mu\nu}(x) \,, \\
\label{eq:amu} a_\mu^{(1)}(\rho) &=&  F_6[\rho]P_\mu^\nu\mathcal{D}_\nu Q(x) +  F_7[\rho]\omega_\mu(x) +  F_8[\rho]\frac{E_\mu(x)}{r_+} +  F_9[\rho]\frac{B_\mu(x)}{r_+} \,.
\end{eqnarray}
We show in Appendix~\ref{ap:Ffunctions} the expressions for the $F$'s functions. $F_5$ writes
\begin{eqnarray}
\nonumber F_5[\rho] &=& -\frac{2 \log{[1+\rho ]}}{-1+M}-\frac{\left(1+\rho_2+\rho_2^2\right) \log{[\rho -\rho_2]}}{(1+\rho_2) \left(1+2 \rho_2^2\right)}+\frac{2 \left(1+\rho_2^3\right) \log{[\rho +\rho_2]}}{-2-2 \rho_2^2+4 \rho_2^4} \\
 &&+\frac{\log{\left[1+\rho ^2+\rho_2^2\right]}}{2+5 \rho_2^2+2 \rho_2^4}+\frac{2 \left(1+\rho_2^2\right)^{3/2}}{2+5 \rho_2^2+2 \rho_2^4}  \,\mathrm{ArcCot}\left[\frac{\rho }{\sqrt{1+\rho_2^2}}\right] \,.
\end{eqnarray}

\section{Second Order Transport Coefficients}
\label{sec:second_order}

The second order coefficients are much more computationally demanding than the first order ones. The parameter $c^{(\bar{2},\epsilon)}$ in~(\ref{eq:Jreno}) can always be chosen to be zero, as it just redefines the charge and mass of the black hole. On the other hand, because we are working in the Landau frame, there is no contribution coming from the scalar sector to the energy-momentum tensor and $(W_2 + W_3)^{(\bar{4},\epsilon)}$ is set to zero. We have checked that this is in fact what happens by using the sources for the scalar sector. So, we will focus in this section on the vector and tensor contributions.

\subsection{Vector sector}

The second order sources in the vector sector are shown in Appendix \ref{ap:Vec_sour}. Again using these expressions and Eqs. (\ref{a2}), (\ref{eq:4Cnu}), (\ref{eq:Dnu}) and (\ref{eq:Jreno}), we can extract the second order transport coefficients. We show first the new non anomalous coefficients
\begin{eqnarray}
\xi_5 &=& \xi_{5,0}(\rho_2) \,, \label{eq:xi5} \\
\xi_6 &=&   \xi_{6,0}(\rho_2) +\frac{3 \left(3+M^2\right)Q^2 \kappa ^2}{4\pi  G M^3 } + \kappa\lambda \xi_{6,\kappa\lambda}(\rho_2) + \lambda^2 \xi_{6,\lambda^2}(\rho_2)  \,, \\
\xi_7 &=& \xi_{7,0}(\rho_2) \,, \\
\xi_8 &=& -\frac{(9 + 12 M + 7 M^2) \pi Q T^3}{128 \sqrt{3} G M^4 (1 + M) r_+^3}   + \kappa^2 \xi_{8,\kappa^2}(\rho_2)+ \kappa\lambda \xi_{8,\kappa\lambda}(\rho_2) + \lambda^2 \xi_{8,\lambda^2}(\rho_2)  \,, \\
\nonumber\xi_9 &=& \frac{Q \left(88+480Q^2M +169Q^6\right)}{512\sqrt{3}\pi  G M^4  r_+} + \frac{1}{r_+}\left( \kappa^2 \xi_{9,\kappa^2}(\rho_2)(\rho_2) + \kappa\lambda \xi_{9,\kappa\lambda}(\rho_2)(\rho_2) + \lambda^2 \xi_{9,\lambda^2}(\rho_2)(\rho_2)  \right) \,, \\
&&\\
\xi_{10} &=& \frac{\left(4+7 Q^2\right)}{64\pi G M  } + \kappa^2 \xi_{10,\kappa^2}(\rho_2) + \kappa\lambda \xi_{10,\kappa\lambda}(\rho_2)+ \lambda^2 \xi_{10,\lambda^2}(\rho_2)   \,. \label{eq:xi10}
\end{eqnarray}
These coefficients had not been computed previously in the literature. The rest of the non anomalous coefficients were obtained in the past without the gravitational anomaly. In this work we have found the $\lambda-$corrected results, which write
\begin{eqnarray}
\xi_1 &=& \frac{\pi  T^3}{8GM^3 (M+1) r_+^2 } \left( Q^2  + \frac{ M^2}{\left(1+2 \rho_2^2\right)} \log{\left[\frac{2+\rho_2^2}{1-\rho_2^2}\right]} \right) \,, \label{eq:xi1} \\
\nonumber \xi_2 &=&  \frac{(3+M) (M (3+M)-6) T^2}{128 G M^3 (M+1) r_+} + \frac{3 \pi  Q^2 T^3 \kappa ^2}{G M^3 (M+1) r_+^2} + r_+\left( \kappa\lambda \xi_{2,\kappa\lambda}(\rho_2)+\lambda^2\xi_{2,\lambda^2}(\rho_2) \right) \,, \\
&&\\
\xi_3 &=& \frac{3 \sqrt{3} Q^3 r_+}{64 \pi G M^2 } \,, \\
\xi_4 &=& \frac{3 \sqrt{3} Q^3 r_+ \kappa ^2}{2\pi  G M^2 } + r_+\kappa\lambda \xi_{4,\kappa\lambda}(\rho_2)+ r_+\lambda^2 \xi_{4,\lambda^2}(\rho_2)  \label{eq:xi4}\,.
\end{eqnarray}
In the anomalous sector, the new coefficients (not computed previously) write
\begin{eqnarray}
\tilde{\xi}_2 &=& \frac{3 \sqrt{3} Q^3 \left(6+M\right) \kappa }{16\pi G M^2   \left(1+2 \rho_2^2\right)^2}+\frac{\sqrt{3} \pi  Q T^2 \kappa  \log{\left[\frac{2+\rho_2^2}{1-\rho_2^2}\right]}}{2G M r_+^2 \left(1+2\rho_2^2\right)^3} + \lambda\tilde \xi_{2,\lambda}(\rho_2) \,, \label{eq:xitilde2} \\
\tilde{\xi}_3 &=& \frac{\sqrt{3}\pi Q T^2}{8r_+^2 M^3G }\left(  Q^2  \kappa +\frac{8  \pi ^2  T^2 \lambda }{ r_+^2}\right) \,, \label{eq:xitilde3}\\
\tilde{\xi}_4 &=& \kappa \tilde \xi_{4,\kappa}(\rho_2) + \lambda\tilde \xi_{4,\lambda}(\rho_2) \,, \\
\tilde{\xi}_5 &=& \kappa \tilde \xi_{5,\kappa}(\rho_2) + \lambda\tilde \xi_{5,\lambda}(\rho_2)\,,  \label{eq:xitilde5}
\end{eqnarray}
while the already known coefficient with the new $\lambda$ contribution writes
\begin{eqnarray}
\tilde{\xi}_1 &=& \frac{3Q^2 r_+ \kappa }{4\pi G M^2  } + \lambda r_+\tilde \xi_{1,\lambda}(\rho_2) \,.  \label{eq:xitilde1}
\end{eqnarray}
The $\xi_{i,(0,\kappa^2,\kappa\lambda,\lambda^2)}(\rho_2)$ and $\tilde\xi_{i,(\kappa,\lambda)}(\rho_2)$ functions are defined in Appendix~\ref{ap:fsecond_order}. These coefficients enter in the constitutive relation for the current through~(\ref{eq:fullconstiJ}) and (\ref{eq:nu2}).

\subsection{Tensor sector}
The second order sources in the tensor sector are shown in Appendix~\ref{ap:Ten_sour}, and again we can extract the transport coefficients at this order after pluging these expressions into Eqs.~(\ref{eq:Treno}) and (\ref{eq:W45}). Due to the length of the expressions, some of them will be shown exactly and the rest are expressed in terms of some functions $\Lambda_{i,(\kappa^2,\kappa\lambda,\lambda^2)}(\rho_2)$ and $\tilde \Lambda_{i,(\kappa,\lambda)}(\rho_2)$ which are presented in the Appendix~\ref{ap:fsecond_order}. Again we split our results in  those non anomalous coefficients which are new,
\begin{eqnarray}
\Lambda_7 &=& -\frac{\sqrt{3} (-3+5 M) Q r_+}{64\pi G M^2  } \,, \label{eq:Lambda7}\\
\Lambda_8 &=& r_+\Lambda_{8,0}(\rho_2) \,, \\
\Lambda_9 &=& \Lambda_{9,0}(\rho_2) \,, \\
\Lambda_{10} &=& \frac{11}{96 \pi G} + \kappa^2 \Lambda_{10,\kappa^2}(\rho_2) + \kappa\lambda \Lambda_{10,\kappa\lambda}(\rho_2) + \lambda^2 \Lambda_{10,\lambda^2}(\rho_2) \,, \\
\Lambda_{11} &=& -\frac{8 \sqrt{3} Q r_+ \kappa  \lambda }{\pi G  } \,, \\
\Lambda_{12} &=& -\frac{\sqrt{3} Q r_+}{16\pi G  } + \frac{3 \sqrt{3} Q^3 r_+ \kappa ^2}{\pi G M } + \kappa\lambda r_+\Lambda_{12,\kappa\lambda}(\rho_2) + \lambda^2r_+ \Lambda_{12,\lambda^2}(\rho_2) \,, \label{eq:lambda12}
\end{eqnarray}
and the rest of the non anomalous ones
\begin{eqnarray}
\Lambda_1 &=& \frac{r_+^2 }{16\pi G  }\left(2+\frac{M }{\sqrt{4 M-3}}\log\left[\frac{3-\sqrt{4 M-3}}{3+\sqrt{4 M-3}}\right]\right) \,, \label{eq:lambda1} \\
\Lambda_2 &=& \frac{r_+^2}{8 \pi G  } \,, \\
\Lambda_3 &=& \frac{r_+^2}{8 G \pi } \left( \frac{ M }{2(1+2 \rho_2^2)}\log{\left[\frac{2+\rho_2^2}{1-\rho_2^2}\right] + 192 Q^2 \kappa  \lambda -\frac{384 (3 M-5) \pi  T \lambda ^2}{r_+}} \right) \,,  \\
\nonumber \Lambda_4 &=& -\frac{Q^2 r_+^2}{16\pi  G } + \frac{3 Q^4 r_+^2 \kappa ^2}{\pi G M} + \frac{18 Q^2 \left(5+Q^2 \left(9 Q^2-16\right)\right) r_+^2 \kappa  \lambda }{5\pi G M  } + \lambda^2r_+^2\Lambda_{4,\lambda^2}(\rho_2) \,, \\
&&\\
\Lambda_5 &=& -\frac{ \pi  QT^3}{16 \sqrt{3} G M^2 (M+1) r_+} \,, \\
\Lambda_6 &=& r_+^2\Lambda_{6,0}(\rho_2) \,. \label{eq:Lambda6} 
\end{eqnarray}
For the anomalous coefficients we get the new ones
\begin{eqnarray}
\tilde\Lambda_4 &=& -\frac{3 Q^2 r_+\kappa }{8\pi G M  }-\frac{ \pi  T^2 \lambda }{G M r_+} \,, \label{eq:Lambdatilde4} \\
\tilde\Lambda_5 &=& \kappa r_+\tilde\Lambda_{5,\kappa}(\rho_2) + \lambda r_+\tilde\Lambda_{5,\lambda}(\rho_2) \,, \\
\tilde\Lambda_6 &=& \kappa \tilde\Lambda_{6,\kappa}(\rho_2) + \lambda\tilde\Lambda_{6,\lambda}(\rho_2) \,, \\
\tilde\Lambda_7 &=& -\frac{2 r_+\lambda }{G \pi } \,, \\
\tilde\Lambda_8 &=& \frac{3 Q^2 \left(Q^2-1\right) r_+ \kappa }{4\pi G M^2  } + \lambda r_+\tilde \Lambda_{8,\lambda}(\rho_2) \,, \label{eq:lambdatilde8}
\end{eqnarray}
and the rest of the anomalous ones
\begin{eqnarray}
\tilde\Lambda_1 &=& -\frac{\sqrt{3} r_+^2  Q^3 \kappa }{4\pi G M  }+\frac{ \sqrt{3}Q r_+(3r_+ + ( Q^2-4)\pi T)}{\pi G M  } \lambda \,, \label{eq:lambdatilde1} \\
\tilde\Lambda_2 &=& \lambda r_+^2 \tilde \Lambda_{2,\lambda}(\rho_2) \,, \\
\tilde\Lambda_3 &=& \frac{2 T^2 \lambda }{G (M+1)} \,. \label{eq:Lambdatilde3}
\end{eqnarray}
These coefficients enter in the constitutive relation for the energy-momentum tensor through (\ref{eq:fullconstiT}) and (\ref{eq:tau2}).

The transport coefficients $\Lambda_1,\ldots,\Lambda_6$, $\tilde{\Lambda}_1,\ldots,\tilde{\Lambda}_3$ and $\xi_1\ldots\xi_4$, $\tilde{\xi}_1$ have been computed in the past in \cite{Erdmenger:2008rm,Banerjee:2008th} without gravitational anomaly. It is interesting to remark that $\Lambda_1, \Lambda_2, \Lambda_5, \Lambda_6, \Lambda_7, \Lambda_8, \Lambda_9, \xi_1, \xi_3, \xi_5$ and $\xi_7$ do not receive $\lambda-$corrections, actually these coefficients do not depend on $\kappa$ either. It is also remarkable that $\tilde{\Lambda}_2$ and $\tilde{\Lambda}_3$ in the presence of gravitational anomaly are not vanishing. The rest of the transport coefficients we have computed are new.

\subsection{Discussion of second order results}
\label{sec:second_order_discussion}

It would be interesting to compare our results with the predictions done in \cite{Kharzeev:2011ds}. Basically the authors tried to fix the anomalous second order transport coefficients using a generalized version of the method developed by Son \& Surowka \cite{Son:2009tf}. The only issue is that they didn't consider the mixed gauge-gravitational anomaly and neglected all the integration constants as the previous authors. Nowadays we know that at least at first order  these integration constants might be related to the anomalous parameter~$\lambda$. The authors presented a set of algebraic and differential constraints. The algebraic ones are
\begin{eqnarray}
\label{eq:Klambda1}\tilde\Lambda_1 &=&\frac{4\eta }{n}\left(\xi_V - TD_B\right)\,,\\
\label{eq:Klambda4}\tilde\Lambda_4 &=&\frac{2\eta }{n}\left(\xi_B - \bar\kappa\mu\right)\,,\\
\label{eq:Kxi3}\tilde\xi_3 &=&\frac{2\sigma}{n}\left(\xi_V - TD_B\right)\,,\\
\label{eq:Kxi5}\tilde\xi_5 &=& 0\,,
\end{eqnarray}
where $D_B=\frac{\kappa T}{4\pi G  }\bar\mu^2$ is the coefficient multiplying the magnetic field in the entropy current computed in \cite{Son:2009tf,Neiman:2010zi,Kharzeev:2011ds} with only pure gauge anomaly, and $\bar{\kappa}=\frac{\kappa}{2\pi G}$ is the anomalous parameter used by the authors of \cite{Kharzeev:2011ds}. Eqs. (\ref{eq:Klambda1}) and (\ref{eq:Kxi3}) are satisfied by our solutions (\ref{eq:lambdatilde1}) and (\ref{eq:xitilde3}) as long as one fixes the anomaly parameter $\lambda$ to zero. However Eq.~(\ref{eq:Klambda4}) is satisfied  with the gravitational anomaly switched on. So far these constraints are satisfied except Eq. (\ref{eq:Kxi5}), as $\tilde{\xi}_5$ is not vanishing in our model even though we fix the anomalous parameter to vanish.

To check the value we get for $\tilde{\xi}_5$~(\ref{eq:lE}), we may proceed by using the Kubo formula formalism. The Kubo formula for $\tilde{\xi}_5$ will relate this coefficient to a two point function at second order in a frequency and momentum expansion. Actually it will appear in the same correlator as the chiral magnetic conductivity. To do so we can switch on a gauge field in the $y$ direction $A_y=A_y(t,z)$. In such a situation the Fourier transformed source $\mathcal{J}^{(5)}_\mu$ reduces to
\begin{equation}
\mathcal{J}^{(5)x} = \omega k_z A_y \,,
\end{equation}
so that using the constitutive relation we can read the two point function
\begin{equation}
\label{eq:jjsecorder}\left\langle \mathcal J^x \mathcal J^y \right\rangle = -i\xi_B k_z + \sigma \tilde l_E \omega k_z \,,
\end{equation}
where we have redefined $\tilde l_E = \tilde \xi_5 / \sigma $ for the reason we will explain below. We have checked that this Kubo formula leads to our result~(\ref{eq:lE}) for $\tilde l_E$ by using the model of section~\ref{sec:holo_model} considered in the probe limit. This is a non trivial check of the non vanishing of $\tilde\xi_5$. We leave a detailed analysis of this and other Kubo formulae for a forthcoming paper~\cite{workinprogress}.  

In order to understand the discrepancy between this result and the prediction done by the authors of \cite{Kharzeev:2011ds}, we can analyze the properties under time reversal of the source associated to $\tilde \xi_5$, which reads in the constitutive relations as
\begin{equation}
J^\mu =  \tilde{\xi}_5\epsilon^{\mu\nu\rho\lambda}u_\nu\mathcal{D}_\rho E_\lambda + \ldots  \,.
\end{equation}
This equation in the local rest frame $u^\mu = (1,0,0,0)$ looks like
\begin{equation}
\vec{J}=\tilde \xi_5 \nabla\times \vec{E} + \ldots = - \tilde \xi_5 \frac{\partial \vec{B}}{\partial t}  + \ldots \,.
\end{equation}
The electric field and the operator $\nabla\times$ are even under time reversal while the current is odd, in consequence the conductivity $\tilde\xi_5$ is $\mathcal{T}-$odd. The fact that this transport coefficient is $\mathcal{T}-$odd tells us that such a source might contribute to the entropy production. For this reason demanding a non contribution to the production of entropy might not be well motivated. The situation would be similar as demanding a vanishing contribution from the usual electric conductivity. One can see also the odd property of $\tilde{\xi}_5$ from Eq. (\ref{eq:jjsecorder}), as $\left\langle \mathcal J^x \mathcal J^y \right\rangle$ is $\mathcal{T}-$even and inverting the time is the same as changing $\omega\to - \omega$. 

We have noticed that the anomalous coefficients associated to sources constructed with the second derivative of the fields can be naturally factorized as
\begin{eqnarray}
\tilde{\Lambda}_1 &=& -2\eta\tilde{l}_\omega  \,, \\
\tilde{\Lambda}_4 &=& -2\eta\tilde{l}_B \,, \\
\tilde{\xi}_5 &=& \sigma\tilde{l}_E \,.
\end{eqnarray}
These expressions make their dissipative nature clear, and they suggest the existence of anomalous relaxation lengths in analogy to the relaxation time $\tau_\pi$. These new $\mathcal{T}-$even quantities write
\begin{eqnarray}
\tilde{l}_\omega &=& \frac{2\pi}{ G  p}\left( \frac{\kappa  \mu ^3}{48\pi^2   } +64\mu\lambda\left( 3 r_+^2 - 2\mu^2 - \frac{\pi  T \mu ^2}{r_+}\right) \right) \,, \\
\tilde{l}_B &=& \frac{1}{2\pi G p}\left( \frac{\kappa  \mu ^2}{8} +  \pi^2 T^2 \lambda  \right) \,,\\
\label{eq:le}\tilde{l}_E &=&  -\frac{8 \bar\mu }{\pi^2 T}(\kappa  \log 2 - 2 \lambda  (1+2\log 2))+  \mathcal{O}(\bar{\mu}^3)\,.
\end{eqnarray}

A last interesting observation comes from the result on the dispersion relation of shear waves in \cite{Kharzeev:2011ds}, where they have found that
\begin{equation}
\label{eq:shear_disp}\omega \approx -i\frac{\eta}{4p}k^2 \mp i C k^3+\ldots \,,
\end{equation}
with $C = -\tilde{\Lambda}_1/(8p)$. It would be interesting to generalize the computation of \cite{Sahoo:2009yq} to the case including the mixed gauge-gravitational anomaly to verify whether the result for $C$ is
\begin{equation}
C = \frac{\eta}{4p} \tilde l_\omega \,.
\end{equation}
%\begin{equation}
%C= \frac{\sqrt{3} r_+^6  Q^3  }{2 M^2  }\kappa-\frac{2 \sqrt{3}Q r_+^5(3r_+ + ( Q^2-4)\pi T)}{ M^2  } \lambda\,.
%\end{equation}  

%\input{renorma}

\section{Discussion and conclusions}
\label{sec:discussion}

We have studied the transport properties of a relativistic fluid affected by gauge and mixed gauge-gravitational anomalies. We have used a holographic bottom up model in 5 dim that implements both anomalies via gauge and mixed gauge-gravitational Chern-Simons terms. This model was used in~\cite{Landsteiner:2011iq} to compute the first order transport coefficients in holography from Kubo formulae. Within the fluid/gravity approach, in this work we have reproduced previous results at first order, and extended the computation up to second order in the hydrodynamical expansion. The computation has been performed in a Weyl covariant formalism, which allows us for a clear classification of terms contributing to second order. 

We have found all the anomalous and non-anomalous transport coefficients of the model up to second order, except the ones associated  to curvature sources. Most of the non-anomalous coefficients receive non trivial contributions coming from the anomaly sector through  terms quadratic in the anomaly coefficients $\kappa$ and $\lambda$. There is a set of coefficients which are not affected by the presence of the anomalies. These are
\begin{eqnarray}
\nonumber T^{\mu\nu} &=& \Lambda_1 u^\alpha\mathcal{D}_\alpha\sigma^{\mu\nu} +\Lambda_2  \sigma^{\langle\mu}\,_\gamma \sigma^{\nu\rangle\gamma}+ \Lambda_5 \mathcal{D}^{\langle\mu}\mathcal{D}^{\nu\rangle} \bar{\mu}  + \Lambda_6 \mathcal{D}^{\langle\mu}\bar{\mu}\mathcal{D}^{\nu\rangle} \bar{\mu} + \Lambda_7 \mathcal{D}^{\langle\mu} E^{\nu\rangle} \,, \\
&&+\Lambda_8 E^{\langle\mu} \mathcal{D}^{\nu\rangle}\bar{\mu} + \Lambda_9 E^{\langle\mu} E^{\nu\rangle}  +\ldots \,,\\
J^\mu &=& \xi_1 \sigma^{\mu\nu}\mathcal{D}_\nu\bar{\mu}  + \xi_3 P^{\mu\nu}\mathcal{D}^\alpha\sigma_{\nu\alpha}  + \xi_5 \sigma^{\mu\nu}E_\nu + \xi_7 u^\nu\mathcal{D}_\nu E^\mu + \ldots\,.
\end{eqnarray}
In particular $\Lambda_1$ is usually redefined in term of the relaxation time $\tau_\pi$, in analogy with the the Israel and Stewart theory, $\Lambda_1=-2\eta\tau_\pi$.

On the other hand, we computed the anomalous coefficients which are $(C,P)=(\pm,-)$, i.e. these contributions are linear in~$\kappa$ or~$\lambda$. These second order transport coefficients are dissipative unlike the first order ones, and in consequence some of them could contribute to entropy production. One example of that is the non vanishing value of $\tilde{\xi}_5$ (\ref{eq:le}). This coefficient was previously predicted to be zero by  using non production of entropy arguments \cite{Kharzeev:2011ds}. It would be interesting to compute within the present model the holographic entropy current and its divergence to study the contributions of such coefficients to the entropy production.

We have defined the $\mathcal T-$even quantities $\tilde{l}_E$, $\tilde{l}_B$ and $\tilde{l}_\omega$  in analogy with the definition of $\tau_\pi$.  A generalization of the Israel and Stewart theory to hydrodynamics with anomalies would give us a better  physical intuition on these parameters. In particular the authors of \cite{Kharzeev:2011ds} have noticed that the chiral coefficient $C$ (\ref{eq:shear_disp}) in the dispersion relation of shear waves is related to $\tilde{\Lambda}_1$, and in consequence to $\tilde{l}_\omega$,
\begin{equation}
\omega \approx -i\frac{\eta}{4p}k^2\left(1\pm \tilde{l}_\omega k\right)\,.
\end{equation}

The role played by the gravitational anomaly in the chiral vortical
effect has shown up for the first time in the calculations of Kubo
formulae involving the two point functions firstly derived
in~\cite{Amado:2011zx}. It is possible to write new Kubo formulae in
terms of two point functions in order to compute the second order
transport coefficients associated to second derivatives of the
background fields. The rest of the transport coefficients at second
order would demand the knowledge of three point functions.  The
advantage of Kubo formulae is that they can be used in either field
theory or holographic computations. This would allow to compare the
weak and strong coupling regimes of second order
coefficients~\cite{workinprogress}.

\appendix

\section{First order solutions}
\label{ap:Ffunctions}
Here we show the exact form of the $F_i[\rho]$ functions defined in Eqs. (\ref{eq:W4}) and (\ref{eq:amu})
{\scriptsize
\begin{eqnarray}
\nonumber F_1[\rho] &=& -Q\Bigg(\frac{\left(9 Q^6 \rho^2+2 M^2 \rho^3 (-4 M+3 (1+M) \rho )-27 Q^4 \left(1+M-\rho^2\right)+6 Q^2 \rho  \left(4 M^2+3 \rho -6 M \rho^4\right)\right) \rho_2 \left(2+5 \rho_2^2 +2 \rho_2^4\right)}{8 M^2 \rho^6 \rho_2 \left(-1+\rho_2^2\right) \left(2+\rho_2^2\right)^2 \left(1+2\rho_2^2 \right)^3} \\
\nonumber && +\frac{-6 \pi  Q \left(-1+\rho^2\right) \left(\rho^2-\rho_2^2\right) \left(-1+\rho_2^2\right)^2 \left(1+\rho ^2+\rho_2^2\right) \left(2+5 \rho_2^2+7 \rho_2^4+5 \rho_2^6+2 \rho_2^8\right)}{8 M^2 \rho ^6 \rho_2 \left(-1+\rho_2^2\right) \left(2+\rho_2^2\right)^2 \left(1+2 \rho_2^2\right)^3}  \Bigg)\\
\nonumber && -\frac{3 \left(Q^2-M \rho ^2+\rho ^6\right) \rho_2 \left(-2-3 \rho_2^2+3 \rho_2^6+2 \rho_2^8\right) \text{ArcTan}\left[\frac{\rho }{\sqrt{1+\rho_2^2}}\right]}{2 M \rho ^6 \left(2+\rho_2^2\right)^2 \left(1+2 \rho_2^2\right)^3} + \frac{3 (1+M) Q \left(Q^2-M \rho ^2+\rho ^6\right) \text{Log}[1+\rho ]}{2 M \left(-2+Q^2\right)^2 \rho ^6}\\
\nonumber && -\frac{3 Q \left(Q^2-M \rho ^2+\rho ^6\right) (-1+\rho_2) \rho_2 \left(2+\rho_2^2\right) \text{Log}[\rho -\rho_2]}{4 \rho ^6 (1+\rho_2)^2 (1+(-1+\rho_2) \rho_2) \left(1+2 \rho_2^2\right)^3} -\frac{3 Q \left(Q^2-M \rho ^2+\rho ^6\right) \rho_2 (1+\rho_2) \left(2+\rho_2^2\right) \text{Log}[\rho +\rho_2]}{4 \rho ^6 (-1+\rho_2)^2 \left(1+\rho_2+\rho_2^2\right) \left(1+2 \rho_2^2\right)^3} \\
 && + \frac{3 Q \left(Q^2-M \rho ^2+\rho ^6\right) \left(-2-\rho_2^2+M \rho_2^4\right) \text{Log}\left[1+\rho ^2+\rho_2^2\right]}{4 M \rho ^6 \left(2+\rho_2^2\right)^2 \left(1+2 \rho_2^2\right)^3}
\end{eqnarray}
}

{\scriptsize
\begin{eqnarray}
\nonumber F_2[\rho] &=& \frac{2 \sqrt{3} \kappa  \rho_2^6 \left(1+3 \rho_2^2 + 2 \rho_2^4\right)^2}{Q \rho ^6 \left(1+2 \rho_2^2\right)^2 \left(Q^2+\rho_2^6\right)} - 2 \sqrt{3} \lambda  \left(\frac{12 M \rho_2^6 \left(1+3 \rho_2^2+2 \rho_2^4\right)^2-2 Q^2 \rho ^2 \rho_2^2 \left(1+3 \rho_2^2+3 \rho_2^4+2 \rho_2^6\right)^2}{Q \rho ^{10} \left(1+2 \rho_2^2\right)^2 \left(Q^2+\rho_2^6\right)}\right.\\
\nonumber && +\frac{+6 \rho ^6 \left(Q^2+\rho_2^6\right)^2 \left(1+3 \rho_2^2+4 \rho_2^4+2 \rho_2^6\right)+4 M^2 \rho ^8 \rho_2^2 \left(2+7 \rho_2^2+9 \rho_2^4+4 \rho_2^6+2 \rho_2^8\right)}{Q \rho ^{10} \left(1+2 \rho_2^2\right)^2 \left(Q^2+\rho_2^6\right)} \\
\nonumber && +\left.\frac{-2 \rho ^4 \rho_2^2 \left(4+Q^2 \left(28+Q^2 \left(60+\left(43+34 Q^2\right) \rho_2^2 \left(1+\rho_2^2\right)\right)\right)\right)}{Q \rho ^{10} \left(1+2 \rho_2^2\right)^2 \left(Q^2+\rho_2^6\right)}\right) +\frac{32 \sqrt{3} M^2 \lambda  \left(Q^2-M \rho ^2+\rho ^6\right) \text{Log}[\rho ]}{Q^3 \rho ^6} \\
\nonumber && - \frac{8 \sqrt{3} Q \lambda  \left(Q^2-M \rho ^2+\rho ^6\right) \left(2+12 \rho_2^2+27 \rho_2^4+35 \rho_2^6+27 \rho_2^8+12 \rho_2^{10}+2 \rho_2^{12}\right) \text{Log}[\rho -\rho_2]}{\rho ^6 \rho_2^4 \left(1+2 \rho_2^2\right)^3}\\
\nonumber && - \frac{8 \sqrt{3} Q \lambda  \left(Q^2-M \rho ^2+\rho ^6\right) \left(2+12 \rho_2^2+27 \rho_2^4+35 \rho_2^6+27 \rho_2^8+12 \rho_2^{10}+2 \rho_2^{12}\right) \text{Log}[\rho +\rho_2]}{\rho ^6 \rho_2^4 \left(1+2 \rho_2^2\right)^3} \\
&& +\frac{8 \sqrt{3} \lambda  \left(Q^2-M \rho ^2+\rho ^6\right) \rho_2 \left(-1-3 \rho_2^2-6 \rho_2^4-7 \rho_2^6-3 \rho_2^8+2 \rho_2^{12}\right) \text{Log}\left[1+\rho ^2+\rho_2^2\right]}{\rho ^6 \left(1+\rho_2^2\right)^{3/2} \left(1+2 \rho_2^2\right)^3}  
\end{eqnarray}
}

{\scriptsize
\begin{eqnarray}
\nonumber F_3[\rho] &=& \sqrt{3} \left(\frac{9 Q^5 +2 M Q \rho ^4  \left(-M+2 \rho  \left(1+2 \rho_2^2\right)^2\right) }{8 M^2 \rho ^6  \left(1+2 \rho_2^2\right)^2}\right. \\
\nonumber && - \left.\frac{\left(\rho_2+\rho_2^3\right) \left(3 M Q \rho ^2 \rho_2 \left(2+\rho_2^2\right)-2 \pi  \left(-1+\rho ^2\right) \left(\rho ^2-\rho_2^2\right) \left(1+\rho ^2+\rho_2^2\right) \left(1+4 \rho_2^2+6 \rho_2^4+5 \rho_2^6+2 \rho_2^8\right)\right)}{8 M^2 \rho ^6 \left(2+\rho_2^2\right) \left(1+2 \rho_2^2\right)^2}\right)\\
\nonumber && \frac{\sqrt{3} \left(Q^2-M \rho ^2+\rho ^6\right) \rho_2 \left(1+\rho_2^2\right)^2 \text{ArcTan}\left[\frac{\rho }{\sqrt{1+\rho_2^2}}\right]}{2 M \rho ^6 \left(2+5 \rho_2^2+2 \rho_2^4\right)} + \frac{\sqrt{3} Q \left(Q^2-M \rho ^2+\rho ^6\right) \text{Log}[1+\rho ]}{2 \rho ^6 \left(-2-\rho_2^2+2 \rho_2^6+\rho_2^8\right)} \\
\nonumber && + \frac{\sqrt{3} Q \left(Q^2-M \rho ^2+\rho ^6\right) \rho_2^2 (4+\rho_2 (-1+4 \rho_2)) \text{Log}[\rho -\rho_2]}{4 \rho ^6 \left(1+2 \rho_2^2\right)^3 \left(1+\rho_2^3\right)} -\frac{\sqrt{3} Q \left(Q^2-M \rho ^2+\rho ^6\right) \rho_2^2 \left(4+\rho_2+4 \rho_2^2\right) \text{Log}[\rho +\rho_2]}{4 \rho ^6 \left(1+2 \rho_2^2\right)^3 \left(-1+\rho_2^3\right)} \\
 && +\frac{\sqrt{3} Q^3 \left(Q^2-M \rho ^2+\rho ^6\right) \left(-1+\rho_2^2\right)^2 \text{Log}\left[1+\rho ^2+\rho_2^2\right]}{4 \rho ^6 \rho_2^2 \left(1+2 \rho_2^2\right)^3 \left(2+3 \rho_2^2+3 \rho_2^4+\rho_2^6\right)}
\end{eqnarray}
}

{\scriptsize
\begin{eqnarray}
\nonumber F_4[\rho] &=& \frac{\kappa  \left(+9 Q^4 \rho ^2-3 M Q^2 \rho ^4+6 M Q^2 \rho ^6\right)}{M \rho ^8 \left(1+2 \rho_2^2\right)^2} - \lambda  \left(\frac{-120 Q^2 \rho ^2-180 Q^4 \rho ^2+18 M Q^4 \rho ^4+72 M Q^2 \rho ^6}{M \rho ^8 \left(1+2 \rho_2^2\right)^2}\right.\\
\nonumber && +\left.\frac{+2 Q^6 \rho ^2 \left(-67+6 \rho ^4\right)+4 \left(-5 \rho ^2+3 \rho ^6+6 \left(\rho_2+2 \rho_2^3\right)^2 \left(1+2 \rho_2^2+2 \rho_2^4+\rho_2^6\right)\right)}{M \rho ^8 \left(1+2 \rho_2^2\right)^2}\right)\\
\nonumber && +\frac{24 M \lambda  \left(Q^2-M \rho ^2+\rho ^6\right) \text{Log}[\rho ]}{Q^2 \rho ^6} +\frac{6 Q^2 \kappa  f[\rho] \text{Log}[\rho -\rho_2]}{\left(1+2 \rho_2^2\right)^3}  -\frac{12 \lambda  f[\rho] \left(1+\rho_2^2\right) \left(1+5 \rho_2^2+9 \rho_2^4+5 \rho_2^6+\rho_2^8\right) \text{Log}[\rho -\rho_2]}{\rho_2^2 \left(1+2 \rho_2^2\right)^3}\\
\nonumber && +\frac{6 \kappa  f[\rho]  Q^2 \text{Log}[\rho +\rho_2]}{\left(1+2 \rho_2^2\right)^3}-\frac{12 \lambda  f[\rho] Q^2\left(1+5 \rho_2^2+9 \rho_2^4+5 \rho_2^6+\rho_2^8\right) \text{Log}[\rho +\rho_2]}{\rho_2^4 \left(1+2 \rho_2^2\right)^3} \\
&& -\frac{6 Q^2 \kappa  f[\rho] \text{Log}\left[1+\rho ^2+\rho_2^2\right]}{\left(1+2 \rho_2^2\right)^3}+\frac{12 \lambda  f[\rho] \rho_2^2 \left(1+2 \rho_2^2-\rho_2^6+\rho_2^8\right) \text{Log}\left[1+\rho ^2+\rho_2^2\right]}{\left(1+\rho_2^2\right) \left(1+2 \rho_2^2\right)^3}
\end{eqnarray}
}
{\scriptsize
\begin{eqnarray}
\nonumber F_6[\rho] &=&\sqrt{3} Q \left(\frac{3 Q \rho_2 \left(2+5 \rho_2^2+2 \rho_2^4\right) \left(-8 M^2 \rho +9 (1+M) \rho_2^2+9 (1+M) \rho_2^4\right)}{8 M^2 \rho ^2 \rho_2 \left(-1+\rho_2^2\right) \left(2+\rho_2^2\right)^2 \left(1+2 \rho_2^2\right)^3}\right.\\
\nonumber && + \left.\frac{2 \pi  \left(3 Q^2-2 M \rho ^2\right) \left(-1+\rho_2^2\right)^2 \left(2+5 \rho_2^2+7 \rho_2^4+5 \rho_2^6+2 \rho_2^8\right)}{8 M^2 \rho ^2 \rho_2 \left(-1+\rho_2^2\right) \left(2+\rho_2^2\right)^2 \left(1+2 \rho_2^2\right)^3}\right) \\
\nonumber && + \frac{\sqrt{3} \left(-3 Q^2+2 M \rho ^2\right) \left(-2-3 \rho_2^2+3 \rho_2^6+2 \rho_2^8\right) \text{ArcTan}\left[\frac{\rho }{\sqrt{1+\rho_2^2}}\right]}{2 M \rho ^2 \sqrt{1+\rho_2^2} \left(2+\rho_2^2\right)^2 \left(1+2 \rho_2^2\right)^3} + \frac{\sqrt{3} (1+M) \left(3 Q^2-2 M \rho ^2\right) \text{Log}[1+\rho ]}{2 M \left(-2+Q^2\right)^2 \rho ^2} \\
\nonumber && +\frac{\sqrt{3} \left(-3 Q^2+2 M \rho ^2\right) (-1+\rho_2) \rho_2 \left(2+\rho_2^2\right) \text{Log}[\rho -\rho_2]}{4 \rho ^2 (1+\rho_2)^2 (1+(-1+\rho_2) \rho_2) \left(1+2 \rho_2^2\right)^3} + \frac{\sqrt{3} \left(-3 Q^2+2 M \rho ^2\right) \rho_2 (1+\rho_2) \left(2+\rho_2^2\right) \text{Log}[\rho +\rho_2]}{4 \rho ^2 (-1+\rho_2)^2 \left(1+\rho_2+\rho_2^2\right) \left(1+2 \rho_2^2\right)^3} \\
 && +\frac{\sqrt{3} \left(3 Q^2-2 M \rho ^2\right) \left(-2-\rho_2^2+\rho_2^4+\rho_2^6+\rho_2^8\right) \text{Log}\left[1+\rho ^2+\rho_2^2\right]}{4 M \rho ^2 \left(2+\rho_2^2\right)^2 \left(1+2 \rho_2^2\right)^3}
\end{eqnarray}
}
{\scriptsize
\begin{eqnarray}
\nonumber F_7[\rho] &=& \frac{6 \kappa  \rho_2^4 \left(1+3 \rho_2^2+2 \rho_2^4\right)^2}{M \rho ^2 \left(1+\rho_2^2\right) \left(\rho_2+2 \rho_2^3\right)^2} + \lambda  \left(\frac{4 Q^4 \left(120+77 Q^2+86 Q^4\right) \rho ^4-4 \left(-8 \rho ^4+9 M \rho_2^4 \left(1+3 \rho_2^2+2 \rho_2^4\right)^2\right)}{M \rho ^6 \left(1+\rho_2^2\right) \left(\rho_2+2 \rho_2^3\right)^2}\right.\\
\nonumber && \left.-\frac{+8 Q^2 \rho ^2 \left(-28 \rho ^2+3 \left(1+3 \rho_2^2+3 \rho_2^4+2 \rho_2^6\right)^2\right)}{M \rho ^6 \left(1+\rho_2^2\right) \left(\rho_2+2 \rho_2^3\right)^2}\right) + \frac{32 M^2 \lambda  \left(3 Q^2-2 M \rho ^2\right) \text{Log}[\rho ]}{Q^4 \rho ^2} \\
\nonumber && - \frac{8 \lambda  \left(3 Q^2-2 M \rho ^2\right) \left(2+12 \rho_2^2+27 \rho_2^4+35 \rho_2^6+27 \rho_2^8+12 \rho_2^{10}+2 \rho_2^{12}\right) \text{Log}[\rho -\rho_2]}{\rho ^2 \rho_2^4 \left(1+2 \rho_2^2\right)^3} \\
\nonumber && - \frac{8 \lambda  \left(3 Q^2-2 M \rho ^2\right) \left(2+12 \rho_2^2+27 \rho_2^4+35 \rho_2^6+27 \rho_2^8+12 \rho_2^{10}+2 \rho_2^{12}\right) \text{Log}[\rho +\rho_2]}{\rho ^2 \rho_2^4 \left(1+2 \rho_2^2\right)^3} \\
&& - \frac{8 \lambda  \left(-3 Q^2+2 M \rho ^2\right) \left(-1-3 \rho_2^2-6 \rho_2^4-7 \rho_2^6-3 \rho_2^8+2 \rho_2^{12}\right) \text{Log}\left[1+\rho ^2+\rho_2^2\right]}{\rho ^2 \left(1+\rho_2^2\right)^2 \left(1+2 \rho_2^2\right)^3}
\end{eqnarray}
}
{\scriptsize
\begin{eqnarray}
\nonumber F_8[\rho] &=& \frac{\left(1+\rho_2^2\right) \left(27 \rho_2^5 \left(1+\rho_2^2\right) \left(2+\rho_2^2\right)+2 \pi  Q \left(-3 Q^2+2 M \rho ^2\right) \left(1+3 \rho_2^2+3 \rho_2^4+2 \rho_2^6\right)\right)}{8 M^2 \rho ^2 \rho_2 \left(2+\rho_2^2\right) \left(1+2 \rho_2^2\right)^2} \\
\nonumber  && + \frac{Q^3 \left(3 Q^2-2 M \rho ^2\right) \text{ArcTan}\left[\frac{\rho }{\sqrt{1+\rho_2^2}}\right]}{2 \rho ^2 \rho_2^3 \left(2+7 \rho_2^2+9 \rho_2^4+7 \rho_2^6+2 \rho_2^8\right)} + \frac{\left(3 Q^2-2 M \rho ^2\right) \text{Log}[1+\rho ]}{2 \rho ^2 \left(-2-\rho_2^2+2 \rho_2^6+\rho_2^8\right)} \\
\nonumber && +\frac{\left(3 Q^2-2 M \rho ^2\right) \rho_2^2 (4+\rho_2 (-1+4 \rho_2)) \text{Log}[\rho -\rho_2]}{4 \rho ^2 \left(1+2 \rho_2^2\right)^3 \left(1+\rho_2^3\right)} + \frac{\left(-3 Q^2+2 M \rho ^2\right) \rho_2^2 \left(4+\rho_2+4 \rho_2^2\right) \text{Log}[\rho +\rho_2]}{4 \rho ^2 \left(1+2 \rho_2^2\right)^3 \left(-1+\rho_2^3\right)} \\
 && \frac{\left(3 Q^2-2 M \rho ^2\right) \left(-1+\rho_2^2\right)^2 \left(1+\rho_2^2\right) \text{Log}\left[1+\rho ^2+\rho_2^2\right]}{4 M \rho ^2 \left(2+\rho_2^2\right) \left(1+2 \rho_2^2\right)^3}
\end{eqnarray}
}
{\scriptsize
\begin{eqnarray}
\nonumber F_9[\rho] &=& \frac{9 \sqrt{3} Q^3 \kappa  \rho_2^2}{\rho ^2 \left(1+2 \rho_2^2\right)^2 \left(Q^2+\rho_2^6\right)} + \frac{2 \sqrt{3} \lambda  \rho_2^2 \left(\left(4+24 Q^2+36 Q^4+43 Q^6\right) \rho ^2-6 \left(\rho_2+2 \rho_2^3\right)^2 \left(1+2 \rho_2^2+2 \rho_2^4+\rho_2^6\right)\right)}{Q \rho ^4 \left(1+2 \rho_2^2\right)^2 \left(Q^2+\rho_2^6\right)} \\
\nonumber &&  +\frac{2 \sqrt{3} Q \kappa  \left(3 Q^2-2 M \rho ^2\right) \text{Log}[\rho -\rho_2]}{\rho ^2 \left(1+2 \rho_2^2\right)^3} -\frac{4 \sqrt{3} Q \lambda  \left(3 Q^2-2 M \rho ^2\right) \left(1+5 \rho_2^2+9 \rho_2^4+5 \rho_2^6+\rho_2^8\right) \text{Log}[\rho -\rho_2]}{\rho ^2 \rho_2^4 \left(1+2 \rho_2^2\right)^3} \\
\nonumber &&  +\frac{2 \sqrt{3} Q \kappa  \left(3 Q^2-2 M \rho ^2\right) \text{Log}[\rho +\rho_2]}{\rho ^2 \left(1+2 \rho_2^2\right)^3}-\frac{4 \sqrt{3} Q \lambda  \left(3 Q^2-2 M \rho ^2\right) \left(1+5 \rho_2^2+9 \rho_2^4+5 \rho_2^6+\rho_2^8\right) \text{Log}[\rho +\rho_2]}{\rho ^2 \rho_2^4 \left(1+2 \rho_2^2\right)^3} \\
\nonumber && -\frac{2 \sqrt{3} \kappa  \left(3 Q^2-2 M \rho ^2\right) \rho_2 \sqrt{1+\rho_2^2} \text{Log}\left[1+\rho ^2+\rho_2^2\right]}{\rho ^2 \left(1+2 \rho_2^2\right)^3}+\frac{4 \sqrt{3} \lambda  \left(3 Q^2-2 M \rho ^2\right) \rho_2 \left(1+2 \rho_2^2-\rho_2^6+\rho_2^8\right) \text{Log}\left[1+\rho ^2+\rho_2^2\right]}{\rho ^2 \left(1+\rho_2^2\right)^{3/2} \left(1+2 \rho_2^2\right)^3}\\
&&-\frac{8 \sqrt{3} M \lambda  \left(-3 Q^2+2 M \rho ^2\right) \text{Log}[\rho ]}{\rho ^2 \rho_2^3 \left(1+\rho_2^2\right)^{3/2}}
\end{eqnarray}
}

\section{Second Order Sources}
\label{ap:sources}

In this appendix we show the second order sources in the vector and tensor sector. The are splitted in terms of the anomalous and non anomalous ones.

\subsection{Second Order Vector Sources}
\label{ap:Vec_sour}
In the vector sector the sources are organized as follow
\begin{equation}
\mathbb{J}_{\mu} = \sum_{a=1}^{10}r_a^{(E)} \mathcal{J}^{(a)}_{\mu} + \sum_{a=1}^{5}\tilde r_a^{(E)} \mathcal{\tilde J}^{(a)}_{\mu}\qquad , \qquad \mathbb{A}_{\mu} = \sum_{a=1}^{10}r_a^{(M)} \mathcal{J}^{(a)}_{\mu} + \sum_{a=1}^{5}\tilde r_a^{(M)} \mathcal{\tilde J}^{(a)}_{\mu} \,,
\end{equation}
where the tildes refer to the anomalous sector. The sources write:

\subsubsection{Non-anomalous vector sources}
{\tiny
\begin{eqnarray}
\nonumber r_+^3 r_1^{(E)} &=& \frac{2 \pi  Q T^2 (-1+\rho )^2 \left(-\rho ^2 (1+\rho  (2+3 \rho ))+Q^2 (3+2 \rho  (3+2 \rho  (2+\rho )))\right)}{\sqrt{3}  M (1+M) \rho ^8 f[\rho ]^2} +\frac{2 \pi  T^2 \rho ^3 \partial_QF_5'[\rho ]}{\sqrt{3}  (1+M)} -\frac{4 \pi  T^2 (-1+\rho ) \left(1+\rho +\rho ^2\right) F_1'[\rho ]}{\sqrt{3}  (1+M) f[\rho ]} \\
 &&  \frac{4 \pi  T^2 (-1+\rho )^2 \left(Q^2 (6+\rho  (12+7 \rho  (2+\rho )))-\rho ^2 (4+\rho  (8+3 \rho  (4+\rho  (3+\rho  (2+\rho )))))\right) F_1[\rho ]}{\sqrt{3}  (1+M) \rho ^7 f[\rho ]^2} \,, \\
 \nonumber r_+^3 r_2^{(E)} &=& -\frac{64 \sqrt{3} \pi  Q T^2 \lambda ^2 \left(-9 Q^4 \left(242+131 Q^2\right)+M Q^2 \left(2654+1603 Q^2\right) \rho ^2-4 M^2 \left(146+127 Q^2\right) \rho ^4\right)}{ M (1+M) \rho ^{17} f[\rho ]}\\
 \nonumber && -\frac{64 \sqrt{3} \pi  Q T^2 \lambda ^2 \left(3 \left(16-498 Q^2-225 Q^4+16 Q^6\right) \rho ^6+220 \left(2+3 Q^2+Q^4\right) \rho ^8\right)}{ M (1+M) \rho ^{17} f[\rho ]}-\frac{2 \sqrt{3} \pi  T^2 \rho ^2 F_1[\rho ]}{ (1+M)}  -\frac{4 \pi  T^2 \rho ^3  F_1'[\rho ]}{\sqrt{3}  (1+M)}\\
\nonumber  && -\frac{768 \sqrt{3} \pi  Q^3 T^2 \kappa  \lambda  \left(-3+\rho ^2+Q^2 \left(-2+\rho ^2\right)\right)}{ M (1+M) \rho ^{11} f[\rho ]}+ \frac{64 \sqrt{3} \pi  T^2 \lambda  \left(\rho ^2+Q^2 \left(-5+\rho ^2\right)\right) \partial_Q  F_7[\rho ]}{ (1+M) \rho ^5}  -\frac{16 \pi  Q T^2 \lambda  \partial_QF_2'[\rho ]}{(1+M)}\\
 && + \frac{48 \pi  T^2 \lambda  \left(2 \rho ^2-2 \rho ^6+Q^4 \left(-5+3 \rho ^2\right)-Q^2 \left(8-5 \rho ^2+\rho ^6\right)\right) F_2'[\rho ]}{ M (1+M) \rho ^6 f[\rho ]} + \frac{2 \pi  Q T^2 F_6'[\rho ]}{\ (1+M) \rho }-\frac{16 \pi  Q T^2 \lambda  \rho  \partial_QF_2''[\rho ]}{ (1+M)}\\
\nonumber && + \frac{8 \sqrt{3} \pi  Q T^2 \lambda  \left(-3 Q^2 \left(2+Q^2\right)+M \left(18+7 Q^2\right) \rho ^2-3 \left(14+9 Q^2\right) \rho ^6+8 M \rho ^8\right) F_7'[\rho ]}{ M (1+M) \rho ^{10} f[\rho ]}-\frac{16 \sqrt{3} \pi  T^2 \lambda  \left(-5 Q^2+2 M \rho ^2\right) \partial_QF_7'[\rho ]}{ (1+M) \rho ^4}, \\
r_+ r_3^{(E)} &=& \rho^3 F_5'(\rho) \,, \\
 r_+ r_4^{(E)} &=& -\frac{24192 Q^4 \lambda ^2-7680 M Q^2 \lambda ^2 \rho ^2+\rho ^{12}}{\rho ^{11}} +\frac{96 \lambda  \left(\rho ^2+Q^2 \left(-5+\rho ^2\right)\right) F_7[\rho ]}{ \rho ^5} -\frac{24 \lambda  \left(-7 Q^2+2 M \rho ^2\right) F_7'[\rho ]}{ \rho ^4} + 32 \sqrt{3} Q \lambda  F_2'[\rho ] \,, \\
\nonumber r_+^2 r_5^{(E)} &=& \frac{\sqrt{3} Q (-1+\rho )^2 \left(\rho ^2 (1+\rho  (2+3 \rho ))-Q^2 (3+2 \rho  (3+2 \rho  (2+\rho )))\right)}{\ M \rho ^8 f[\rho ]^2} -\frac{2 (-1+\rho ) \left(1+\rho +\rho ^2\right) F_3'[\rho ]}{ f[\rho ]}-\frac{2 (-1+\rho ) \left(1+\rho +\rho ^2\right) F_3'[\rho ]}{ f[\rho ]} \\
&& + \frac{2Q^2 (-1+\rho )^2  (6+\rho  (12+7 \rho  (2+\rho ))) F_3[\rho ]}{ \rho ^7 f[\rho ]^2}- \frac{2\rho ^2 (-1+\rho )^2  (4+\rho  (8+3 \rho  (4+\rho  (3+\rho  (2+\rho ))))) F_3[\rho ]}{ \rho ^7 f[\rho ]^2} \,, 
\end{eqnarray}
}
{\tiny
\begin{eqnarray}
\nonumber r_+^2 r_6^{(E)} &=& -\frac{96 \sqrt{3} Q \lambda  \left(-999 Q^6 \lambda +1813 M Q^4 \lambda  \rho ^2-978 M Q^2 \lambda  \rho ^4+2 M Q^2 (4 \kappa +341 \lambda ) \rho ^8-120 M^2 \lambda  \rho ^{10}\right)}{ M \rho ^{17} f[\rho ]}- 2 \rho ^3 F_3'[\rho ] + \frac{\sqrt{3} Q F_8'[\rho ]}{ \rho }\\
\nonumber && -\frac{96 \sqrt{3} Q \lambda  \left(\left(152 \lambda +456 Q^2 \lambda +152 Q^6 \lambda -3 Q^4 (4 \kappa +121 \lambda )\right) \rho ^6\right)}{ M \rho ^{17} f[\rho ]} -\frac{24 \lambda  \left(9 Q^4-8 M Q^2 \rho ^2+M^2 \rho ^4+3 Q^2 \rho ^6-M \rho ^8\right) F_2'[\rho ]}{ M \rho ^6 f[\rho ]}\\
&& + \frac{4 \sqrt{3} Q \lambda  \left(9 Q^4-19 M Q^2 \rho ^2+6 M^2 \rho ^4+45 Q^2 \rho ^6-22 M \rho ^8\right) F_7'[\rho ]}{ M (-1+\rho ) \rho ^4 (1+\rho ) \left(Q^2-\rho ^2 \left(1+\rho ^2\right)\right)} -3 \rho ^2 F_3[\rho ]\,, \\
r_7^{(E)} &=& 0 \,, \\
\nonumber r_+^4 r_8^{(E)} &=& -\frac{384 \pi  T^2 \lambda  \left(5 Q^4 \left(14+Q^2\right) \lambda -4 M Q^2 \left(22+Q^2\right) \lambda  \rho ^2-M^2 \left(-18+Q^2\right) \lambda  \rho ^4\right)}{ M (1+M) \rho ^{15} f[\rho ]}-\frac{32 \sqrt{3} \pi  Q T^2 \lambda  \left(\rho ^2+Q^2 \left(-5+\rho ^2\right)\right) F_9[\rho ]}{ M (1+M) \rho ^5}\\
\nonumber && -\frac{384 \pi  T^2 \lambda  \left(Q^2 \left(\left(6+4 Q^2\right) \kappa +5 \left(14+Q^2\right) \lambda \right) \rho ^6+M \left(-18 \lambda +Q^2 (-2 \kappa +\lambda )\right) \rho ^8\right)}{ M (1+M) \rho ^{15} f[\rho ]}- \frac{64 \sqrt{3} \pi  T^2 \lambda  \left(\rho ^2+Q^2 \left(-5+\rho ^2\right)\right) \partial_Q F_9[\rho ]}{ (1+M) \rho ^5}\\
\nonumber &&  -\frac{16 \pi  T^2 \lambda  \left(-Q^2 \left(24+13 Q^2\right)+M \left(6+7 Q^2\right) \rho ^2-\left(6+Q^2\right) \rho ^6\right) F_4'[\rho]}{ M (1+M) \rho ^6 f[\rho ]}\\
\nonumber && -\frac{64 \pi  Q T^2 \lambda  \partial_QF_4'[\rho]}{ (1+M)} -\frac{2 \pi  T^2 \rho  F_6'[\rho ]}{\sqrt{3}  (1+M)} + \frac{16 \sqrt{3} \pi  T^2 \lambda  \left(-7 Q^2+2 M \rho ^2\right) \partial_Q F_9'[\rho ]}{ (1+M) \rho ^4}-\frac{16 \sqrt{3} \pi  Q^3 T^2 \lambda  \left(-1+4 \rho ^2+2 \rho ^4-8 \rho ^6+3 \rho ^8\right) F_9'[\rho ]}{ M (1+M) \rho ^{10} f[\rho ]}\\
\nonumber && -\frac{16 \sqrt{3} \pi  Q T^2 \lambda  \left(7 \rho ^2+\rho ^4-19 \rho ^6+3 \rho ^8+Q^4 \left(4-3 \rho ^2+\rho ^4\right)\right) F_9'[\rho ]}{ M (1+M) \rho ^{10} f[\rho ]}  \,, \\
\nonumber r_+^3 r_{9}^{(E)} &=& -\frac{48 \sqrt{3} Q \lambda  \left(\rho ^2+Q^2 \left(-5+\rho ^2\right)\right) F_9[\rho ]}{ M \rho ^5}  + \rho  F_8'[\rho ] -\frac{24 \lambda  \left(\rho ^4-\rho ^8-Q^2 \rho ^2 \left(-5+\rho ^2\right) \left(-2+\rho ^4\right)+Q^4 \left(11-10 \rho ^2+\rho ^4\right)\right) F_4'[\rho ]}{ M \rho ^6 f[\rho ]} \\
 && -\frac{8 \sqrt{3} Q \lambda  \left(15 Q^4+2 M \rho ^4 \left(3 M-7 \rho ^4\right)+Q^2 \left(-23 M \rho ^2+33 \rho ^6\right)\right) F_9'[\rho ]}{ M \rho ^{10} f[\rho ]}\\
 \nonumber r_+^2 r_{10}^{(E)} && \frac{8\sqrt{3} \lambda ^2 \left(-480  Q^3+96  Q \rho ^2+96  Q^3 \rho ^2\right)}{ \rho ^9}+\frac{96 \lambda  \left(\rho ^6+Q^2 \rho ^4 \left(-5+\rho ^2\right)\right) F_9[\rho ]}{ \rho ^9}+32 \sqrt{3} Q \lambda  F_4'[\rho ]+\frac{8 \lambda  \left(21 Q^2 \rho ^5-6 \rho ^7-6 Q^2 \rho ^7\right) F_9'[\rho ]}{ \rho ^9} \,,\\
\end{eqnarray}
}
{\tiny
\begin{eqnarray}
\nonumber r_+^3 r_1^{(M)} &=& \frac{\pi  T^2 (-1+\rho ) \left(8 \rho ^2 \left(1+\rho +\rho ^2\right)+Q^4 \left(3-\rho  \left(-3+4 \rho  (1+2 \rho )^2\right)\right)\right)}{2  M^2 (1+M) \rho ^8 f[\rho ]} + \frac{\pi  T^2Q^2 (-1+\rho )  \left(12+\rho  \left(12+\rho  \left(13+\rho +\rho ^2+9 \rho ^3\right)\right)\right)}{2  M^2 (1+M) \rho ^8 f[\rho ]}\\
&& + \frac{4 \pi  T^2 F_6'[\rho ]}{\sqrt{3}  (1+M) \rho ^2}  -\frac{8 \pi  Q T^2 (-1+\rho ) \left(1+\rho +\rho ^2\right) F_1[\rho ]}{ (1+M) \rho ^5 f[\rho ]} -\frac{3 \pi  Q T^2 \rho ^3 f[\rho ] \partial_QF_5'[\rho ]}{2  M (1+M)} \,, \\
 \nonumber r_+^3 r_2^{(M)} &=&  \frac{\pi  T^2 \left(Q^6 \left(-3+4 \rho ^2\right)+8 \left(\rho ^4+\rho ^6\right)+Q^4 \left(6+\rho ^2 \left(-1+\rho  \left(-3-\rho +8 \rho ^3\right)\right)\right)\right)}{4  M^2 (1+M) \rho ^5 \left(Q^2-\rho ^2 \left(1+\rho ^2\right)\right)}\\
\nonumber  && +\frac{\pi  T^2 Q^2 \rho ^2 (14+\rho  (12+\rho  (2+\rho  (9+\rho  (-16+9 \rho ))))))}{4  M^2 (1+M) \rho ^5 \left(Q^2-\rho ^2 \left(1+\rho ^2\right)\right)} -\frac{4 \pi  Q T^2 F_1[\rho ]}{ (1+M) \rho ^2} -\frac{32 \pi  Q T^2 \kappa  \partial_QF_7[\rho ]}{ (1+M) \rho ^3} + \frac{\pi  Q T^2 \left(3 Q^2-7 M \rho ^2+3 \rho ^6\right) F_1'[\rho ]}{2  M (1+M) \rho ^3}+\\
\nonumber && \frac{24 \sqrt{3} \pi  Q T^2 \lambda  \left(-4-3 Q^2+2 M \rho ^2\right) F_2'[\rho ]}{ M (1+M) \rho ^4} -\frac{16 \sqrt{3} \pi  T^2 \lambda  \left(-5 Q^2+2 M \rho ^2\right) \partial_QF_2'[\rho ]}{ (1+M) \rho ^4} -\frac{3 \pi  Q T^2 \rho ^3 f[\rho ] \partial_QF_5'[\rho ]}{4  M (1+M)} -\frac{3 \sqrt{3} \pi  Q^2 T^2 f[\rho ] F_6'[\rho ]}{2  M (1+M) \rho }\\
&& + \frac{8 \pi  T^2 \left(-6 Q^2 \left(4+3 Q^2\right) \lambda +12 M Q^2 \lambda  \rho ^2-\left(2+Q^2\right) \kappa  \rho ^6\right) F_7'[\rho ]}{ M (1+M) \rho ^8} +\frac{384 \pi  Q^2 T^2 \lambda ^2 \left(-4-3 Q^2+2 M \rho ^2\right) \left(-63 Q^2+20 M \rho ^2\right)}{ M (1+M) \rho ^{15}}\,, \\
r_+ r_3^{(M)} &=& -\frac{3 \sqrt{3} Q}{2  M \rho ^2} \,, \\
\nonumber r_+ r_4^{(M)} &=& -\frac{\sqrt{3} Q \left(\rho ^{12}-3840 \lambda ^2 \rho ^4 \left(-1+\rho ^4\right)\right)}{ \rho ^{15}}  -\frac{48 \sqrt{3}  Q \lambda  \left(-5 Q^2+2 M \rho ^2\right)F_7[1]}{ \rho ^9}  + \frac{16 \sqrt{3} Q \left(-15 Q^2 \lambda +6 M \lambda  \rho ^2-\kappa  \rho ^6\right) F_7[\rho ]}{ \rho ^9}\\
&& \frac{24  \lambda  \left(5 Q^2-2 M \rho ^2\right)F_2'[1]}{ \rho ^9}-\frac{\sqrt{3} Q \left(384 Q^2 \lambda ^2 \rho ^2 \left(-46+20 \rho ^2+25 \rho ^4+\rho ^6\right)+192 Q^4 \lambda ^2 \left(105-92 \rho ^2+20 \rho ^4-55 \rho ^6+22 \rho ^8\right)\right)}{ \rho ^{15}} \,, \\
\nonumber r_+^2 r_5^{(M)} &=& -\frac{(-1+\rho ) \left(8 \rho ^4 \left(1+\rho +\rho ^2\right)+Q^4 \left(-27+\rho  \left(-27+4 \rho  \left(1+\rho +\rho ^2\right) \left(-9+2 \rho ^2\right)\right)\right)\right)}{4  M^2 \rho ^8 f[\rho ]}-F_8[\rho ] -\frac{2 \left(-1+\rho ^3\right) F_8'[\rho ]}{ \rho ^2} \\
 && -\frac{(-1+\rho ) \left(Q^2 \rho ^2 (-9+\rho  (-9+\rho  (7+\rho  (43+16 \rho ))))\right)}{4  M^2 \rho ^8 f[\rho ]} -\frac{4 \sqrt{3} Q (-1+\rho ) \left(1+\rho +\rho ^2\right) F_3[\rho ]}{ \rho ^5 f[\rho ]} \,,
 \end{eqnarray}
 }
 {\tiny
 \begin{eqnarray}
\nonumber r_+^2 r_6^{(M)} &=& \frac{3 Q^2 \left(512 M \lambda ^2 \left(-3 Q^2+2 M \rho ^2\right) \left(-63 Q^2+20 M \rho ^2\right)+9 Q^2 \rho ^{10}-6 M \rho ^{12}-9 \rho ^{13}\right)}{8 \ M^2 \rho ^{15}} -\frac{3 \sqrt{3}  QF_3'[1]}{8  M \rho ^2} -\frac{9 Q^2F_8[1] }{4  M \rho ^2}\\
\nonumber &&  -\frac{2 \sqrt{3} Q F_3[\rho ]}{ \rho ^2} -\frac{1}{2 }F_8[\rho ] + \frac{12 \sqrt{3} Q \lambda  \left(-3 Q^2+2 M \rho ^2\right) F_2'[\rho ]}{ M \rho ^4}+ \frac{\sqrt{3} Q \left(3 Q^2-7 M \rho ^2+3 \rho ^6\right) F_3'[\rho ]}{4  M \rho ^3} \\
&&+ \frac{4 \left(-3 Q^2+2 M \rho ^2\right) \left(6 Q^2 \lambda +\kappa  \rho ^6\right) F_7'[\rho ]}{ M \rho ^8} + \frac{\left(-4 \rho ^8+9 Q^4 \left(-1+\rho ^2\right)+Q^2 \rho ^2 \left(9-9 \rho ^4-4 \rho ^6\right)\right) F_8'[\rho ]}{4  M \rho ^7}\, ,\\
r_7^{(M)} &=& -\frac{3 \sqrt{3}  QF_3'[1]}{4 \rho ^2}-\frac{9 Q^2F_8[1]}{2 M \rho ^2} - F_8[\rho ] - 2 \rho  F_8'[\rho ]\, ,\\
\nonumber r_+^4 r_8^{(M)} &=&-\frac{3 \sqrt{3} \pi  Q T^2 \left(\pi  T \rho ^{10}+512 r_+ \lambda ^2 \left(35 Q^4+15 Q^6+2 M^2 \rho ^2 \left(-2+M \rho ^2\right)+Q^2 \left(20-13 M^2 \rho ^2\right)\right)\right)}{2 r_+ M^2 (1+M) \rho ^{13}}  + \frac{\pi  T^2 \left(9 Q^2+2 M \rho ^2\right) F_1[\rho ]}{\sqrt{3} \text{b0}^4 M (1+M) \rho ^2}\\
\nonumber && + \frac{16 \pi  Q^2 T^2 \kappa  F_9[\rho ]}{ M (1+M) \rho ^3} + \frac{32 \pi  Q T^2 \kappa  \partial_QF_9[\rho ]}{ (1+M) \rho ^3} + \frac{2 \pi  T^2 \rho  F_1'[\rho ]}{\sqrt{3}  (1+M)} -\frac{32 \sqrt{3} \pi  Q T^2 \lambda  \left(-3+\rho ^2+Q^2 \left(-1+\rho ^2\right)\right) F_4'[\rho ]}{ M (1+M) \rho ^4} \\
&& +\frac{16 \sqrt{3} \pi  T^2 \lambda  \left(-5 Q^2+2 M \rho ^2\right) \partial_Q F_4'[\rho ]}{ (1+M) \rho ^4} + \frac{3 \pi  Q T^2 \left(Q^2-M \rho ^2+\rho ^6\right) F_6'[\rho ]}{2  M (1+M) \rho ^5} \,,\\
\nonumber r_+^3 r_{9}^{(M)} &=& \frac{3 \sqrt{3} Q \left(1024 M \lambda ^2 \left(15 Q^4-13 M Q^2 \rho ^2+2 M^2 \rho ^4\right)+9 Q^2 \rho ^{10}-2 M \rho ^{12}\right)}{8  M^2 \rho ^{13}} - \rho  F_3'[\rho ]  -\frac{\left(9 Q^2+2 M \rho ^2\right) F_3[\rho ]}{2  M \rho ^2} + \frac{24 Q^2 \kappa  F_9[\rho ]}{ M \rho ^3}\\
&&   + \frac{48 \sqrt{3} Q \lambda  \left(\rho ^2+Q^2 \left(-2+\rho ^2\right)\right) F_4'[\rho ]}{ M \rho ^4} -\frac{3 \sqrt{3} Q \rho  f[\rho ] F_8'[\rho ]}{4  M} + \frac{4 \left(-3 Q^2+2 M \rho ^2\right) \left(6 Q^2 \lambda +\kappa  \rho ^6\right) F_9'[\rho ]}{ M \rho ^8} \, ,\\
r_+^2 r_{10}^{(M)} &=& -\frac{1}{ \rho }-\frac{16 \sqrt{3} Q \kappa  F_9[\rho ]}{ \rho ^3}-\frac{24 \lambda  \left(2 \rho ^2+Q^2 \left(-5+2 \rho ^2\right)\right) F_4'[\rho ]}{ \rho ^4}\,,
\end{eqnarray}
}
\subsubsection{Anomalous vector sources}
{\tiny
\begin{eqnarray}
\nonumber r_+\tilde r_1^{(E)} &=& \frac{16 \sqrt{3} Q \lambda  \left(2 \rho ^2 (-3+\rho  (-3+\rho  (-3+2 \rho )))+Q^2 (17+\rho  (17+\rho  (11-10 \rho  (1+\rho ))))\right)}{\rho ^5 (1+\rho ) \left(-Q^2+\rho ^2+\rho ^4\right)}-\frac{16 \sqrt{3} Q \lambda  \left(-63 Q^2+20 M \rho ^2\right) F_5[\rho]}{ \rho^7} \\
\nonumber && -\frac{2 \rho ^5 \left(-Q^2 (6+\rho  (12+7 \rho  (2+\rho )))+\rho ^2 (4+\rho  (8+3 \rho  (4+\rho  (3+\rho  (2+\rho )))))\right) F_2[\rho ]}{(1+\rho )^2 \left(-Q^2+\rho ^2+\rho ^4\right)^2} -\frac{2 (-1+\rho ) \left(1+\rho +\rho ^2\right) F_2'[\rho ]}{\ f[\rho ]} \,, \\
&&\\
\nonumber r_+^2 \tilde r_2^{(E)} &=& -\frac{48 \lambda  \left(\rho ^2+\rho ^3+\rho ^4-\rho ^5+Q^2 (-4+\rho  (-4+\rho  (-3+2 \rho  (1+\rho ))))\right)}{ \rho ^3 (1+\rho ) \left(-Q^2+\rho ^2+\rho ^4\right)}-\frac{96 \lambda  \left(\rho ^2+Q^2 \left(-5+\rho ^2\right)\right) F_5[\rho ]}{ \rho ^5}\\
\nonumber && -\frac{2 \rho ^5 \left(-Q^2 (6+\rho  (12+7 \rho  (2+\rho )))+\rho ^2 (4+\rho  (8+3 \rho  (4+\rho  (3+\rho  (2+\rho )))))\right) F_4[\rho ]}{ (1+\rho )^2 \left(-Q^2+\rho ^2+\rho ^4\right)^2}  -\frac{2 (-1+\rho ) \left(1+\rho +\rho ^2\right) F_4'[\rho ]}{ f[\rho ]} \,, \\
&&\\
r_+^2\tilde r_3^{(E)} &=& \frac{-120 Q^2 \lambda +\frac{\sqrt{3} Q F_9'[\rho ]}{ \rho }+48 \lambda  \rho ^2+48 Q^2 \lambda  \rho ^2}{ \rho ^6}-3 \rho ^2 F_4[\rho ]-2 \rho ^3 F_4'[\rho ]-\rho  F_7'[\rho ] \,, \\
\nonumber r_+^4 \tilde r_4^{(E)} &=& -\frac{8 \sqrt{3} \pi  Q T^2 \lambda  \left(2 \rho ^2 \left(-5+\rho ^2\right)+Q^2 \left(-21+7 \rho ^2+2 \rho ^4\right)\right)}{ M (1+M) \rho ^8 f[\rho ]} -\frac{16 \sqrt{3} \pi  T^2 \lambda  \left(9 Q^4-8 M Q^2 \rho ^2+M^2 \rho ^4+3 Q^2 \rho ^6-M \rho ^8\right) F_1'[\rho ]}{ M (1+M) \rho ^6 f[\rho ]}\\
\nonumber && -\frac{64 \sqrt{3} \pi  T^2 \lambda  \left(\rho ^2+Q^2 \left(-5+\rho ^2\right)\right) \partial_Q F_8[\rho ]}{ (1+M) \rho ^5} -\frac{32 \sqrt{3} \pi  Q T^2 \lambda  \left(\rho ^2+Q^2 \left(-5+\rho ^2\right)\right) F_8[\rho ]}{ M (1+M) \rho ^5}-\frac{64 \pi  Q T^2 \lambda  \partial_Q F_3'[\rho ]}{ (1+M)} \\
\nonumber &&+ \frac{16 \pi  T^2 \lambda  \left(Q^4 \left(13-7 \rho ^2\right)+6 \rho ^2 \left(-1+\rho ^4\right)+Q^2 \left(24-13 \rho ^2+\rho ^6\right)\right) F_3'[\rho ]}{ M (1+M) \rho ^6 f[\rho ]}+ \frac{16 \sqrt{3} \pi  T^2 \lambda  \left(-7 Q^2+2 M \rho ^2\right) \partial_Q F_8'[\rho ]}{ (1+M) \rho ^4}\\
\nonumber  &&  -\frac{8 \pi  Q T^2 \lambda  \left(9 Q^4-19 M Q^2 \rho ^2+6 M^2 \rho ^4+45 Q^2 \rho ^6-22 M \rho ^8\right) F_6'[\rho ]}{ M (1+M) \rho ^{10} f[\rho ]}-\frac{16 \sqrt{3} \pi  Q^3 T^2 \lambda   \left(-1+4 \rho ^2+2 \rho ^4-8 \rho ^6+3 \rho ^8\right) F_8'[\rho ]}{ M (1+M) \rho ^{10} f[\rho ]}\\
&& -\frac{16 \sqrt{3} \pi  Q T^2 \lambda  \left(7 \rho ^2+\rho ^4-19 \rho ^6+3 \rho ^8+Q^4 \left(4-3 \rho ^2+\rho ^4\right)\right) F_8'[\rho ]}{ M (1+M) \rho ^{10} f[\rho ]}\,, \\
\nonumber r_+^2\tilde r_5^{(E)} &=& 32 \sqrt{3} Q \lambda  F_3'[\rho ] + \frac{96 \lambda  \left(\rho ^2+Q^2 \left(-5+\rho ^2\right)\right) F_8[\rho ]}{ \rho ^5}-\frac{24 \lambda  \left(2 \rho ^2+Q^2 \left(-7+2 \rho ^2\right)\right) F_8'[\rho ]}{ \rho ^4} \,, \\
\end{eqnarray}
}

{\tiny
\begin{eqnarray}
\nonumber r_+\tilde r_1^{(M)} &=& \frac{12 \lambda  \left(-63 Q^4 \rho +8 M^2 \rho ^2+2 M Q^2 \left(-8+15 \rho ^3\right)\right)}{ M \rho ^9} -\frac{4 \sqrt{3} Q (-1+\rho ) \left(1+\rho +\rho ^2\right) F_2[\rho ]}{ \rho ^5 f[\rho ]}-\frac{3 \sqrt{3} Q \rho ^3 F_2'[\rho ]}{2 M} + \frac{2 \left(1-2 \rho ^3\right) F_7'[\rho ]}{\ \rho ^2}\\
 && + \frac{48 \left(15 Q^4 \lambda +4 M^2 \lambda  \rho ^4+Q^2 \left(-16 M \lambda  \rho ^2+\kappa  \rho ^6\right)\right) F_5[\rho ]}{ \rho ^{11}} -\frac{\left(9 Q^2+2 M \rho ^2\right) F_7[\rho ]}{ M \rho ^2} \,,  \\
\nonumber r_+^2 \tilde r_2^{(M)} &=& -\frac{4 \sqrt{3} Q (-1+\rho ) \left(1+\rho +\rho ^2\right) F_4[\rho ]}{ \rho ^5 f[\rho ]} + \frac{16 \sqrt{3} Q \kappa  F_5[\rho ]}{ \rho ^3} -\frac{3 \sqrt{3} Q \rho ^3 F_4'[\rho ]}{2  M}+ \frac{2 \left(1-2 \rho ^3\right) F_9'[\rho ]}{ \rho ^2}\\
 && \frac{36 \sqrt{3} Q \lambda  \left(-5 Q^2+2 M \rho ^2\right)-\left(9 Q^2 \rho ^4+2 M \rho ^6\right) F_9[\rho ]}{ M \rho ^6}  \,, \\
\nonumber r_+^2\tilde r_3^{(M)} &=& -\frac{9 Q^2F_2'[1] }{4  M^2 \rho ^2} -\frac{9 \sqrt{3}  Q^3F_7[1]}{2 M^2 \rho ^2} + \frac{3 \sqrt{3} F_4'[1] Q}{2  M \rho ^2} +  \frac{9 Q^2F_9[1]}{ M \rho ^2}+ \left(1+\frac{9 Q^2}{2M \rho ^2}\right) F_2[\rho ] -\frac{2 \sqrt{3} Q F_4[\rho ]}{ \rho ^2} + \frac{9 \sqrt{3} Q^3 F_7[\rho ]}{4  M^2 \rho ^2}\\
\nonumber && \frac{9 \sqrt{3} Q \lambda  \left(10 MQ^2 \left(Q^2-2 \rho ^6\right)+\rho ^4 \left(4+12 Q^2+33 Q^4+4 Q^6-2 \left(4-12 Q^2+5 Q^4\right) \rho ^6\right)\right)}{ M^2 \rho ^{12}}\\
\nonumber && \frac{9 \sqrt{3} Q \lambda  M^2 \left(-14 Q^2 \rho ^2+4 \rho ^8\right)}{ M^2 \rho ^{12}}  -\frac{9 Q^2 F_9[\rho ]}{ M \rho ^2} + \frac{\rho  \left(8 M^2+9 Q^2 \rho ^2\right) F_2'[\rho ]}{8  M^2} -\frac{\sqrt{3} Q \left(-3 Q^2+7 M \rho ^2+3 \rho ^6\right) F_4'[\rho ]}{4 M \rho ^3}\\
&& \frac{3 \sqrt{3} Q \left(Q^2-M \rho ^2+\rho ^6\right) F_7'[\rho ]}{4  M \rho ^5} -\frac{9 Q^2 f[\rho ] F_9'[\rho ]}{4  M \rho }  \,, \\
\nonumber r_+^4 \tilde r_4^{(M)} &=&  -\frac{18 \pi ^2 Q^2 T^3 \lambda  \left(-21 Q^2+10 M \rho ^2\right)}{r_+ M^3 (1+M) \rho ^8} + \frac{\pi ^2 T^3 \left(9 Q^2+2 M \rho ^2\right) F_7[\rho ]}{2 r_+ M^3 (1+M) \rho ^2} + \frac{16 \pi  Q^2 T^2 \kappa  F_8[\rho ]}{ M (1+M) \rho ^3}+ \frac{32 \pi  Q T^2 \kappa  \partial_Q F_8[\rho ]}{ (1+M) \rho ^3} \\
\nonumber && + \frac{24 \pi  Q T^2 \lambda  \left(-3 Q^2+2 M \rho ^2\right) F_1'[\rho ]}{ M (1+M) \rho ^4} + \frac{3 \sqrt{3} \pi ^2 Q T^3 \rho ^3 F_2'[\rho ]}{4 r_+ M^3 (1+M)}-\frac{32 \sqrt{3} \pi  Q T^2 \lambda  \left(-3+\rho^2+Q^2 \left(-1+\rho^2\right)\right) F_3'[\rho]}{ (1+M) \rho ^4}  \\
\nonumber && + \frac{16 \sqrt{3} \pi  T^2 \lambda  \left(-5 Q^2+2 M \rho^2\right) \partial_Q F_3'[\rho ]}{ (1+M) \rho ^4}+ \frac{8 \pi  T^2 \left(-3 Q^2+2 M \rho ^2\right) \left(6 Q^2 \lambda +\kappa  \rho ^6\right) F_6'[\rho]}{\sqrt{3}  M (1+M) \rho ^8} + \frac{2 \pi ^2 T^3 \rho  F_7'[\rho]}{r_+ M^2 (1+M)} \\
&& + \frac{8 \pi  T^2 \left(6 Q^2 \left(4+3 Q^2\right) \lambda -12 M Q^2 \lambda  \rho ^2+\left(2+Q^2\right) \kappa  \rho ^6\right) F_8'[\rho]}{M (1+M) \rho^8} \,,\\
\nonumber  r_+^2\tilde r_5^{(M)} &=&  -\frac{9 Q^2F_2'[1] }{8 M^2 \rho ^2} -\frac{9 \sqrt{3} Q^3F_7[1] }{4 M^2 \rho ^2} + \frac{3 \sqrt{3} QF_4'[1]}{4 M \rho ^2} + \frac{9 Q^2F_9[1] }{2  M \rho ^2} -\frac{9 \sqrt{3} Q \left(4-12 Q^2+5 Q^4\right) \lambda }{ M^2 \rho ^2}\\
 && -\frac{\sqrt{3} Q F_7[\rho ]}{2  M} -\frac{16 \sqrt{3} Q \kappa  F_8[\rho ]}{ \rho ^3} + F_9[\rho ] -\frac{24 \lambda  \left(-5 Q^2+2 M \rho ^2\right) F_3'[\rho ]}{ \rho ^4}  -\frac{\sqrt{3} Q \rho  F_7'[\rho ]}{M} + 2 \rho  F_9'[\rho ] \,.
\end{eqnarray}
}

\subsection{Second Order Tensorial Sources}
\label{ap:Ten_sour}

In the tensor sector the sources are also splitted in terms of the anomalous and non anomalous, the tildes refer to the anomalous sector
\begin{equation}
\mathbb P_{\mu\nu} = \sum_{a=1}^{12}P_a \mathcal{T}^{(a)}_{\mu\nu} + \sum_{a=1}^{8}\tilde P_a \mathcal{\tilde T}^{(a)}_{\mu\nu} \,.
\end{equation}

\subsubsection{Non-anomalous tensorial sources}
{\tiny
\begin{eqnarray}
P_1 &=& 2 \rho -3 \rho ^2 F_5[\rho ]-2 \rho ^3 F_5'[\rho ] \,, \\
P_2 &=& 4 \rho -6 \rho ^2 F_5[\rho ]+\frac{\left(Q^2-M \rho ^2+\rho ^6\right) F_5'[\rho ]^2}{\rho } \,, \\
\nonumber P_3 &=& \frac{768\lambda Q^4 \left(\rho ^2 \left(-5 M \lambda -3 (\kappa -29 \lambda ) \rho ^4\right)+2 Q^2 \lambda \left(9+4 \rho ^6\right)\right)}{\rho ^{11} \left(Q^2-M \rho ^2+\rho ^6\right)} + 3 \rho ^2 F_5[\rho ] + 2 \rho ^3 F_5'[\rho ]+ \frac{32 \sqrt{3} Q \lambda  \left(Q^2-2 \rho ^6\right) F_2'[\rho ]}{Q^2-M \rho ^2+\rho ^6} \\
 \nonumber && +\frac{2 Q^2  \left(9216 \lambda ^2 (\rho ^2 - M^2) +768 M (\kappa -10 \lambda ) \lambda  \rho ^4+\rho ^8\right)+\rho ^2 \left(3072 \lambda ^2-M \rho ^8+\rho ^{12}\right)}{\rho ^{7} \left(Q^2-M \rho ^2+\rho ^6\right)}  +\frac{32 \lambda  \left(-3 Q^2+2 M \rho ^2\right) \left(Q^2-2 \rho ^6\right) F_7'[\rho ]}{\rho ^4 \left(Q^2-M \rho ^2+\rho ^6\right)} \,, \\
 &&\\
\nonumber  P_4 &=&  \lambda ^2 \left(Q^4 \left(\frac{644352 M}{\rho ^{15}}-\frac{396288}{\rho ^{11}}\right)+Q^6 \left(-\frac{549504}{\rho ^{17}}-\frac{3072}{\rho ^{11}}\right) -\frac{3072}{\rho ^{11}}\right) +\frac{2 Q^2}{\rho ^5}+\frac{M}{\rho ^3}+\rho  \lambda^2 Q^2 \left(-\frac{130560 M^2}{\rho ^{13}}-\frac{9216}{\rho ^{11}}+\frac{107520 M}{\rho ^9}\right) \\
 &&+ f[\rho]\left(192 \sqrt{3} Q \lambda  F_2'[\rho ] -\rho ^3  F_7'[\rho ]^2\right)  +\frac{16 \lambda  \left(27 Q^2+4 M \rho ^2\right) f[\rho ] F_7'[\rho ]}{\rho ^4} -\rho ^5 F_2'[\rho ]^2 \,, \\
r_+^2 P_5 &=& -\frac{4 \pi  T^2}{\sqrt{3}  (1+M)}\left(\rho^3F_1[\rho ]\right)' \,, \\
 P_6 &=& -\left(1+\frac{ \pi T}{ r_+(1+M)}\right)\frac{8\pi  Q T^3\left(\rho^3F_1[\rho ]\right)'}{3 r_+^3 (1+M)^2} -\frac{3 Q \rho ^2 F_1[\rho ]}{M} -2  \partial_Q\left(\rho^3F_1[\rho ]\right)' -\rho ^5 F_1'[\rho ]^2  +\frac{\sqrt{3} \left(2+Q^2\right) F_6'[\rho ]}{M \rho } -\rho ^3 f[\rho ] F_6'[\rho ]^2 \,, \\
r_+ P_7 &=& -2 \left(\rho^3F_3[\rho ]\right)'  \,, \\
 r_+^3 P_8 &=& \frac{6 \pi  Q T^2 \rho ^2 F_1[\rho ]}{r_+^2 M (1+M)} -\frac{4 \sqrt{3} \pi  Q T^2 \rho ^2 F_3[\rho ]}{ M (1+M)} -\frac{4 \sqrt{3} \pi  T^2 \rho ^2 \partial_QF_3[\rho ]}{ (1+M)} -\frac{2 \pi  Q T^2 \rho ^3 F_3'[\rho ]}{\sqrt{3}  M (1+M)}\\
\nonumber && -\frac{4 \pi  T^2 \rho ^5 F_1'[\rho ] F_3'[\rho ]}{\sqrt{3}  (1+M)} -\frac{4 \pi  T^2 \rho ^3 \partial_QF_3'[\rho ]}{\sqrt{3}  (1+M)} -\frac{4 \pi  T^2 \rho ^3 f[\rho ] F_6'[\rho ] F_8'[\rho ]}{\sqrt{3}  (1+M)} + \frac{2 \pi  \left(2+Q^2\right) T^2 F_8'[\rho ]}{ M (1+M) \rho } \,, \\
 r_+^2 P_9 &=& \frac{6 \sqrt{3} Q \rho ^2 F_3[\rho ]}{ M} + \frac{\sqrt{3} Q \rho ^3 F_3'[\rho ]}{ M} -\rho ^5 F_3'[\rho ]^2 + \frac{\left(3 Q^2-2 M \rho ^2\right) F_8'[\rho ]}{ M \rho } - \rho ^3 f[\rho ] F_8'[\rho ]^2 \,, \\
 \nonumber r_+^2 P_{10} &=& \frac{\lambda ^2 \left(-57600 Q^4+59904 M Q^2 \rho ^2-8448 M^2 \rho ^4-34560 Q^2 \rho ^6+5376 M \rho ^8\right)}{ \rho ^{13}} +\frac{1}{ \rho }-\frac{768 Q^2 \kappa  \lambda }{ \rho ^7} + 48 \lambda  \rho ^2 f[\rho ] F_4'[\rho ] - \rho ^5 F_4'[\rho ]^2\\
 &&  + \frac{112 \sqrt{3} Q \lambda  f[\rho ] F_9'[\rho ]}{\rho ^2} - \rho ^3 f[\rho ] F_9'[\rho ]^2 \,, \\
 \nonumber r_+ P_{11} &=& -\frac{256 \sqrt{3} Q \lambda  \left(30 Q^4 \lambda -36 M Q^2 \lambda  \rho ^2+6 M^2 \lambda  \rho ^4-3 Q^2 (\kappa -10 \lambda ) \rho ^6+2 M (\kappa -3 \lambda ) \rho ^8\right)}{\rho ^{15} f[\rho ]} -\frac{32 \sqrt{3} Q \lambda  \left(Q^2-2 \rho ^6\right) F_4'[\rho ]}{\rho ^6 f[\rho ]}\\
 &&  -\frac{32 \lambda  \left(-3 Q^2+2 M \rho ^2\right) \left(Q^2-2 \rho ^6\right) F_9'[\rho ]}{\text{b0} \rho ^{10} f[\rho ]} \,, \\
  \nonumber r_+ P_{12} &=& -\frac{2 \sqrt{3} Q \lambda ^2 \left(112704 Q^4-127296 M Q^2 \rho ^2+24320 M^2 \rho ^4+71232 Q^2 \rho ^6-16128 M \rho ^8\right)}{ \rho ^{15}} -\frac{2 \sqrt{3} Q \lambda  \left(576 Q^2 \kappa  \rho ^6+128 M \kappa  \rho ^8\right)}{\rho ^{15}}\\
\nonumber  &&  + \frac{2 \sqrt{3} Q}{ \rho ^3} + 48 \lambda  \rho ^2 f[\rho ] F_2'[\rho ] + \frac{2 \left(96 \sqrt{3} Q \lambda  \left(Q^2-M \rho ^2+\rho ^6\right)-\rho ^{11} F_2'[\rho ]\right) F_4'[\rho ]}{ \rho ^6} + \frac{112 \sqrt{3} Q \lambda  f[\rho ] F_7'[\rho ]}{ \rho ^2} \\
&& +  \frac{16 \lambda  \left(27 Q^2+4 M \rho ^2\right) f[\rho ] F_9'[\rho ]}{ \rho ^4} - 2 \rho ^3 f[\rho ] F_7'[\rho ] F_9'[\rho ] \,, 
\end{eqnarray}
}

\subsubsection{Anomalous tensorial sources}
{\tiny
\begin{eqnarray}
\tilde P_1 &=& \frac{64 \sqrt{3} Q \lambda }{\rho ^5} - 2 \left(\rho ^3 F_2[\rho ]\right)' \,,  \\
\nonumber r_+^2 \tilde P_2 &=& \frac{32 \pi  T^2 \lambda  \left(-\pi  T \left(51 Q^2+4 M \rho ^2\right)+12 r_+ M Q \left(F_1'[1]+2 \sqrt{3} F_6[1] Q\right) \rho ^3 f[\rho ]\right)}{r_+ M (1+M) \rho ^8} -\frac{2 \sqrt{3} \pi  Q T^2 \rho ^2 F_2[\rho ]}{ M\left(1+M\right)} -\frac{4 \sqrt{3} \pi  T^2 \rho ^2 \partial_QF_2[\rho ]}{(1+M)} \\
 && -\frac{768 \sqrt{3} \pi  Q^2 T^2 \lambda  f[\rho ] F_6[\rho ]}{ (1+M) \rho ^5} -\frac{2 \pi  T^2 \left(2 F_1'[1] M+4 \sqrt{3} F_6[1] M Q+5 Q \rho ^3\right) F_2'[\rho ]}{\sqrt{3}  M\left(1+M\right)} + \frac{8 \pi  Q T^2 F_6[\rho ] F_2'[\rho ]}{ (1+M)}\\
\nonumber&& -\frac{4 \pi  T^2 \rho ^3 \partial_Q F_2'[\rho ]}{\sqrt{3}  (1+M)} + \frac{32 \pi  T^2 \lambda  \left(27 Q^2+4 M \rho ^2\right) f[\rho ] F_6'[\rho ]}{\sqrt{3}  (1+M) \rho ^4} + \frac{4 \pi ^2 T^3 F_7'[\rho ]}{r_+ M\left(1+M\right) \rho }-\frac{4 \pi  T^2 \rho ^3 f[\rho ] F_6'[\rho ] F_7'[\rho ]}{\sqrt{3}  (1+M)} -\frac{2 \pi  Q T^2 \rho ^4 F_2''[\rho ]}{\sqrt{3} M \left(1+M\right)} \,, \\
\nonumber r_+^2\tilde P_3 &=& \frac{16 \pi  Q^6 T^2 \lambda  \left(-1+\rho ^2 \left(3-5 \rho +7 \rho ^3-4 \rho ^5\right)\right)}{ M (1+M) \rho ^{17} f[\rho ]^2} +\frac{64 \pi  T^2 \lambda  \left(2+\rho ^4 \left(-3+\rho ^3 \left(-5+5 \rho +\rho ^4\right)\right)\right)}{ M (1+M) \rho ^{13} f[\rho ]^2}\\
\nonumber &&+\frac{16 \pi  Q^4 T^2 \lambda  \left(-6+\rho ^2 \left(13+\rho  \left(-6+\rho  \left(8+\rho -33 \rho ^2-8 \rho ^3+14 \rho ^4+39 \rho ^5-22 \rho ^7\right)\right)\right)\right)}{ M (1+M) \rho ^{17} f[\rho ]^2}\\
\nonumber &&+\frac{32 \pi  Q^2 T^2 \lambda  \left(5+\rho ^2 \left(8+\rho  \left(-3+\rho  \left(-23+\rho  \left(-2+\rho +29 \rho ^2-21 \rho ^4+5 \rho ^5+\rho ^8\right)\right)\right)\right)\right)}{ M (1+M) \rho ^{15} f[\rho ]^2}\\
\nonumber &&+ \frac{128 \sqrt{3} \pi  T^2 \lambda  Q^2\left(Q^2-2  \rho ^6\right) F_6[\rho ]}{ (1+M) \rho ^{11} f[\rho ]} + \frac{32 \pi  Q T^2 \lambda  \left(-5 Q^2+3 M \rho ^2+\rho ^6\right) \partial_QF_5'[\rho ]}{ (1+M) \rho ^6}\\
\nonumber &&+ \frac{8 \pi  T^2 \lambda  \left(4 \rho ^6+3 Q^4 \left(-5+4 \rho ^2\right)+2 Q^2 \left(-5+6 \rho ^2+\rho ^6\right)\right) F_5'[\rho ]}{ M (1+M) \rho ^6}+ \frac{64 \pi  T^2 \lambda  \left(3 Q^2-2 M \rho ^2\right) \left(Q^2-2 \rho^6\right) F_6'[\rho ]}{\sqrt{3}  (1+M) \rho ^{10} f[\rho ]} \\
&&-\frac{64 F_1'[1] \pi  T^2 \lambda  Q\left(Q^2-2  \rho ^6\right)}{ (1+M) \rho ^{11} f[\rho ]} -\frac{128 \sqrt{3} F_6[1] \pi  T^2 \lambda Q^2 \left(Q^2-2  \rho ^6\right)}{ (1+M) \rho ^{11} f[\rho ]} + \frac{32 \pi  Q T^2 \lambda  \rho  f[\rho ] \partial_Q F_5''[\rho ]}{(1+M)} \,, \\
 r_+ \tilde P_4 &=& - 2 \left(\rho ^3 F_4[\rho ]\right)' \,, \\
\nonumber r_+^3\tilde P_5 &=& \frac{16 \sqrt{3} \pi  Q T^2 \lambda  \left(-18-11 Q^2+4 M \rho ^2\right)}{ M (1+M) \rho ^6} -\frac{4 \sqrt{3} \pi  Q T^2 \rho ^2 F_4[\rho ]}{ M\left(1+M\right)} -\frac{4 \sqrt{3} \pi  T^2 \rho ^2 \partial_QF_4[\rho ]}{ (1+M)}+ \frac{32 \sqrt{3} \pi  T^2 \lambda  \rho ^2 f[\rho ] F_1'[\rho ]}{(1+M)} -\frac{2 \pi  Q T^2 \rho ^3 F_4'[\rho ]}{\sqrt{3} M\left(1+M\right)} \\
&&-\frac{4 \pi  T^2 \rho ^5 F_1'[\rho ] F_4'[\rho ]}{\sqrt{3} (1+M)} -\frac{4 \pi  T^2 \rho ^3 \partial_QF_4'[\rho ]}{\sqrt{3}  (1+M)} +\frac{224 \pi  Q T^2 \lambda  f[\rho ] F_6'[\rho ]}{ (1+M) \rho ^2} + \frac{2 \pi  \left(2+Q^2\right) T^2 F_9'[\rho ]}{M \left(1+M\right) \rho } -\frac{4 \pi  T^2 \rho ^3 f[\rho ] F_6'[\rho ] F_9'[\rho ]}{\sqrt{3}  (1+M)} \,, \\
\nonumber r_+^2\tilde P_6 &=& \frac{56 \sqrt{3} Q \lambda  \left(-3 Q^2+2 M \rho ^2\right)}{ M \rho ^6} + \frac{6 \sqrt{3} Q \rho ^2 F_4[\rho ]}{M} + 48 \lambda  \rho ^2 f[\rho ] F_3'[\rho ] + \rho ^3 \left(\frac{\sqrt{3} Q}{M}-2 \rho ^2 F_3'[\rho ]\right) F_4'[\rho ]\\
&& +\frac{112 \sqrt{3} Q \lambda  f[\rho ] F_8'[\rho ]}{\rho ^2} + \frac{\left(3 Q^2-2 M \rho ^2\right) F_9'[\rho ]}{ M \rho } -2 \rho ^3 f[\rho ] F_8'[\rho ] F_9'[\rho ]  \,,\\
\nonumber r_+\tilde P_7 &=& -\frac{32 \sqrt{3} Q \lambda  \left(Q^2-2 \rho ^6\right) F_3'[\rho ]}{ \rho ^6 f[\rho ]} -\frac{12 \lambda  \left(4 \rho ^8+Q^4 \left(-5+2 \rho ^2\right)+2 Q^2 \left(\rho ^2-\rho ^6+2 \rho ^8\right)\right) F_5'[\rho ]}{ M \rho ^6}\\
\nonumber &&  -\frac{32 \lambda  \left(-3 Q^2+2 M \rho ^2\right) \left(Q^2-2 \rho ^6\right) F_8'[\rho ]}{ \rho ^{10} f[\rho ]} + \frac{12 \lambda  \left(-4 \rho ^8+Q^4 \left(-5+4 \rho ^2\right)+4 Q^2 \rho ^2 \left(1+\rho ^4-\rho ^6\right)\right) F_5''[\rho ]}{ M \rho ^5}\\
&& -\frac{4 \lambda  \left(-3 Q^2+2 M \rho ^2\right) \left(-18 Q^2 \rho ^2-4 M \rho ^4+\left(Q^2-M \rho ^2+\rho ^6\right)^2 F_5^{(3)}[\rho ]\right)}{ M \rho ^{10} f[\rho ]} \,, \\
\nonumber r_+ \tilde P_8 &=&\frac{3 \sqrt{3} Q \rho ^2 F_2[\rho ]}{ M}-2 \left(-96 \sqrt{3} Q \lambda  f[\rho ]+\rho ^5 F_2'[\rho ]\right) F_3'[\rho ]  -\frac{\left(-8 \lambda  \left(27 Q^2+4 M \rho ^2\right)+\rho ^7 F_7'[\rho ]\right) \left(-3 Q^2+2 M \rho ^2+2 M \rho ^4 f[\rho ] F_8'[\rho ]\right)}{ M \rho ^8} \,,\\
\end{eqnarray}
}

\section{Transport coefficients at second order}
\label{ap:fsecond_order}
In this appendix we will write the expressions for transport coefficients up to second order.
\subsection{Vector sector}

The solutions for the non anomalous coefficients $\xi_1 \,, \dots \,, \xi_{10}$  as written in (\ref{eq:xi5})-(\ref{eq:xi4}), are given in terms of functions $\xi_{i,(0,\kappa^2,\kappa\lambda,\lambda^2)}$  whose expressions are
{\footnotesize
\begin{eqnarray}
\xi_{2,\kappa\lambda}(\rho_2)  &=&  \frac{2 \pi T^3}{5 G M^3 (M + 1) Q^4 (1 + 2 \rho_2^2)^4r_+^3}  \Bigg( \bigg( 2430 - 14121 M + 32625 M^2 - 36279 M^3 + 17151 M^4  \nonumber \\
&&+ 286 M^5 -2648 M^6 + 656 M^7 \bigg) + \frac{10 M^3 (1 + 2 \rho_2^2)^4}{Q^2} (9 + 6 M - 7 M^2 + 2 M^3)  \log[1-\rho_2^2] \nonumber \\
&&+ \frac{10 M^2 \rho_2^6 (2 + \rho_2^2)}{Q^2 (1 + 2 \rho_2^2)} \Big( 9(1-\rho_2^2)  + 3 M (5 + 13 \rho_2^2)  - 6 M^2 (13 - 7 \rho_2^2) + 8 M^3 (17 - 14 \rho_2^2) \nonumber \\
&&- 32 M^4 (4 - \rho_2^2) +   48 M^5 \Big)  \log\left[\frac{2 + \rho_2^2}{1-\rho_2^2}\right]
\Bigg)   \,, \\
\xi_{2,\lambda^2}(\rho_2) &=&   \frac{\pi T^3}{210 G M^3 (M+1) \rho_2^8 (1 + 2 \rho_2^2)^4 r_+^3} 
\Bigg(
\frac{1}{ M^2 + Q^2 + \rho_2^2(2M-1)} \Big(-1837080 + 15089571 M \nonumber \\
&& - 54047817 M^2 + 109739475 M^3 - 136610865 M^4 + 102222345 M^5 - 35693475 M^6 \nonumber \\
&&- 7547847 M^7 + 12206741 M^8 - 3911944 M^9 + 427856 M^{10} \Big) 
+\frac{1680(1-\rho_2^2)}{\rho_2^2 (1 + 2 \rho_2^2)} \Big( 81(1+2\rho_2^2) \nonumber \\
&& - M (351 + 783 \rho_2^2) + 
 M^2 (540 + 1674 \rho_2^2)    + M^3 (225 - 1719 \rho_2^2) - M^4 (1497 - 672 \rho_2^2) \nonumber \\
&&+ M^5 (1332 - 12 \rho_2^2) - M^6 (340 - 32 \rho_2^2)  +  32 M^7  \Big) \log[1-\rho_2^2]
- \frac{1680 M^2 \rho_2^8(2+\rho_2^2)}{(1 + \rho_2^2)^5(1 + 2\rho_2^2)} \nonumber \\
&&\times\Big(  81 (1 + 2 \rho_2^2)-  27 M (16 + 29 \rho_2^2) + 54 M^2 (21 + 31 \rho_2^2) - 9 M^3 (216 + 191 \rho_2^2) \nonumber \\
&&+ 3 M^4 (723 + 224 \rho_2^2) - 12 M^5 (112 + \rho_2^2) + 4 M^6 (93 + 8 \rho_2^2)  - 32 M^7 \Big) \log\left[2 + \rho_2^2\right]
\Bigg)  \,, \\
\xi_{4,\kappa\lambda}(\rho_2)  &=& -\frac{ 2 \sqrt{3} }{5 \pi G M^2 Q^3 (1 + 2 \rho_2^2)^3} \Bigg( (1+2\rho_2^2) \big( 405 - 1944 M + 3708 M^2 - 3612 M^3 + 1903 M^4 - 507 M^5 \nonumber \\
&&+ 57 M^6 \big) 
+\frac{5 M^3 (M^2-3) (1 + 2 \rho_2^2)^2 }{Q^2 \rho_2^8 }\bigg(2 + \rho_2^2 - 
    5M + M^2 (3 - 2 \rho_2^2) \bigg)\log[1-\rho_2^2] \nonumber \\
&&+  \frac{5 M^3 \rho_2^6 (2 + \rho_2^2)^3}{Q^2} \bigg(  3 \rho_2^2  - 3M (1 + 2 \rho_2^2)  + 2M^2 \bigg)    \log\left[\frac{2 + \rho_2^2}{1-\rho_2^2}\right]
\Bigg)   \,, \\
\xi_{4,\lambda^2}(\rho_2) &=& \frac{1}{280\sqrt{3}\pi G Q^7 M^2 (1 + 2\rho_2^2)^2} \Bigg( 
\big( 1224720 - 8421651 M + 24887169 M^2 - 41179203 M^3 \nonumber \\
&&+ 41512749 M^4 - 25998933 M^5 + 10023123 M^6 - 2364497 M^7 + 330895 M^8 - 21092 M^9 \big) \nonumber \\
&&- \frac{3360 M^4 (1+2\rho_2^2)^2}{Q^2}\big(9 - 60 M + 105 M^2 - 70 M^3 + 20 M^4 - 2 M^5 \big)  \log[1-\rho_2^2] \nonumber \\
&&+ \frac{3360 M^4 \rho_2^8 (2 + \rho_2^2)^3 ( M \rho_2^2 - 3Q^2) }{Q^2 (1 + 2 \rho_2^2)} ( -3\rho_2^2  + 3 M (1 + 2 \rho_2^2) - 2 M^2 ) \log\left[\frac{2 + \rho_2^2}{1-\rho_2^2}\right]
\Bigg)  \,, \\
\xi_{5,0}(\rho_2) &=& \frac{3Q^2}{256 \pi G M^3(1 + 2 \rho_2^2)^2}  \Bigg(
(5 M-3 )(11 M - 15) + \frac{8 M^3 (1 + 2 \rho_2^2)^2}{3 Q^2} \log[1-\rho_2^2] \nonumber \\
&&-  \frac{2 M^3 \rho_2^2 (2 + \rho_2^2)^2 }{3 Q^2  (Q^2 + \rho_2^6) (1 + 2 \rho_2^2) }\big(9 \rho_2^2 + M (1 - 16 \rho_2^2)\big) \log\left[\frac{2 + \rho_2^2}{1-\rho_2^2}\right]
\Bigg)  \,,  \\
\xi_{6,0}(\rho_2)  &=&  -\frac{1}{512 \pi G (1 + Q^2)^3 (1 + 2 \rho_2^2)^2} \Bigg(
\big( 243 - 513 M + 189 M^2 - 39 M^3 + 152 M^4 \big)  \nonumber \\
\nonumber &&+ 8 M^3 (1 + 2 \rho_2^2)^2  \log[1-\rho_2^2]
- \frac{2M^2}{(1 + 2 \rho_2^2)} \big( 3\rho_2^2 + M(4 + \rho_2^2) \big) \big(15 (1 + \rho_2^2) - 16M \big)   \log\left[\frac{2 + \rho_2^2}{1-\rho_2^2}\right]
\Bigg)   \,,\\ 
\end{eqnarray}
}
{\footnotesize
\begin{eqnarray}
\xi_{6,\kappa\lambda}(\rho_2)  &=& \frac{1}{10 \pi G M^3 Q^2 (1 + 2 \rho_2^2)^2}    
\Bigg( \big( 2430 - 9477 M + 13824 M^2 - 9576 M^3 + 3444 M^4 - 591 M^5 - 14 M^6
\big) \nonumber \\
&&+\frac{20 M^3 (1 + 2 \rho_2^2)^2 }{Q^2} (18 - 27 M + 12 M^2 - M^3) \log[1-\rho_2^2]
+\frac{10 M^2 \rho_2^4 (2 + \rho_2^2)^3}{Q^2 (1 + 2 \rho_2^2)} \times \nonumber \\
&&\big(3 \rho_2^2 - 3 M (1 + 2 \rho_2^2) + 2 M^2 \big) \big(3 + 6 \rho_2^2 - M (4 + 5 \rho_2^2) \big)  \log\left[\frac{2 + \rho_2^2}{1-\rho_2^2}\right]
\Bigg)  \,, \\
\xi_{6,\lambda^2}(\rho_2) &=&  - \frac{1}{280 \pi G M^3 Q^6 (1 + 2 \rho_2^2)^2} \Bigg(
\Big( 612360 - 3485349 M + 7477191 M^2 - 6516837 M^3 - 444309 M^4 \nonumber \\
&&+ 5355933 M^5 - 4409403 M^6 + 1747097 M^7 - 359455 M^8 + 29492 M^9 \Big) 
+ \frac{3360 M^3 (1 + 2 \rho_2^2)^2 }{Q^2} \times \nonumber \\
&&\big(81 - 315 M + 453 M^2 - 303 M^3 + 101 M^4 - 16 M^5 + M^6\big) \log[1-\rho_2^2]
+\frac{1680 M^2 \rho_2^{10} (2 + \rho_2^2)^3}{Q^2 (1 + 2 \rho_2^2)}\times \nonumber \\
&&  \Big(3 \rho_2^2 -3 M (1 + 2 \rho_2^2) + 2 M^2 \Big) \Big(  9 (1 + 2 \rho_2^2) - 12 M (1 + 2 \rho_2^2)  + 3 M^2 (1 + \rho_2^2)  - M^3 \Big)   \log\left[\frac{2 + \rho_2^2}{1-\rho_2^2}\right]
\Bigg)
 \,, \nonumber\\
 &&\\
\xi_{7,0}(\rho_2) &=&  -\frac{1}{256 \pi G M^3} \frac{1}{(1+2\rho_2^2)^3} \Bigg( (32 + 224 Q^2 + 426 Q^4 + 443 Q^6 + 128 Q^8) (1 + 2 \rho_2^2)  \nonumber \\
&&+ M^2 \big(90  - 168 M + 82 M^2 + 4 M (1 + 2 \rho_2^2)^3 \big) \log\left[\frac{2 + \rho_2^2}{1-\rho_2^2}\right]  \Bigg)\,, \\
\xi_{8,\kappa^2}(\rho_2) &=& -\frac{\sqrt{3} \pi Q T^3}{2 G M^3 (1 + M) (1 + 2 \rho_2^2)^5r_+^3} \Bigg( (18 -69M + 63M^2 - 28M^3) (1 + 2 \rho_2^2)  \nonumber \\
&&+2 M (3 - M)(3 - M  + 4 M^2) \log\left[\frac{2 + \rho_2^2}{1-\rho_2^2}\right]    \Bigg) \,, \\
\xi_{8,\kappa\lambda}(\rho_2)  &=&  \frac{\sqrt{3}\pi T^3}{  G M^3 (1 + M) Q^5 (1 + 2\rho_2^3)^5r_+ ^3}  \Bigg( Q^2 \Big(216 - 1233 M + 2691 M^2 - 2823 M^3 + 1379 M^4 \nonumber \\
&&- 202 M^5 - 24 M^6\Big)(1 + 2 \rho_2^2)  +2 (3 - 4 M)^2 M^3 (1 + 2M - M^2 ) (1 + 2 \rho_2^2) \log[1-\rho_2^2] \nonumber \\
&&- M\Big(216  - 1062 M + 2094 M^2 - 1950 M^3 + 634 M^4 + 264 M^5 - 
 244 M^6 + 48 M^7 \nonumber \\
&&- 18 M^2 \rho_2^2 + 12 M^3 \rho_2^2 + 
 82 M^4 \rho_2^2 - 112 M^5 \rho_2^2 + 32 M^6 \rho_2^2\Big)  \log\left[\frac{2 + \rho_2^2}{1-\rho_2^2}\right]  \Bigg)\,,  \\
\xi_{8,\lambda^2}(\rho_2)  &=&  \frac{\pi T^3}{35 \sqrt{3} G M^3 (1 + M) Q^7(1 + 2 \rho_2^2)^4r_+^3} \Bigg( \Big(-102060 + 846855 M - 2955555 M^2 + 5599602 M^3 \nonumber \\
&&- 5989080 M^4 + 
 3085503 M^5 + 60033 M^6 - 797012 M^7 + 279586 M^8 - 32072 M^9\Big) \nonumber \\
&&+ \frac{420M^2}{Q^2(1+2\rho_2^2)}(3 - 4 M)^2 (18 - 93 M + 207 M^2 - 191 M^3 + 54 M^4 - 6 M^5 + M^6) \log[1-\rho_2^2] \nonumber \\
&&+  \frac{420M}{Q^2(1+2\rho_2^2)} \Big( 243 + 162 M (-11 + \rho_2^2)  - 27 M^2 (-207 + 47 \rho_2^2)
- M^3 (9243 - 4383 \rho_2^2) \nonumber \\
&&- 
 15 M^4 (-497 + 545 \rho_2^2)  - M^5 (372 - 8382 \rho_2^2) - 2M^6 (2243 + 2203 \rho_2^2) + 
 9 M^7 (404 + 113 \rho_2^2) \nonumber \\
&&- M^8 (1219 + 120 \rho_2^2)   + 
 M^9 (191 + 16 \rho_2^2)  - 12 M^{10}  \Big)\log\left[\frac{2 + \rho_2^2}{1-\rho_2^2}\right]   \Bigg) \,, 
\end{eqnarray}
\begin{eqnarray}
\xi_{9,\kappa^2}(\rho_2) &=&  \frac{-\sqrt{3}Q}{4 \pi G M^3 (1 + 2 \rho_2^2)^2} \Bigg(
9 (3 - 4 M + M^3) + 3 (3 - 4 M) M^2 \log[1-\rho_2^2]  \nonumber \\
&&-\frac{M}{2 (1 + 2 \rho_2^2)}\bigg( 54 -9 M (10 + 2 \rho_2^2) + 24 M^2 (2 + \rho_2^2)  + 4 M^3 )\bigg)  \log\left[\frac{2 + \rho_2^2}{1-\rho_2^2}\right]
 \Bigg) \,, \\
\xi_{9,\kappa\lambda}(\rho_2) &=& \frac{2 \sqrt{3}}{M^3 (-3 + 4 M) Q}  \Bigg(
\big(648 - 2079 M + 2475 M^2 - 1365 M^3 + 357 M^4 - 52 M^5 \big) \nonumber \\
&&+ \frac{M(-3 + 4 M)}{Q^2 (1 + 2 \rho_2^2)^3} \Big(2 (1 - \rho_2^2)^3 (1 + \rho_2^2) \big(-36 + 129 M - 
      129 M^2 + 28 M^3 \nonumber \\
&&+ \rho_2^2(36 -39M - 12 M^2 + 8 M^3) \big) + 
  2M \big(324 -1269M + 1935 M^2 - 1443 M^3 \nonumber \\
&&+ 537 M^4 - 92 M^5 + 
         8 M^6 + \rho_2^2M(-3 + 4 M) (-18 + 9 M + M^3) \big) \Big)   \log[1-\rho_2^2] \nonumber \\
&&+ \frac{2M}{Q^2 (1 + 2 \rho_2^2)^3)} \Big(-972 + 
   5103M - 10881 M^2 + 12069 M^3 - 7383 M^4 + 2424 M^5 \nonumber \\
&&- 392 M^6 + 32 M^7 + \rho_2^2 M(3 - 4 M)^2 (-18 + 9 M + M^3) \Big)  \log\left[\frac{2 + \rho_2^2}{1-\rho_2^2}\right] \Bigg)  \,, \\
\xi_{9,\lambda^2}(\rho_2)  &=& \frac{ \sqrt{3} \rho_2^2}{140\pi G M^3 Q^7 (1 + 2 \rho_2^2)^3}  \Bigg(  (-1 + 2 M + \rho_2^2) \Big(102060 - 496935 M + 963855 M^2 \nonumber \\
&&- 954012 M^3 + 518949 M^4 - 159036 M^5 + 27094 M^6 - 2411 M^7 + 16 M^8 \Big)  \nonumber \\
&&+\frac{420 M^2 (1 + 2 \rho_2^2) }{\rho_2^2} \Big(243 - 1269 M + 2673 M^2 - 2919 M^3 + 1779 M^4 - 610 M^5 \nonumber \\
&&+ 110 M^6 - 8 M^7\Big)  \log[1-\rho_2^2] + \nonumber \\
&&+ \frac{420 M}{\rho_2^2} \Big(243 - 243 M (6 - \rho_2^2)  + 27 M^2 (134 - 47 \rho_2^2) - 9 M^3 (529 - 297 \rho_2^2)  \nonumber \\
&&+ M^4 (3513 - 2919 \rho_2^2)  - 3 M^5 (456 - 593 \rho_2^2) + 
   M^6 (181 - 610 \rho_2^2) + 10 M^7 (5 + 11 \rho_2^2) \nonumber \\
&& -4 M^8 (5 + 2 \rho_2^2)    + 2 M^9  \Big)  \log\left[\frac{2 + \rho_2^2}{1-\rho_2^2}\right]   \Bigg)  \,, \\
\xi_{10,\kappa^2}(\rho_2)  &=&  -\frac{9 \rho_2^2 }{4 \pi G M^2  (1 + 2 \rho_2^2)^3} \Bigg( 
 (3 - 4 M + 3 M^2) (-1 + 2 M + \rho_2^2) 
 -\frac{2 M^2 (1 + 2 \rho_2^2)^3}{3 \rho_2^2}  \log[1-\rho_2^2] \nonumber \\
&&-\frac{2}{3} M^2 \Big(3 + 6 \rho_2^2 + M (5 + \rho_2^2) \Big)  \log\left[\frac{2 + \rho_2^2}{1-\rho_2^2}\right]
\Bigg) \,, \\
\xi_{10,\kappa\lambda}(\rho_2)  &=& \frac{3}{2 \pi G M^2 (1 + 2 \rho_2^2)^2} \Bigg(
\big(-108 + 333 M - 372 M^2 + 163 M^3 - 16 M^4\big)  \\
\nonumber&&+6 (3 - 4 M) M^2   \log[1-\rho_2^2]
+\frac{6 M^2 (6 - 5 M - M^2 + (3 - 4 M) \rho_2^2)}{1 + 2 \rho_2^2}   \log\left[\frac{2 + \rho_2^2}{1-\rho_2^2}\right]
 \Bigg)  \,, \\
\xi_{10,\lambda^2}(\rho_2) &=&  \frac{4 Q^{10}}{35 M^2(1 + 2 \rho_2^2)^2} \Bigg(
\Big( 51030 - 308205 M + 782505 M^2 - 1075848 M^3 + 866631 M^4 \nonumber \\
&&-421824 M^5 + 126626 M^6 - 22369 M^7 + 1874 M^8 \Big)
+  \frac{420 M^3 (1 - \rho_2^2)^3 (1 + \rho_2^2)^4}{\rho_2^2(1 + 
  2 \rho_2^2)} \nonumber \\
&&\times\big(9 - 3 M (7 + \rho_2^2)  + 6 M^2 (2 + \rho_2^2)  + M^3 \big) \log[1-\rho_2^2]
+ \frac{420 M^3 \rho_2^{10}}{Q^2 (1 + 2 \rho_2^2)} \big(27 + 54 \rho_2^2  \\
&&- 27 M (1 + 4 \rho_2^2) + 3 M^2 (-7 + 16 \rho_2^2)  + M^3 (6 - 12 \rho_2^2) -M^4 (1 - \rho_2^2)   \big)   \log\left[2 + \rho_2^2\right]
\Bigg) \,, \nonumber
\end{eqnarray}
}
and for the anomalous coefficients $\tilde{\xi}_1 \,, \dots \,, \tilde{\xi}_5$, (\ref{eq:xitilde2})-(\ref{eq:xitilde1}), one has functions $\tilde \xi_{i,(\kappa,\lambda)}$ that write
{\footnotesize
\begin{eqnarray}
\tilde \xi_{1,\lambda}(\rho_2) &=& \frac{2 \pi T^2}{G M^2 Q^2 (1 + 2 \rho_2^2)^2) r_+^2 } \Bigg(
\Big( 9 - 24 M + 14 M^2 + 3 M^3 \Big)
+ \frac{2 M^3}{Q^2} (1 + 2 \rho_2^2) \log[1-\rho_2^2] \nonumber \\
&&-\frac{M^2 \rho_2^2}{(1 + \rho_2^2)(1 + 2\rho_2^2)^2} \big( 3 \rho_2^2 - 
    3 M (1 + 2 \rho_2^2) + 2 M^2 \big)  \log\left[\frac{2 + \rho_2^2}{1-\rho_2^2}\right]
\Bigg) \,, \\
\tilde \xi_{2,\lambda}(\rho_2)  &=&  \frac{\sqrt{3}\pi T^2}{2 G M^2 Q (1 + 2 \rho_2^2)^2 r_+^2} \Bigg( \big(-6 + 7 M + M^2\big)
+ \frac{2 M^2}{Q^2} (1 + 2 \rho_2^2)^2 \log[1-\rho_2^2] \nonumber \\
&&-\frac{2 M \rho_2^4}{Q^2 (1 + 2 \rho_2^2)} (M^2 - 3 Q^2 \rho_2^2) \log\left[\frac{2 + \rho_2^2}{1-\rho_2^2}\right]
\Bigg)  \,, \\
\tilde \xi_{4,\kappa}(\rho_2) &=& \frac{3 \pi Q^2T^3 }{8 G M^4 (M + 1) (1 + 2 \rho_2^2)^4 r_+^3} \Bigg( 
(5 M-3) (15 - 14 M +  4 M^2) \nonumber \\
&&-\frac{2 M^2(M-3)}{3 Q^2 (1 + 2 \rho_2^2)} (3 - 16 M + 12 M^2)  \log\left[\frac{2 + \rho_2^2}{1-\rho_2^2}\right]
\Bigg)  \,, \\
\tilde \xi_{4,\lambda}(\rho_2)  &=& \frac{T^3}{4GM^4(M + 1) Q^2 (1 + 2 \rho_2^2)^4r_+^3 } \Bigg( 
\Big( 1215 - 6075 M + 11880 M^2 - 11367 M^3 + 5355 M^4 \nonumber \\
&&- 1096 M^5 + 100 M^6 \Big)
-\frac{2 \pi M^3(M -3) (2M + 1) (1 + 2 \rho_2^2)^4}{Q^2} \log[1-\rho_2^2]  \\
&&+\frac{2 \pi (M-3) M^3 \rho_2^4}{Q^2 (1 + 2 \rho_2^2)} \big( 9 (1 + \rho_2^2) + 3 M (1 + 6 \rho_2^2)  - 14 M^2 (1 + 2 \rho_2^2) + 4 M^3 \big) \log\left[\frac{2 + \rho_2^2}{1-\rho_2^2}\right]
\Bigg)\,,  \nonumber\\
\tilde \xi_{5,\kappa}(\rho_2) &=&  \frac{-3 \sqrt{3} \pi Q^3 T^2}{4 G M^3 (1 + 2 \rho_2^2)^2r_+^2} \Bigg(
1 + \frac{2 M^2}{3 Q^2 (1 + 2 \rho_2^2)}  \log\left[\frac{2 + \rho_2^2}{1-\rho_2^2}\right]
\Bigg) \,,  \label{eq:xi5tkappa}\\
\tilde \xi_{5,\lambda}(\rho_2)  &=& \frac{\sqrt{3} \pi T^2 }{2 G M^3 Q (1 + 2 \rho_2^2)^2 r_+^2} \Bigg(
\big(27 - 72 M + 54 M^2 - 7 M^3\big) \nonumber \\
&&+\frac{2 M^3 (4 M-3) (Q^2 - \rho_2^2)}{Q^2 \rho_2^4} \log[1-\rho_2^2]  
- \frac{2 M^3 \rho_2^2 ( M\rho_2^2 -3 Q^2 )}{Q^2 (1 + 2 \rho_2^2)} \log\left[\frac{2 + \rho_2^2}{1-\rho_2^2}\right]
\Bigg) \,. 
\end{eqnarray}
}

\subsection{Tensor sector}

In this sector the non anomalous coefficients $\Lambda_1 \,, \dots \,, \Lambda_{12}$  written in (\ref{eq:Lambda7})-(\ref{eq:Lambda6}) and the anomalous ones $\tilde{\Lambda}_1 \,, \dots \,, \tilde{\Lambda}_8$ (\ref{eq:Lambdatilde4})-(\ref{eq:Lambdatilde3}), are given in terms of functions $\Lambda_{i,(0,\kappa^2,\kappa\lambda,\lambda^2)}$ and $\Lambda_{i,(\kappa,\lambda)}$ respectively.  The expressions for these functions are in general very complicated, and we present here the result as an expansion at footnotesize $\rho_2$ up to order ${\cal O}(\rho_2^6)$, which is equivalent to order ${\cal O}(\bar{\mu}^6)$. For the non anomalous coefficients we get
{\footnotesize
\begin{eqnarray}
\Lambda_{4,\lambda^2}(\rho_2) &=&  \frac{64}{15\pi G} \Big( (-4 + 15 \log{2}) - \frac{1}{16} (557 + 840 \log{2}) \rho_2^2 - 
 \frac{3}{112} (2789 - 9660 \log{2}) \rho_2^4  + {\cal O}(\rho_2^6) \Big) \,, \nonumber\\
 &&\\
\Lambda_{6,0}(\rho_2)  &=& \frac{ 1}{384 \pi ^3G} \Big(  2(47-66 \log 2)\rho_2^2 -4 (1-3 \log 2)- 3(89-101 \log 2)\rho_2^4 + {\cal O}(\rho_2^6)\Big)  \,, \\
\Lambda_{8,0}(\rho_2) &=&  -\frac{1}{128\pi^2 G} \Big(4 + 2(1 - 14\log{2}) \rho_2^2 - (137 - 224 \log{2}) \rho_2^4 + {\cal O}(\rho_2^6) \Big) \,, \\
\Lambda_{9,0}(\rho_2)  &=& \frac{1}{768\pi G} \Big( 8 (11 + 3 \log{2}) + 12 (7 - 8\log{2}) \rho_2^2 - 3 (91 - 226 \log{2}) \rho_2^4 + {\cal O}(\rho_2^6)  \Big) \,, \\
\Lambda_{10,\kappa^2}(\rho_2) &=& \frac{1}{2\pi G} \Big( -12 (1 - 2 \log{2}) \rho_2^2 + 3 (25 - 36 \log{2}) \rho_2^4 + {\cal O}(\rho_2^6)  \Big) \,, \\
\Lambda_{10,\kappa\lambda}(\rho_2)  &=& \frac{1}{2\pi G}  \Big( 8 (5 - 12 \log{2}) \rho_2^2 - 3 (91 - 144 \log{2}) \rho_2^4   + {\cal O}(\rho_2^6) \Big)\,, \\
\Lambda_{10,\lambda^2}(\rho_2) &=& \frac{1}{5\pi G} \Big( -90 + 4 (29 + 60 \log{2}) \rho_2^2 + (617 - 1080 \log{2}) \rho_2^4   + {\cal O}(\rho_2^6)  \Big) \,, \\
\Lambda_{12,\kappa\lambda}(\rho_2)  &=&  \frac{2}{5\sqrt{3}\pi G} \Big( 
20(-5 + 12 \log{2}) \rho_2 + 5(71 - 192 \log{2}) \rho_2^3 - (2713 -3930\log{2})\rho_2^5 + {\cal O}(\rho_2^7) \Big) \,, \\
\Lambda_{12,\lambda^2}(\rho_2) &=& \frac{2}{35\sqrt{3} \pi G} \Big(
-56 (29 + 60 \log{2}) \rho_2 + 840 (1 + 16 \log{2}) \rho_2^3  + (37403 -51660  \log{2}) \rho_2^5 + {\cal O}(\rho_2^7)
\Big) \,, \nonumber\\
\end{eqnarray}
}
and for the anomalous coefficients
{\footnotesize
\begin{eqnarray}
\tilde \Lambda_{2,\lambda}(\rho_2) &=& \frac{1}{12\pi^2 G} \Big(-
24 (1 - \log{2}) + 4 (26 - 45 \log{2}) \rho_2^2 - (527 - 822 \log{2}) \rho_2^4 + {\cal O}(\rho_2^6)
 \Big) \,, \\
\tilde \Lambda_{5,\kappa}(\rho_2)  &=& \frac{\sqrt{3}}{32\pi^2 G} \Big(
-8(1 - 2 \log{2}) \rho_2 + 16 (5 - 4 \log{2}) \rho_2^3 - (389 - 566\log{2})\rho_2^5 + {\cal O}(\rho_2^7)
\Big) \,, \\
\tilde \Lambda_{5,\lambda}(\rho_2) &=& \frac{\sqrt{3}}{48\pi^2 G} \Big(-
  24(1 + 2 \log{2}) \rho_2 - 8(17 - 48 \log{2}) \rho_2^3 +3( 379 - 550\log{2} )\rho_2^5 + {\cal O}(\rho_2^7)
\Big) \,, \\
\tilde \Lambda_{6,\kappa}(\rho_2)  &=& \frac{\sqrt{3}}{16\pi G} \Big( -
8 (\log{2}) \rho_2 - 12(2 - 5 \log{2}) \rho_2^3 +(174 - 257\log{2})\rho_2^5 + {\cal O}(\rho_2^7) \Big) \,, \\
\tilde \Lambda_{6,\lambda}(\rho_2) &=&  \frac{\sqrt{3}}{48\pi G} \Big( 
24 (5 + 2 \log{2}) \rho_2 - 4(23 + 90 \log{2}) \rho_2^3 -  (991-1494\log{2})\rho_2^5 + {\cal O}(\rho_2^7) 
 \Big) \,, \\
\tilde \Lambda_{8,\lambda}(\rho_2)  &=& \frac{1}{12\pi G} \Big( 
-24 \log{2} + 4 (25 + 42 \log{2}) \rho_2^2 + (199 - 750 \log{2}) \rho_2^4 + {\cal O}(\rho_2^6)
\Big) \,.  
\end{eqnarray}
}
Note that in some cases the order ${\cal O}(\rho_2^6)$ vanishes, so that the corresponding expressions are valid up to ${\cal O}(\rho_2^7)$.

\hfill\newpage

\begin{acknowledgments}
We would like to thank K.~Landsteiner and L.~Melgar for enlightening
discussions and collaboration on related topics. We are also grateful
to A.~Buchel, S.~Dutta, J.~Erdmenger, D.~Grumiller, M.~Kaminski,
A.~Karch, D.~Kharzeev, R.~Loganayagam, D.~Mateos, S.~Minwalla,
D.T.~Son, P.~Surowka, A.~Yarom and H.~Yee for discussions. This work
has been supported by Plan Nacional de Altas Energ\'{\i}as
(FPA2009-07908, FPA2008-01430 and FPA2011-25948), Spanish MICINN
Consolider-Ingenio 2010 Program CPAN (CSD2007-00042), Comunidad de
Madrid HEP-HACOS S2009/ESP-1473. E.M. would like to thank the
Instituto de F\'{\i}sica Te\'orica UAM/CSIC, Spain, the Institute for
Nuclear Theory at the University of Washington, USA, and the Institut
f\"ur Theoretische Physik at the Technische Universit\"at Wien,
Austria, for their hospitality and partial support during the
completion of this work. E.M.~thanks also the organizers and all the
participants in the INT Program 11-3 ``Frontiers in QCD'' of the
University of Washington, for the stimulating atmosphere and fruitful
discussions. F.P. thanks the Max Planck Institute
(Werner-Heisenberg-Institut), Munich, for their hospitality during his
research visit in the first stage of this work, and the Institut de
F\'{\i}sica d'Altes Energies, Universitat Aut\`onoma de Barcelona, Spain, for
hospitality. F.P. also acknowledges Brookhaven National Laboratory and
the organizers of the workshop ``P and CP odd effect in hot and dense
matter 2012" for financial support, and for the stimulating atmosphere
and fruitful discussions.  The research of E.M. is supported by the
Juan de la Cierva Program of the Spanish MICINN. F.P. has been
supported by fellowship CPI Comunidad de Madrid.
\end{acknowledgments}

\bibliographystyle{JHEP}
\bibliography{HolofluidGravAnomal}

\end{document}